\documentclass[11pt,english]{article}
\usepackage[T1]{fontenc}
\usepackage[latin9]{inputenc}
\usepackage{amsmath}
\usepackage{amssymb}
\usepackage{geometry}
\makeatletter
\usepackage{amsmath}
\usepackage{amsthm}
\usepackage{amssymb}

\theoremstyle{plain}
\newtheorem{theorem}{Theorem}[section]

\theoremstyle{plain}
\newtheorem{theorem-seq}{Theorem}

\newenvironment{myproof}[1][Proof.]{\par
	\pushQED{\qed}%
	\normalfont \topsep6\p@\@plus6\p@\relax
	\trivlist
		\item[\hskip\labelsep
		\bfseries
	#1]\ignorespaces
	}{%
	\popQED\endtrivlist\@endpefalse
}


\newcommand{\eqdef}{\stackrel{\rm{def}}{=}}
\DeclareMathOperator{\poly}{poly}

\newcommand{\N}{\mathbb{N}}

\newcommand{\R}{\mathbb{R}}
\newcommand{\F}{\mathbb{F}}

\ifx \C \undefined
\newcommand{\C}{\mathbb{C}}
\else
\renewcommand{\C}{\mathbb{C}}
\fi

\newcommand{\B}{\left\{ 0,1 \right \}}

\renewcommand{\P}{\mathbf{P}}

\renewcommand{\L}{\mathbf{L}}

\newcommand{\NC}{\mathbf{NC}}
\newcommand{\NCo}{\NC^1}

\usepackage{crossreftools}
\usepackage{cleveref}
\usepackage{aliascnt}
\crefalias{theorem-seq}{theorem}
\newcommand{\oldref}{\labelcref}
\renewcommand{\ref}{\Cref}
\crefname{enumi}{}{}
\crefname{enumii}{}{}
\crefname{enumiii}{}{}

\newaliascnt{conjecture-cnt}{theorem}
    \theoremstyle{plain}    
    \newtheorem{conjecture}[conjecture-cnt]{Conjecture} 
\crefname{conjecture-cnt}{conjecture}{conjectures}
\newaliascnt{definition-cnt}{theorem}
\theoremstyle{definition}
\newtheorem{definition}[definition-cnt]{Definition}
\crefname{definition-cnt}{definition}{definitions}
\newaliascnt{fact-cnt}{theorem}
    \theoremstyle{plain}    
    \newtheorem{fact}[fact-cnt]{Fact}
\crefname{fact-cnt}{fact}{facts}
\newaliascnt{claim-cnt}{theorem}
    \theoremstyle{plain}    
    \newtheorem{claim}[claim-cnt]{Claim}
\crefname{claim-cnt}{claim}{claims}
\newaliascnt{corollary-cnt}{theorem}
	\theoremstyle{plain}    
\newtheorem{corollary}[corollary-cnt]{Corollary}  %%Delete [thm] to re-start numbering
	\crefname{corollary-cnt}{corollary}{corollaries}
\newaliascnt{remark-cnt}{theorem}
    \theoremstyle{definition}
    \newtheorem{remark}[remark-cnt]{Remark}
\crefname{remark-cnt}{remark}{remarks}
\newaliascnt{proposition-cnt}{theorem}
    \theoremstyle{plain}    
    \newtheorem{proposition}[proposition-cnt]{Proposition} %%Delete [theorem] to re-start numbering
\crefname{proposition-cnt}{proposition}{propositions}
\newaliascnt{lemma-cnt}{theorem}
\theoremstyle{plain}    
\newtheorem{lemma}[lemma-cnt]{Lemma}  %%Delete [theorem] to re-start numbering
	\crefname{lemma-cnt}{lemma}{lemmas}
\theoremstyle{plain}
\newtheorem*{restated@theorem}{\rep@title}

\newenvironment{restated}[1]{%
	 \def\rep@title{#1}%
	 \begin{restated@theorem}}%
	{\end{restated@theorem}}

\usepackage{bbm}

\makeatother

\usepackage{babel}
\begin{document}
\title{KRW Composition Theorems via Lifting
\global\long\def\d{\diamond}%
\global\long\def\dm{\circledast}%
\global\long\def\fg{f\d g}%
\global\long\def\KW{\textit{KW}}%
\global\long\def\mKW{\textit{mKW}}%
\global\long\def\pnc{\P\not\subseteq\NCo}%
\global\long\def\MX{\textit{MUX}}%
\global\long\def\C{\mathsf{CC}}%
\global\long\def\Q{\mathsf{Q}}%
\global\long\def\Dpt{\mathsf{D}}%
\global\long\def\L{\mathsf{L}}%
\global\long\def\mD{\mathsf{mD}}%
\global\long\def\mL{\mathsf{mL}}%
\global\long\def\cI{\mathcal{I}}%
\global\long\def\cO{\mathcal{O}}%
\global\long\def\D{\Lambda}%
\global\long\def\gd{\mathrm{gd}}%
\global\long\def\eq{\mathrm{eq}}%
\global\long\def\ip{\mathrm{ip}}%
\global\long\def\cA{\mathcal{A}}%
\global\long\def\cB{\mathcal{B}}%
\global\long\def\cX{\mathcal{X}}%
\global\long\def\cY{\mathcal{Y}}%
\global\long\def\cU{\mathcal{U}}%
\global\long\def\cV{\mathcal{V}}%
\global\long\def\cW{\mathcal{W}}%
\global\long\def\cV{\mathcal{V}}%
\global\long\def\cT{\mathcal{T}}%
\global\long\def\rout{R_{\text{out}}}%
\global\long\def\rz{\mu_{\F_{2}}}%
\global\long\def\rzf{\mu_{\F}}%
\global\long\def\rk{\mathrm{rank}_{\F_{2}}}%
\global\long\def\rkf{\mathrm{rank}_{\F}}%
\global\long\def\NS{NS_{\F_{2}}}%
\global\long\def\NSF{NS_{\F}}%
\global\long\def\disc{{\rm disc}}%
\global\long\def\Hm{H_{\infty}}%
\global\long\def\cE{\mathcal{E}}%
\global\long\def\sgd{S_{\gd}}%
\global\long\def\fsgd{\mKW_{f}\dm\sgd}%
\global\long\def\seq{S_{\phi}\d\eq}%
\global\long\def\xy{\mKW_{\cX\times\cY}}%
\global\long\def\uxy{U_{m}\d\mKW_{\cX\times\cY}}%
\global\long\def\AD{\mathrm{AvgDeg}}%
\global\long\def\rAD{\mathrm{rAvgDeg}}%
\global\long\def\dns{\mathrm{density}}%
\global\long\def\x{\boldsymbol{x}}%
\global\long\def\y{\boldsymbol{y}}%
\global\long\def\z{\boldsymbol{z}}%
\global\long\def\v{\boldsymbol{v}}%
\global\long\def\w{\boldsymbol{w}}%
\global\long\def\fix{\mathrm{fix}}%
\global\long\def\free{\mathrm{free}}%
\global\long\def\one{\mathbbm{1}}%
\global\long\def\csat{\textsc{CspSat}}%
\global\long\def\gr{\mathrm{graph}}%
\global\long\def\stc{\textsc{stConn}_{n}}%
\global\long\def\clq{\textsc{Clique}_{n,k}}%
\global\long\def\gen{\textsc{Gen}_{n}}%
\global\long\def\ind{\mathrm{Ind}_{\ell}}%
\global\long\def\php{\mathrm{bitPHP}_{d}}%
\global\long\def\peb{\mathrm{Peb}_{\Delta_{h}}}%
}
\author{Susanna F. de Rezende\thanks{Institute of Mathematics of the Czech Academy of Sciences. Research
supported by the European Research Council under the European Union's
Seventh Framework Programme (FP7/2007--2013) ERC grant agreement
no.~279611, as well as by the Knut and Alice Wallenberg grants KAW
2016.0066 and KAW 2018.0371.}\and Or Meir\thanks{Department of Computer Science, University of Haifa, Haifa 3498838,
Israel. \texttt{ormeir@cs.haifa.ac.il}. Research supported by the
Israel Science Foundation (grant No. 1445/16).}\and Jakob Nordstr\"{o}m\thanks{University of Copenhagen and Lund University. Research supported by
the Swedish Research Council grant \mbox{2016-00782}, the Knut and
Alice Wallenberg grant \mbox{KAW 2016.006}, and the Independent Research
Fund Denmark grant \mbox{9040-00389B}. }\and Toniann Pitassi\thanks{Department of Computer Science, University of Toronto, Canada, and
Institute of Advanced Study, Princeton, USA. \texttt{toni@cs.toronto.edu}.
Research supported by NSERC and by NSF CCF grant 1900460}\and Robert Robere\thanks{McGill University, Canada. This research was performed while Robert
Robere was a postdoctoral researcher at DIMACS and the Institute for
Advanced Study. Robert Robere was supported by NSERC, the Charles
Simonyi Endowment, and indirectly supported by the National Science
Foundation Grant No. CCF-1900460. Any opinions, findings and conclusions
or recommendations expressed in this material are those of the author(s)
and do not necessarily reflect the views of the National Science Foundation. }}
\maketitle
\begin{abstract}
One of the major open problems in complexity theory is proving super-logarithmic
lower bounds on the depth of circuits (i.e., $\P\not\subseteq\NCo$).
Karchmer, Raz, and Wigderson \cite{KRW95} suggested to approach this
problem by proving that depth complexity behaves ``as expected''
with respect to the composition of functions~$f\d g$. They showed
that the validity of this conjecture would imply that $\P\not\subseteq\NCo$.

Several works have made progress toward resolving this conjecture
by proving special cases. In particular, these works proved the KRW
conjecture for every outer function~$f$, but only for few inner
functions~$g$. Thus, it is an important challenge to prove the KRW
conjecture for a wider range of inner functions.

In this work, we extend significantly the range of inner functions
that can be handled. First, we consider the \emph{monotone} version
of the KRW conjecture. We prove it for every monotone inner function~$g$
whose depth complexity can be lower bounded via a query-to-communication
lifting theorem. This allows us to handle several new and well-studied
functions such as the $s\text{\textbf{-}}t$-connectivity, clique,
and generation functions.

In order to carry this progress back to the \emph{non-monotone} setting,
we introduce a new notion of \emph{semi-monotone} composition, which
combines the non-monotone complexity of the outer function~$f$ with
the monotone complexity of the inner function~$g$. In this setting,
we prove the KRW conjecture for a similar selection of inner functions~$g$,
but only for a specific choice of the outer function~$f$.\newpage{}
\end{abstract}
\tableofcontents{}

\newpage{}

\section{\label{sec:Introduction}Introduction}

A major frontier of the research on circuit complexity is proving
super-logarithmic lower bounds on the depth complexity of an explicit
function, i.e., proving that $\pnc$. This question is an important
milestone toward proving lower bounds on general circuits, and also
captures the natural question of whether there are tractable computational
tasks that cannot be parallelized. The state of the art is the work
of H{\aa}stad~\cite{H98}, which proved a lower bound of $(3-o(1))\cdot\log n$,
following a long line of work~\cite{S61,K72,A87,PZ93,IN93}. This
lower bound has not been improved for more than two decades except
for the lower order terms~\cite{T14}, and it is an important problem
to break this barrier.

Karchmer, Raz, and Wigderson~\cite{KRW95} proposed to approach this
problem by studying the (block-)composition of Boolean functions,
defined as follows: if $f:\B^{m}\to\B$ and $g:\B^{n}\to\B$ are Boolean
functions, then their composition $\fg$ takes inputs in $\left(\B^{n}\right)^{m}$
and is defined by
\begin{equation}
\fg(x_{1},\ldots,x_{m})=f\left(g(x_{1}),\ldots,g(x_{m})\right).\label{eq:composition}
\end{equation}
Let us denote by $\Dpt(f)$ the minimal depth of a circuit with fan-in~$2$
that computes~$f$. The circuit that computes $\fg$ using \ref{eq:composition}
has depth $\Dpt(f)+\Dpt(g)$. Karchmer et al.~\cite{KRW95} conjectured
that this upper bound is roughly optimal:
\begin{conjecture}[The KRW conjecture]
\label{KRW}Let $f:\B^{m}\to\B$ and $g:\B^{n}\to\B$ be non-constant
functions. Then 
\begin{equation}
\Dpt(\fg)\approx\Dpt(f)+\Dpt(g).\label{eq:KRW}
\end{equation}
\end{conjecture}

\noindent Karchmer et al. observed that their conjecture, if proved,
would imply that $\pnc$. They also successfully used this approach
to give an alternative proof for $\pnc$ in the monotone setting.
The meaning of ``approximate equality'' in \ref{eq:KRW} is intentionally
left vague, since there are many variants that would imply the separation.

While we are still far from resolving the KRW conjecture, several
works \cite{KRW95,EIRS01,HW93,H98,GMWW17,DM18,KM18} have made progress
toward it by proving special cases. The state of the art is that the
KRW conjecture is known to hold for every outer function~$f$, but
only when combined with two specific choices of the inner function~$g$:
the parity function, and the universal relation. There are no results
proving the KRW conjecture for a broader family of inner functions.

In this work, we prove the KRW conjecture for a rich family of inner
functions~$g$, namely, those functions whose depth complexity can
be lower bounded using \emph{lifting theorems}. This includes functions
that are considerably more interesting than previous composition theorems
could handle. We prove these results in the \emph{monotone} setting,
and in a new setting which we call the \emph{semi-monotone} setting.
Below, we discuss the background to this work and present our results. 

\paragraph*{Karchmer-Wigderson relations.}

It is useful to study the KRW conjecture through the lens of communication
complexity, and in particular, using the framework of \emph{Karchmer-Wigderson
relations}. Let us denote the (deterministic) communication complexity
of a problem~$R$ by $\C(R)$. The \emph{Karchmer-Wigderson relation}
of a function~$f:\B^{n}\to\B$, denoted $\KW_{f}$, is the communication
problem in which the inputs of Alice and Bob are $x\in f^{-1}(1)$
and $y\in f^{-1}(0)$ respectively, and their goal is to find a coordinate~$i$
such that $x_{i}\ne y_{i}$. Karchmer and Wigderson~\cite{KW90}
observed that~$\Dpt(f)=\C(\KW_{f})$. This connection between functions
and communication problems allows us to study the depth complexity
of functions using techniques from communication complexity.

\paragraph*{The KRW conjecture from the KW perspective.}

Let $f:\B^{m}\to\B$ and $g:\B^{n}\to\B$ be non-constant functions.
It will be useful to denote the KW relation~$\KW_{\fg}$ of the composed
function by $\KW_{f}\d\KW_{g}$. In this relation, Alice and Bob get
$X\in(\fg)^{-1}(1)$ and $Y\in(\fg)^{-1}(0)$, viewed as $m\times n$
matrices, and their goal is to find an entry $(i,j)$ such that $X_{i,j}\ne Y_{i,j}$.
The KRW conjecture can be restated as:
\[
\C(\KW_{f}\d\KW_{g})\approx\C(\KW_{f})+\C(\KW_{g}).
\]
It is worth noting the obvious protocol for solving $\KW_{f}\d\KW_{g}$:
Let $a,b$ be the column vectors that are obtained from applying $g$
to the rows of~$X,Y$, and observe that they constitute an instance
of~$\KW_{f}$. The players begin by solving $\KW_{f}$ on $a$ and~$b$,
thus obtaining a coordinate~$i\in\left[m\right]$ such that $a_{i}\ne b_{i}$.
Then, they solve $\KW_{g}$ on the rows $X_{i},Y_{i}$, which constitute
an instance of~$\KW_{g}$, thus obtaining a coordinate $j\in\left[n\right]$
where $X_{i,j}\ne Y_{i,j}$. The communication complexity of this
protocol is $\C(\KW_{f})+\C(\KW_{g})$, and the KRW conjecture says
that this obvious protocol is roughly optimal.

\paragraph*{Previous work on the KRW conjecture.}

The KRW conjecture has been studied extensively, and a long line of
papers have made progress on important restricted cases. These papers
can be broadly divided into two categories.

The first category involves proving the KRW conjecture for a simplified
communication problem. Specifically, Karchmer et al. \cite{KRW95}
proposed a simplification of KW relations called the \emph{universal
relation} (denoted $U_{n}$) which is the following communication
problem: Alice and Bob get two \emph{distinct} strings $x,y\in\B^{n}$,
and their goal is to find a coordinate on which they disagree. The
universal relation is harder to solve than KW relations, since the
inputs of Alice and Bob are not assumed to come from the preimage
of some function~$f$, and so the protocol cannot take advantage
of any properties of~$f$. Just as the universal relation is a simplified
version of KW relations, one can define simplified versions of $\KW_{f}\d\KW_{g}$,
such as the composition $U_{m}\d U_{n}$ of two universal relations
and the composition $\KW_{f}\d U_{n}$ of a KW relation and a function.
Several works have studied this type of compositions \cite{KRW95,EIRS01,HW93,GMWW17,KM18},
and the state of the art is that the KRW conjecture holds for $\KW_{f}\d U_{n}$
for every non-constant function~$f:\B^{m}\to\B$ \cite{GMWW17,KM18}.

The second category where important progress was made is for $\KW_{f}\d\KW_{\bigoplus}$
where $f$ can be any non-constant function and $\bigoplus$ is the
parity function. The KRW conjecture for this case has been proved
implicitly by H{\aa}stad~\cite{H98}, and an alternative proof was
recently given by Dinur and Meir~\cite{DM18}.

The papers discussed so far are able to handle an arbitrary choice
of the outer relation~$\KW_{f}$, but only very specific choices
of the inner relation~$\KW_{g}$. This seems to suggest that the
crux of the difficulty in proving the KRW conjecture lies in having
to deal with an arbitrary choice of~$\KW_{g}$. In order to bypass
this difficulty, Meir \cite{M20} recently observed that in order
to prove that $\pnc$, it suffices to prove a version of the KRW conjecture
in which $\KW_{g}$ is replaced with a specific communication problem,
namely, the \emph{multiplexor relation}~$\MX$ of \cite{EIRS01}.
Specifically, he defined a composition of the form $\KW_{f}\d\MX$,
and showed that if a variant of the KRW conjecture for $\KW_{f}\d\MX$
holds for every non-constant outer function~$f$, then $\pnc$.

\paragraph*{Motivation.}

Following the above discussion, our goal is to ``replace'' the relations
$U_{n}$ and $\KW_{\bigoplus}$ in the known results with $\MX$.
Unfortunately, this seems to be very difficult --- in particular,
the relation $\MX$ seems to be significantly more complicated than
$U_{n}$ and $\KW_{\bigoplus}$.

In order to make progress, we propose that a good intermediate goal
would be to try to prove the KRW conjecture for the composition $\KW_{f}\d\KW_{g}$
\emph{for inner functions~$g$ that are as complex and expressive
as possible}. Ideally, by extending the range of inner functions~$g$
that we can handle, we will develop stronger techniques, which would
eventually allow us to prove the conjecture for $\KW_{f}\d\MX$.

An additional motivation for proving the KRW conjecture for harder
inner functions is that it may allow us to improve the state of the
art lower bounds on depth complexity. The best known lower bound of
$(3-o(1))\cdot\log n$ \cite{A87,PZ93,IN93,H98} was achieved by implicitly
proving the KRW conjecture for $\KW_{f}\d\KW_{\bigoplus}$, and it
may be improved by proving the KRW conjecture for new inner functions.

The question is, which inner functions~$g$ would be good candidates
for such a program? Ideally, a good candidate for~$g$ would be such
that the KW relation $\KW_{g}$ is more interesting than $U_{n}$
and $\KW_{\bigoplus}$, but less complicated than $\MX$. Unfortunately,
there are not too many examples for such relations: in fact, the relations
$U_{n}$, $\KW_{\bigoplus}$, and $\MX$ are more or less the only
relations that are well-understood. Thus, we have a shortage of good
candidates~$g$ for this program.

As a way out of this shortage, we propose to consider \emph{monotone
depth complexity} in the study of inner functions. Given a \emph{monotone}
function~$f$, the \emph{monotone depth complexity} \emph{of~}$f$,
denoted~$\mD(f)$, is the minimal depth of a \emph{monotone }circuit
that computes~$f$. The \emph{monotone KW relation} of a monotone
function~$f$, denoted $\mKW_{f}$, is defined similarly to $\KW_{f}$,
but this time the goal of Alice and Bob is to find a coordinate~$i$
such that $x_{i}>y_{i}$ (rather than $x_{i}\ne y_{i}$). Karchmer
and Wigderson~\cite{KW90} observed that $\mD(f)=\C(\mKW_{f})$.

Fortunately, there are many monotone KW relations that are well-understood,
and which are significantly more interesting than $U_{n}$ and $\KW_{\bigoplus}$.
We would like to study compositions in which these monotone KW relations
serve as the ``inner part'', in the hope that such study would lead
us to discover new techniques.

\subsection{\label{subsec:Intro-Our-results}Our results}

\subsubsection{\label{subsec:Intro-Monotone-composition-theorem}The monotone composition
theorem}

Motivated by considerations discussed above, our first result concerns
the \emph{monotone KRW conjecture}. This conjecture says that for
every two non-constant monotone functions $f,g$ it holds that 
\[
\C(\mKW_{f}\d\mKW_{g})\approx\C(\mKW_{f})+\C(\mKW_{g})
\]
(where $\mKW_{f}\d\mKW_{g}\eqdef\mKW_{\fg}$). This conjecture was
studied in the original paper of Karchmer et al.~\cite{KRW95}, who
proved it for the case where both $f$ and~$g$ are the set-cover
function, and used the latter result to prove that $\pnc$ in the
monotone setting. However, this conjecture received far less attention
than the non-monotone conjecture, perhaps because the monotone analogue
of $\pnc$ has been known to hold for a long time, and monotone depth
complexity is considered to be very well understood in general.

Nevertheless, we believe that this conjecture is interesting for several
reasons: First, it is a very natural question in its own right. Second,
if we cannot prove the KRW conjecture in the monotone setting, what
hope do we have to prove it in the non-monotone setting, which is
far less understood? Finally, proving the monotone KRW conjecture
might prove useful for tackling other important questions on monotone
depth complexity, such as proving lower bounds on slice functions
(which in particular would imply non-monotone lower bounds).

Our first main result is a proof of the monotone KRW conjecture for
every non-constant monotone function~$f$, and for a wide range of
monotone functions~$g$. Specifically, our result holds for every
function~$g$ whose monotone depth complexity can be lower bounded
using a ``lifting theorem'': A \emph{lifted search problem} $S\d\gd$
is obtained by composing a search problem~$S$ with an appropriate
``gadget'' function $\gd$. A \emph{lifting theorem} is a theorem
that translates a lower bound for $S$ in a weak model of computation
to a lower bound for~$S\d\gd$ in a strong model. 

Here, the relevant weak model of computation is query complexity.
Informally, the \emph{query complexity} of a search problem~$S$,
denoted~$\Q(S)$, is the number of queries one should make to the
input in order to find a solution (see \ref{subsec:Preliminaris-Decision-trees}
for a formal definition). Fix a gadget~$\gd:\B^{t}\times\B^{t}\to\B$
of input length~$t$. Several lifting theorems~\cite{RM99,CKLM19,WYY17,CFKMP19}
establish that if the gadget $\gd$ satisfies certain conditions,
then $\C(S\d\gd)=\Omega(\Q(S)\cdot t)$. In this work, we use a lifting
theorem of Chattopadhyay et al. \cite{CFKMP19}, which holds for every
gadget $\gd$ that has sufficiently low discrepancy and sufficiently
large input length (see \ref{lifting-query-thm} for the formal statement).

Our result says that the monotone KRW conjecture holds whenever the
lower bound on $\mKW_{g}$ can be proved using the theorem of~\cite{CFKMP19}.
More specifically, there should exist a reduction to $\mKW_{g}$ from
a lifted search problem~$S\d\gd$ that satisfies the conditions of~\cite{CFKMP19}.
This is a much wider family of inner functions than what previous
composition theorems could handle (i.e., universal relation and parity),
though we are now working in the monotone rather than the non-monotone
setting. Informally, the composition theorem can be stated as follows
(see \ref{monotone-composition-theorem} for the formal statement):
\begin{theorem}[monotone composition theorem, informal]
\label{monotone-composition-informal}Let $f:\B^{m}\to\B$ and $g:\B^{n}\to\B$
be non-constant monotone functions. If there is a lifted search problem
$S\d\gd$ that reduces to $\mKW_{g}$ and satisfies the conditions
of the theorem of~\cite{CFKMP19}, then 
\[
\C(\mKW_{f}\d\mKW_{g})\ge\C(\mKW_{f})+\Omega(\Q(S)\cdot t).
\]
In particular, if $\C(\mKW_{g})=\tilde{O}\left(\Q(S)\cdot t\right)$,
then
\begin{equation}
\C(\mKW_{f}\d\mKW_{g})\ge\C(\mKW_{f})+\tilde{\Omega}(\C(\mKW_{g})).\label{eq:monotone-composition-informal-in-particular}
\end{equation}
\end{theorem}

We would like to note that the theorem is applicable to many interesting
inner functions, including the classical $s\text{-}t$-connectivity
function~\cite{KW90,GS91}, clique function~\cite{GH92,RW92}, and
generation function \cite{RM99} (see \ref{sec:classic-functions}
for details). Moreover, we would like to mention that the bound of
\ref{eq:monotone-composition-informal-in-particular} is good enough
for the purposes of the KRW conjecture.

We would also like to stress that while the statement of our monotone
composition theorem refers to the lifting theorem of \cite{CFKMP19},
we believe it can be adapted to work with similar lifting theorems
such as the ones of \cite{RM99,CKLM19,WYY17} (in other words, the
specific choice of the lifting theorem is not particularly crucial).
Finally, it should be mentioned that the formal statement of the monotone
composition theorem actually refers to formula complexity rather than
depth complexity.

In order to prove \ref{monotone-composition-informal}, we introduce
a generalization of the lifting theorem of~\cite{CFKMP19}, which
may be of independent interest. Roughly, our generalization shows
a lower bound for the lifted problem~$S\d\gd$ even when restricted
to a subset of its inputs, as long as this subset satisfies a certain
condition. See \ref{subsec:Intro-Our-techniques-monotone} for further
discussion.

\subsubsection{\label{subsec:Intro-Semi-monotone-composition}The semi-monotone
composition theorem}

Recall that our end goal is to gain insight into the \emph{non-monotone}
setting. To this end, we define a new form of composition, called
\emph{semi-monotone composition}, which composes a \emph{non-monotone}
outer KW relation with a \emph{monotone }inner KW relation. The purpose
of this new composition is to enjoy the best of both worlds: On the
one hand, this notion allows us to use candidates for the inner function~$g$
that come from the monotone setting. On the other hand, we believe
that this notion is much closer to the non-monotone setting. Thus,
by studying semi-monotone composition we can tackle issues that come
up in the non-monotone setting but not in the monotone setting.

In order to gain intuition for the definition of this composition,
consider the obvious protocol for the non-monotone composition $\KW_{f}\d\KW_{g}$.
Recall that the inputs to this protocol are matrices $X,Y\in\B^{m\times n}$,
and that we denote by $a,b$ the column vectors that are obtained
by applying~$g$ to the rows of those matrices. Observe that there
are two key properties of $\KW_{f}\d\KW_{g}$ that allow the obvious
protocol to work:
\begin{itemize}
\item The players can find a row $i\in\left[m\right]$ such that $a_{i}\ne b_{i}$
by solving $\KW_{f}$ on~$a,b$.
\item For every $i\in\left[m\right]$ such that $a_{i}\ne b_{i}$, the players
can find a solution for $\KW_{f}\d\KW_{g}$ by solving $\mKW_{g}$
on the rows~$X_{i},Y_{i}$.
\end{itemize}
Note that, while the obvious protocol always finds a solution in a
row~$i$ where $a_{i}\ne b_{i}$, the rows where $a_{i}=b_{i}$ might
contain solutions as well.

We define the semi-monotone composition of $\KW_{f}$ and $\mKW_{g}$
as a communication problem that is identical to $\KW_{f}\d\KW_{g}$,
except that in the second property above, the non-monotone relation
$\KW_{g}$ is replaced with the monotone relation $\mKW_{g}$. Formally,
we define semi-monotone composition as follows.
\begin{definition}[Semi-monotone composition]
\label{semi-monotone-composition}Let $f:\B^{m}\to\B$ be a non-constant
(possibly non-monotone) function, and let $g:\B^{n}\to\B$ be a non-constant
monotone function. The \emph{semi-monotone composition} $\KW_{f}\d\mKW_{g}$
is the following communication problem. Alice and Bob get as inputs
$m\times n$~binary matrices $X$ and~$Y$ respectively. Let $a,b\in\B^{m}$
denote the column vectors that are obtained by applying~$g$ to each
row of~$X$ and~$Y$ respectively. Then, $f(a)=1$ and $f(b)=0$,
and the goal of the players is to find an entry $(i,j)$ that satisfies
one of the following three options:
\begin{itemize}
\item $a_{i}>b_{i}$ and $X_{i,j}>Y_{i,j}$.
\item $a_{i}<b_{i}$ and $X_{i,j}<Y_{i,j}$.
\item $a_{i}=b_{i}$ and~$X_{i,j}\ne Y_{i,j}$.
\end{itemize}
\end{definition}

Note that this communication problem has the desired structure: Indeed,
it is not hard to see that when $a_{i}\ne b_{i}$, finding a solution
in the $i$-th row is equivalent to solving $\mKW_{g}$ on~$X_{i},Y_{i}$.
It is also not hard to show that $\C(\KW_{f}\d\mKW_{g})\le\C(\KW_{f})+\C(\mKW_{g})$
bits, by using an appropriate variant of the obvious protocol of~$\KW_{f}\d\KW_{g}$.
Therefore, a natural ``semi-monotone variant'' of the KRW conjecture
would be the following.
\begin{conjecture}[Semi-monotone KRW conjecture]
\label{semi-monotone-KRW}For every non-constant function $f:\B^{m}\to\B$
and non-constant monotone function~$g:\B^{n}\to\B$, 
\[
\C(\KW_{f}\d\mKW_{g})\gtrapprox\C(\KW_{f})+\C(\mKW_{g}).
\]
\end{conjecture}

\paragraph*{Our result.}

Ideally, we would have liked to prove \ref{semi-monotone-KRW} for
every outer function $f$ and for a wide range of inner functions~$g$.
Unfortunately, we are only able to prove it for the case where the
outer relation~$\KW_{f}$ is replaced with the (non-monotone) universal
relation, i.e., the composition $U_{m}\d\mKW_{g}$. This composition
is defined similarly to \ref{semi-monotone-composition}, with the
following difference: instead of promising that $f(a)=1$ and $f(b)=0$,
we only promise that $a\ne b$. The natural conjecture in this case
would be that
\begin{equation}
\C(U_{m}\d\mKW_{g})\gtrapprox\C(U_{m})+\C(\mKW_{g})\ge m+\C(\mKW_{g}),\label{eq:semi-monotone-conjecture-universal}
\end{equation}
where the second inequality holds since $\C(U_{m})=m+\Theta(1)$ (see
\cite{KRW95,TZ97}). Our semi-monotone composition theorem proves
such a result for every monotone inner function~$g$ for which a
lower bound on $\C(\mKW_{g})$ can be proved using a lifting theorem
of \cite{RMNPRV20}.

Before describing our result, we briefly describe the lifting theorem
of \cite{RMNPRV20}. Given an unsatisfiable CNF formula~$\phi$,
its \emph{associated search problem}~$S_{\phi}$ is the following
task: given an assignment~$z$ to~$\phi$, find a clause of~$\phi$
that is violated by~$z$. The \emph{Nullstellensatz degree }of $\phi$,
denoted $\NSF(\phi)$, is a complexity measure that reflects how hard
it is prove that $\phi$ is unsatisfiable in the Nullstellensatz proof
system over a field~$\F$ (see \ref{subsec:Preliminaries-Nullstellensatz}
for a formal definition). Fix a gadget~$\gd:\B^{t}\times\B^{t}\to\B$
of input length~$t$. The lifting theorem of~\cite{RMNPRV20} says
that $\C(S_{\phi}\d\gd)\ge\Omega(\NS(\phi)\cdot t)$ provided that
the gadget~$\gd$ has sufficiently large rank. 

Our result says that \ref{eq:semi-monotone-conjecture-universal}
holds whenever there is a reduction from such a lifted problem $S_{\phi}\d\gd$
to $\mKW_{g}$. We require the gadget $\gd$ to be the equality function~$\eq$,
and require the reduction to be \emph{injective} (see \ref{reduction}
for the definition of injective reduction). Informally, our semi-monotone
composition theorem can be stated as follows (see \ref{semi-monotone-composition-thm}
for the formal statement):
\begin{theorem}[semi-monotone composition theorem, informal]
\label{semi-monotone-composition-informal}Let $g:\B^{n}$ be a non-constant
monotone function, and let $\eq$~be the equality function on strings
of length~$t$. Suppose there exists a lifted search problem $S_{\phi}\d\eq$
that reduces to $\mKW_{g}$ via an injective reduction and satisfies
the conditions of the theorem of~\cite{RMNPRV20}. Then 
\[
\C(U_{m}\d\mKW_{g})\ge m+\Omega(\NS(\phi)\cdot t).
\]
In particular, if $\C(\mKW_{g})=\tilde{O}(\NS(\phi)\cdot t)$, then
\[
\C(U_{m}\d\mKW_{g})\ge m+\tilde{\Omega}(\C(\mKW_{g})).
\]
\end{theorem}

As in the case of the monotone composition theorem, the semi-monotone
theorem is applicable to many interesting inner functions, including
the classical $s\text{-}t$-connectivity, clique, and generation functions
mentioned above (see \ref{sec:classic-functions} for details), and
the bound that it gives is good enough for the purposes of the KRW
conjecture.

\paragraph*{Comparison to monotone composition.}

Recall that our goal in defining semi-monotone composition is to captures
issues that arise in the non-monotone setting but are not captured
by the monotone setting. We claim that our definition succeeds in
this task for at least one significant issue, to be discussed next.

Recall that the KRW conjecture says that the obvious protocol for
$\KW_{f}\d\KW_{g}$ is essentially optimal. Intuitively, this should
be the case since it seems that the best strategy for the players
is to work on a row where $a_{i}\ne b_{i}$, and to do so, they must
first find such a row. While it seems reasonable that the best strategy
is to work on a row where $a_{i}\ne b_{i}$, it is not clear how to
prove it: indeed, this is a central challenge in the proofs of known
composition theorems (though not the only challenge).

On the other hand, Karchmer et al. \cite{KRW95} observed that in
the monotone setting, the players can be forced to solve the problem
on a row where $a_{i}>b_{i}$. This means that in the monotone setting,
we can easily bypass a central challenge of the non-monotone case.
An important feature of semi-monotone composition is that the observation
of \cite{KRW95} fails for this composition. Hence, we believe that
the semi-monotone setting is much closer to the non-monotone KRW conjecture
than the monotone setting.

\subsection{\label{subsec:Intro-Our-techniques}Our techniques}

\subsubsection{\label{subsec:Intro-Our-techniques-monotone}The monotone composition
theorem}

We use the high level proof strategy that was introduced by \cite{EIRS01},
and further developed in \cite{DM18,M17_derandomized_composition,KM18}.
The main technical lemma is a structure theorem, formalizing that
any correct protocol must first solve $\mKW_{f}$, and then solve
$\mKW_{g}$. A bit more formally, we show that for any partial transcript
$\pi_{1}$ of $\Pi$, if $\mKW_{f}$ has not yet been solved at $\pi_{1}$,
then $\Pi$ must send $\approx\C(\mKW_{g})$ additional bits before
it can find a solution for $\mKW_{f}\d\mKW_{g}$.

To accomplish this, at $\pi_{1}$, we partition the rows of $X,Y$
into two types: (1) ``revealed'' rows where $\pi_{1}$ reveals a
lot of information, and (2) ``unrevealed'' rows, where $\pi_{1}$
reveals only a small amount of information. We then show that the
revealed rows can be forced to be useless (that is, we can ensure
that there is no solution $(i,j)$ where $i$ is a revealed row).
It follows that in order for the protocol to finish after $\pi_{1}$,
it has to solve $\mKW_{g}$ on one of the unrevealed rows.

The remaining step is therefore to show that in order to solve $\mKW_{g}$
on one of the unrevealed rows, the protocol must transmit $\approx\C(\mKW_{g})$
additional bits. While this claim sounds intuitive, proving it is
non-trivial since some (small amount of) information has been learned
about each unrevealed row, and this revealed information can be highly
dependent. Moreover, the protocol is allowed to choose on which unrevealed
row it solves $\mKW_{g}$, and this could in principle make the task
significantly easier. In previous works, these issues are dealt with
in a way that is tailored to the particular choice of~$g$. Specifically,
one takes a known lower bound proof for $\KW_{g}$, and shows that
it still goes through even after accounting for the aforementioned
complications.

In our case, we do not know the particular choice of~$g$, but we
do know that the lower bound for $\mKW_{g}$ is proved using the lifting
theorem of \cite{CFKMP19}. Hence, our goal is show that this lower
bound proof still goes through. To this end, we prove a generalization
of this lifting theorem which may be of independent interest (see
\ref{generalized-lifting-thm}). Informally, our generalization shows
that $S\d\gd$ remains hard even if we restrict it to a subset~$\cX\times\cY$
of its inputs, as long as the coordinates remain \emph{unpredictable}.
Since this is the case for the unrevealed rows, we get the lower bound
that we desire.

The notion of unpredictability required by our lifting theorem is
based on \emph{average degree} as defined by \cite{EIRS01,RM99}:
given a set of strings $\cW\in\D^{\ell}$ and a subset of coordinates~$I\subseteq\left[\ell\right]$,
the \emph{average degree} $\AD_{I}(\cW)$ is the average number of
ways to complete a string in $\cW|_{\left[\ell\right]-I}$ to a string
in~$\cW$. Informally, our generalized lifting theorem says the following
(see \ref{generalized-lifting-thm} for the formal statement): 
\begin{theorem}[informal]
\label{generalized-lifting-informal}Let $S\d\gd$ be a lifted search
problem that satisfies the conditions of~\cite{CFKMP19}. Let $\cX\times\cY$
be a subset of the inputs of $S\d\gd$ such that $\AD_{I}(\cX)$ and
$\AD_{I}(\cY)$ are sufficiently large for every set of coordinates~$I$.
Then, the communication complexity of solving $S\d\gd$ on the inputs
in $\cX\times\cY$ is at least $\Omega\left(\Q(S)\cdot t\right)$. 
\end{theorem}

Our proof of the generalized lifting theorem mostly follows the proof
of \cite{CFKMP19}, but uses a different potential argument to bound
the communication complexity: whereas in the original proof of \cite{CFKMP19}
the potential function is the min-entropy deficiency with respect
to the uniform distribution \emph{over all the inputs}, the potential
function in our proof measures the deficiency with respect to the
uniform distribution \emph{over the restricted} \emph{set of inputs}.
The latter distribution is less structured, and hence the potential
argument requires a more refined analysis.

\subsubsection{\label{subsec:Intro-Our-techniques-semi-monotone}The semi-monotone
composition theorem}

We prove the lower bound on $U_{m}\d\mKW_{g}$ using Razborov's rank
method (see \ref{subsec:Preliminaries-Razborov}). Basically, in order
to use this method to prove a lower bound on a communication problem
$S\subseteq\cX\times\cY\times\cO$, one needs to construct a matrix~$A$
of order $\left|\cX\right|\times\left|\cY\right|$ such that $A$
has high rank, but its restriction to every $S$-monochromatic rectangle
has low rank. Roughly, the lifting theorem of~\cite{RMNPRV20} gives
such a matrix~$A$ for~$\mKW_{g}$, and we use this matrix to construct
a corresponding matrix~$M$ for $U_{m}\d\mKW_{g}$.

The matrix $M$ for $U_{m}\d\mKW_{g}$ is constructed as follows.
The rows and columns of $M$ are indexed by matrices $X$ and~$Y$
respectively. We view the matrix $M$ as a block matrix that consists
of $2^{m}\cdot2^{m}$ blocks --- a block for each value of $a$ and
$b$. For every $a,b$ such that $a=b$, the corresponding block is
the all-zeros matrix. For every other choice of $a,b$, the corresponding
block is formed by taking the Kronecker product, for every $i\in\left[m\right]$,
of either $A$ (if $a_{i}\ne b_{i}$) or the identity matrix~$I$
(if $a_{i}=b_{i}$).

The matrix $M$ is constructed in this way in order to guarantee that
all its restrictions to monochromatic rectangles have low rank. Very
roughly, setting blocks to~$A$ where $a_{i}\ne b_{i}$ guarantees
that monochromatic rectangles that solve $\mKW_{g}$ on~$X_{i},Y_{i}$
have low rank. On the other hand, setting blocks to the identity matrix~$I$
where $a_{i}=b_{i}$ guarantees that monochromatic rectangles that
find different entries $X_{i,j}\ne Y_{i,j}$ are all-zeros rectangles.

An important part of the proof is the observation that when the theorem
of~\cite{RMNPRV20} is applied with the equality gadget over~$\F_{2}$
(as we do), it gives a matrix~$A$ that satisfies $A^{2}=I$. This
property creates a connection between $A$ and~$I$ that allows us
to analyze the rank of $M$ and of its sub-matrices using Gaussian
elimination.

\paragraph*{Organization of this paper.}

We cover the necessary preliminaries in \ref{sec:Preliminaries}.
Then, we prove the monotone composition theorem in \ref{sec:Monotone-composition},
and the semi-monotone composition theorem in \ref{sec:Semi-monotone-composition}.
We prove our generalization of the lifting theorem of \cite{CFKMP19}
in \ref{sec:generalized-lifting-thm}. Next, in \ref{sec:classic-functions},
we show how to apply our theorems to the classical functions $s\text{-}t$-connectivity,
clique, and generation. Finally, in \ref{sec:Intro-Open-problems}
we discuss open problems for future research.

\section{\label{sec:Preliminaries}Preliminaries}

Throughout the paper, we use bold letters to denote random variables.
For any $n\in\N$, we denote by $\left[n\right]$ the set $\left\{ 1,\ldots,n\right\} $.
We denote by $\F_{2}$ the finite field of size~$2$. We say that
a CNF formula~$\phi$ is a \emph{CNF contradiction} if and only if
it is unsatisfiable.

Given two strings $x,y\in\B^{n}$, we write $x\ge y$ if $x_{i}\ge y_{i}$
for every $i\in\left[n\right]$. We say that a Boolean function~$f:\B^{n}\to\B$
is \emph{monotone} if for every $x,y\in\B^{n}$ such that $x\ge y$
it holds that $f(x)\ge f(y)$. 

Given an alphabet~$\D$ and a set~$I\subseteq\left[n\right]$, we
denote by $\D^{I}$ the set of strings of length~$\left|I\right|$
whose coordinates are indexed by~$I$. Given a string~$w\in\D^{n}$
and a set $I\subseteq\left[n\right]$, we denote by $w|_{I}\in\D^{I}$
the projection of~$w$ to the coordinates in~$I$ (in particular,
$w_{\emptyset}$ is defined to be the empty string). Given a set of
strings $\cW\subseteq\D^{n}$ and a set $I\subseteq\left[n\right]$,
we denote by $\cW|_{I}$ the set of projections of strings in~$\cW$
to~$I$. We will sometimes omit the projection symbol~$|$ when
it is clear from the context.

We denote by $\D^{m\times n}$ the set of $m\times n$~matrices with
entries in~$\D$, and for sets $I\subseteq\left[m\right]$ and $J\subseteq\left[n\right]$,
we denote by $\D^{I\times J}$ the set of $\left|I\right|\times\left|J\right|$
matrices whose entries are indexed by $I\times J$. Given a matrix
$X\in\D^{m\times n}$ and a rectangle $R\eqdef I\times J\subseteq\left[m\right]\times\left[n\right]$,
we denote by $X|_{R}$ the projection of~$X$ to $R$. Here, too,
we extend this notation to sets of matrices~$\cW\subseteq\D^{m\times n}$,
and sometimes omit the projection symbol when it is clear from the
context. We denote by $X_{i}\in\D^{n}$ the $i$-th row of~$X$.
Given a matrix $A\in\F^{m\times n}$ over a finite field~$\F$, we
denote its rank by $\rkf(A)$.

\paragraph*{Search problems.}

Given a finite set of inputs $\cI$ and a finite set of outputs~$\cO$,
a \emph{search problem}~$S\subseteq\cI\times\cO$ is a relation between
$\cI$ and~$\cO$. Given $z\in\cI$, we denote by $S(z)$ the set
of outputs $o\in\cO$ such that $(z,o)\in S$. Intuitively, a search
problem~$S$ represents the following task: given an input $z\in\cI$,
find a solution $o\in S(z)$. Without loss of generality, we may assume
that $S(z)$ is always non-empty, since otherwise we can set $S(z)=\left\{ \bot\right\} $
where $\bot$ is some special failure symbol that does not belong
to~$\cO$.

\subsection{\label{subsec:Preliminaries-Communication-complexity}Communication
complexity}

We assume familiarity with basic definitions of communication complexity
(see, e.g., \cite{KN_book}). In what follows, we highlight some important
standard definitions and facts that we will use, and define one less-standard
notion. As usual, we define a \emph{(deterministic) protocol~$\Pi$
}as a binary tree. We identify the vertices of a protocol with the
transcripts that they represent. Given sets $\cX$ and~$\cY$, we
say that the protocol has \emph{domain $\cX\times\cY$} if the inputs
of Alice and Bob are taken from the sets $\cX$ and~$\cY$ respectively.
We say that the \emph{range} of the protocol is a set~$\cO$ if the
protocol outputs elements in~$\cO$. 
\begin{definition}
A transcript~$\pi$ is a \emph{full transcript} if it corresponds
to a leaf of the protocol tree, and otherwise it is a \emph{partial
transcript}. Given a pair of inputs $\left(x,y\right)\in\cX\times\cY$,
we define the transcript of~$(x,y)$, denoted $\Pi(x,y)$, as the
full transcript of the protocol when Alice and Bob get the inputs
$x$ and~$y$ respectively.
\end{definition}

\begin{definition}
Two protocols $\Pi$ and~$\Pi'$ over the same domain and range are
\emph{equivalent} if they have the same output on every pair of inputs.
\end{definition}

\begin{definition}
A \emph{communication problem} $S\subseteq\cX\times\cY\times\cO$
is the search problem in which Alice and Bob get inputs $x\in\cX$
and $y\in\cY$ respectively, and would like to find a solution~$o\in S(x,y)$.
A protocol \emph{solves~$S$} if on every pair of inputs $(x,y)\in\cX\times\cY$
it outputs some $o\in S(x,y)$. 
\end{definition}

\begin{definition}
The \emph{communication complexity} of a protocol $\Pi$, denoted
$\C(\Pi)$, is the depth of the protocol tree. For a search problem~$S$,
the \emph{(deterministic) communication complexity} of $S$, denoted
$\C(S)$, is the minimal communication complexity of a protocol that
solves $S$. 
\end{definition}

\begin{definition}
The \emph{size} of a protocol~$\Pi$, denoted $\L(\Pi)$, is the
number of leaves in the protocol tree. The \emph{protocol size }of
a search problem~$S$, denoted $\L(S)$, is the size of the smallest
protocol that solves~$S$ (this is also known as the \emph{protocol
partition number} of~$S$).
\end{definition}

It is not hard to see that for every protocol~$\Pi$ it holds that
$\C(\Pi)\ge\log\L(\Pi)$ --- informally, every ``shallow'' protocol
is ``small''. The following folklore fact establishes a connection
in the other direction: namely, every ``small'' protocol can be
transformed into a ``shallow'' one. This transformation is sometimes
called \emph{protocol balancing}.
\begin{fact}[{protocol balancing, see, \cite[Lemma 2.8]{KN_book}}]
\label{protocol-balancing}For every protocol $\Pi$ there is an
equivalent protocol $\Pi'$ such that $\C(\Pi')\le4\log\L(\Pi)$.
In particular, for every communication problem~$S$ it holds that
\[
\log\L(S)\le\C(S)\le4\log\L(S)
\]
and hence $\C(S)=\Theta(\log\L(S))$.
\end{fact}

Let $\Pi$ be a protocol with domain $\cX\times\cY$ and let $\pi$
be a transcript of $\Pi$. It is a standard fact that the set of inputs~$(x,y)\in\cX\times\cY$
on which the protocol reaches the vertex~$\pi$ is a combinatorial
rectangle. We denote this rectangle by~$\cX_{\pi}\times\cY_{\pi}$.
Finally, we use the following definition, which generalizes the notion
of rectangular reduction~\cite{BNS92} to search problems.
\begin{definition}
\label{reduction}Let $S\subseteq\cX\times\cY\times\cO$ and $S'\subseteq\cX'\times\cY'\times\cO'$
be communication problems. A \emph{reduction} from $S$ to $S'$ consists
of functions $R_{A}:\cX\to\cX'$, $R_{B}:\cY\to\cY'$, and $\rout:\cO'\to\cO$
that satisfy the following condition: for every $x\in\cX$, $y\in\cY$,
and $o'\in\cO'$, if $o'$ is a solution for $S'$ on inputs $R_{A}(x)$
and $R_{B}(y)$, then $\rout(o')$ is a solution for $S$ on $(x,y)$.

We say that the reduction is \emph{injective} if the functions $R_{A}$
and~$R_{B}$ are injective (but the function $\rout$ is not required
to be injective).
\end{definition}

An important aspect of \ref{reduction} is that the function $\rout$
is required not to depend on the inputs~$x,y$. This stands in contrast
to other definitions of reductions for search problems (e.g. a Levin
reduction), which do allow their analogue of $\rout$ to depend on
the inputs. We note that this requirement is used in the proof of
the semi-monotone composition theorem (\ref{semi-monotone-composition-thm}),
but not in the proof of the monotone composition theorem~(\ref{monotone-composition-theorem}).

\subsection{Subadditive measures on trees}

We use the following notions of a subadditive measure and a separating
set of a tree.
\begin{definition}
Given a binary tree~$T=(V,E)$, we say that a function $\gamma:V\to\N$
is a \emph{subadditive measure} \emph{on $T$} if for every internal
vertex $v$ with children $v_{0}$ and~$v_{1}$ it holds that $\gamma(v)\le\gamma(v_{0})+\gamma(v_{1})$.
\end{definition}

\begin{definition}
\label{separating-set}Given a binary tree~$T=(V,E)$, we say that
a set of vertices $M\subseteq V$ is a \emph{separating set} of $T$
if every path from the root of~$T$ to its leaves passes through~$M$.
\end{definition}

We use the following fact about subadditive measures.
\begin{claim}
\label{subadditive-measure-averaging-on-cutset}Let $T=(V,E)$ be
a binary tree with root~$r$, let $\gamma$ be a subadditive measure
on~$T$, and let $M$ be a\emph{ }separating set of~$T$. Then,
there exists a vertex $v\in M$ such that $\gamma(v)\ge\gamma(r)/\left|M\right|$.
\end{claim}

\begin{myproof}[Proof sketch.]
Let $T$, $r$, $\gamma$, and $M$ be as in the claim. By applying
the definition of subadditive measure inductively, it is not hard
to show that 
\[
\gamma(r)\le\sum_{v\in M}\gamma(v).
\]
The claim now follows by averaging.
\end{myproof}

\subsection{\label{subsec:Preliminaries-Karchmer-Wigderson}Monotone formulas
and Karchmer-Wigderson relations}

In this section, we define monotone formulas and KW relations formally,
and state the connections between them.
\begin{definition}
A \emph{monotone formula}\textsf{~$\phi$} is a binary tree, whose
leaves are labeled with input variables~$x_{i}$, and whose internal
vertices are labeled as AND ($\wedge$) or OR ($\vee$) gates. We
note that a single input variable~$x_{i}$ can be associated with
many leaves. The \emph{size} of a monotone formula is the number of
its \emph{leaves} (which up to a factor of~$2$ is the same as the
number of edges or vertices of the tree).
\end{definition}

\begin{definition}
\label{formulas-computing-functions}A monotone formula $\phi$ over
$n$~variables computes a monotone Boolean function~$f:\B^{n}\to\B$
in the natural way. The \emph{monotone formula complexity} of a monotone
function~$f:\B^{n}\to\B$, denoted $\mL(f)$, is the size of the
smallest monotone formula that computes~$f$. The \emph{monotone
depth complexity} of~$f$, denoted~$\mD(f)$, is the smallest depth
of a formula that computes~$f$.
\end{definition}

Note that we define here the monotone depth complexity of a function
as the depth of a monotone \emph{formula} that computes $f$, whereas
in the introduction we defined it as the depth of a monotone \emph{circuit}
that computes~$f$. However, it is not hard to see that the two definitions
are equivalent. Next, we generalize the above definitions from functions
to promise problems, which will be useful when we discuss Karchmer-Wigderson
relations. 
\begin{definition}
\label{formulas-separating-sets}Let $\cX,\cY\subseteq\B^{n}$. A
monotone formula~$\phi$ \emph{separates} $\cX$ and $\cY$ if $\phi(x)=1$
for every $x\in\cX$ and $\phi(y)=0$ for every $y\in\cY$.
\end{definition}

It is not hard to prove that two sets $\cX,\cY\subseteq\B^{n}$ are
separated by some monotone formula if and only if they satisfy the
following property: for every $x\in\cX$ and $y\in\cY$ it holds that
$x_{i}>y_{i}$ for some coordinate~$i\in\left[n\right]$. We denote
this property by $\cX\succ\cY$.
\begin{definition}
Let $\cX,\cY\subseteq\B^{n}$ be sets such that $\cX\succ\cY$. The
\emph{monotone formula complexity of the rectangle $\cX\times\cY$},
denoted~$\mL(\cX\times\cY)$, is the size of the smallest monotone
formula that separates $\cX$ and~$\cY$. The \emph{monotone depth
complexity} \emph{of} \emph{the rectangle}~$\cX\times\cY$, denoted~$\mD(\cX\times\cY)$,
is the smallest depth of a formula that separates $\cX$ and~$\cY$.
If the rectangle~$\cX\times\cY$ is empty, we define $\mL(\cX\times\cY)=\mD(\cX\times\cY)=0$.
\end{definition}

Note that \ref{formulas-computing-functions} is indeed a special
case of \ref{formulas-separating-sets} where $\cX=f^{-1}(1)$ and~$\cY=f^{-1}(0)$.
We turn to defining monotone KW relations. We first define them for
general rectangles, and then specialize the definition to functions. 
\begin{definition}
Let $\cX,\cY\subseteq\B^{n}$ be two sets such that $\cX\succ\cY$.
The \emph{monotone KW relation} $\mKW_{\cX\times\cY}$ is the communication
problem in which Alice's input is~$x\in\cX$, Bob's input is~$y\in\cY$,
and they would like to find a coordinate $i\in\left[n\right]$ such
that $x_{i}>y_{i}$. Note that such a coordinate always exists by
the assumption that $\cX\succ\cY$. 
\end{definition}

\begin{definition}
Let $f:\B^{n}\to\B$ be a non-constant monotone function. The \emph{monotone
KW relation}\textsf{ }\emph{of}\textsf{~$f$}, denoted $\mKW_{f}$,
is defined by $\mKW_{f}\eqdef\mKW_{f^{-1}(1)\times f^{-1}(0)}$.
\end{definition}

We are now ready to state the connection between monotone KW relations
and monotone formulas.
\begin{theorem}[\cite{KW90}, see also \cite{R90}]
\label{KW-connection}For every two sets $\cX,\cY\subseteq\B^{n}$
such that $\cX\succ\cY$ it holds that $\mD(\cX\times\cY)=\C(\mKW_{\cX\times\cY})$
and $\mL(\cX\times\cY)=\L(\mKW_{\cX\times\cY})$. In particular, for
every non-constant $f:\B^{n}\to\B$, it holds that $\mD(f)=\C(\mKW_{f})$
and $\mL(f)=\L(\mKW_{f})$. 
\end{theorem}

\noindent Due to \ref{KW-connection}, in the rest of the paper we
use the notations $\mL(\cX\times\cY)$ and $\L(\mKW_{\cX\times\cY})$
interchangeably.

Given a protocol $\Pi$ that solves $\mKW_{\cX\times\cY}$, we can
view the complexity measure $\mL$ as a subadditive measure over the
protocol tree. Namely, this measure assigns to each vertex~$v$ of~$\Pi$
the value $\mL(v)\eqdef\mL(\cX_{v}\times\cY_{v})$, where $\cX_{v}\times\cY_{v}$
is the rectangle that is associated with~$v$.

To see that this is indeed a subadditive measure, let $v$ be an internal
vertex of~$\Pi$, and let $v_{0}$ and~$v_{1}$ be its children.
Without loss of generality, assume that at the vertex~$v$ it is
Alice's turn to speak. Then, $\cX_{v}=\cX_{v_{0}}\cup\cX_{v_{1}}$
and $\cY_{v}=\cY_{v_{0}}=\cY_{v_{1}}$. It holds that
\begin{align}
\mL(v) & =\mL(\cX_{v}\times\cY_{v})\nonumber \\
 & \le\mL(\cX_{v_{0}}\times\cY_{v})+\mL(\cX_{v_{1}}\times\cY_{v})\label{eq:subadditivity-of-formula-complexity}\\
 & =\mL(\cX_{v_{0}}\times\cY_{v_{0}})+\mL(\cX_{v_{1}}\times\cY_{v_{1}}) & \text{(Since \ensuremath{\cY_{v}=\cY_{v_{0}}=\cY_{v_{1}}})}\nonumber \\
 & =\mL(v_{0})+\mL(v_{1}).\nonumber 
\end{align}
To see why Inequality~\oldref{eq:subadditivity-of-formula-complexity}
holds, consider the following protocol for $\mKW_{\cX_{v}\times\cY_{v}}$:
Alice starts by saying whether her input belongs to~$\cX_{v_{0}}$
or to~$\cX_{v_{1}}$. Then, the players proceed by invoking the optimal
protocol for either $\mKW_{\cX_{v_{0}}\times\cY_{v}}$ or $\mKW_{\cX_{v_{1}}\times\cY_{v}}$
respectively. It is easy to see that the size of this protocol is
at most $\mL(\cX_{v_{0}}\times\cY)+\mL(\cX_{v_{1}}\times\cY)$. Hence,
$\mL$ is a subadditive measure, as required.

\subsection{\label{subsec:Preliminaris-Decision-trees}Decision trees}

Informally, a decision tree is an algorithm that solves a search problem~$S\subseteq\B^{\ell}\times\cO$
by querying the individual bits of its input. The tree is computationally
unbounded, and its complexity is measured by the number of bits it
queried. Formally, a decision tree is defined as follows.
\begin{definition}
A \emph{(deterministic) decision tree}~$T$ with domain $\B^{\ell}$
and range~$\cO$ is a binary tree in which every internal node is
labeled with a coordinate in~$\left[\ell\right]$ (which represents
a query), every edge is labeled by a bit (which represents the answer
to the query), and every leaf is labeled by an output in~$\cO$.
Such a tree computes a function from~$\B^{\ell}$ to~$\cO$ in the
natural way, and with a slight abuse of notation, we identify this
function with~$T$. The \emph{query complexity} of~$T$ is the depth
of the tree.
\end{definition}

\begin{definition}
We say that a decision tree~$T$ \emph{solves a search problem }$S\subseteq\B^{\ell}\times\cO$
if for every $z\in\B^{\ell}$ it holds that $T(z)\in S(z)$. The \emph{(deterministic)
query complexity of}~$S$, denoted~$\Q(S)$, is the minimal query
complexity of a deterministic decision tree that solves~$S$.
\end{definition}

\subsection{\label{subsec:Preliminaries-Razborov}The Razborov rank measure}

The Razborov rank measure \cite{R90} is a complexity measure that
can be used to prove lower bounds on communication complexity. In
order to introduce this measure, we first establish some notation.
Let $S\subseteq\cX\times\cY\times\cO$ be a communication problem.
For some $o\in\cO$, we say that a rectangle $R\subseteq\cX\times\cY$
is \emph{$o$-monochromatic (for }$S$) if $o\in S(x,y)$ for every
$(x,y)\in R$. We say that $R$ is \emph{$S$-monochromatic} if it
is $o$-monochromatic for some~$o\in\cO$. Let $\mathcal{R}$ denote
the set of $S$-monochromatic rectangles.

Now, let $\F$ be a field. Given a matrix $A\in\F^{\cX\times\cY}$,
the \emph{Razborov $\F$-rank measure of $S$ with respect to} $A$
is
\[
\rzf(S,A)\eqdef\frac{\rkf(A)}{{\displaystyle \max_{R\in\mathcal{R}}}\left\{ \rkf(A|_{R})\right\} }.
\]
The \emph{Razborov $\F$-rank measure of~$S$}, denoted $\rzf(S)$,
is the maximum of $\rzf(S,A)$ over all matrices $A\in\F^{\cX\times\cY}$.
We have the following result.
\begin{fact}[\cite{R90}]
For every field~$\F$, it holds that $\L(S)\ge\rzf(S)$, and hence
$\C(S)\ge\log\rzf(S)$.
\end{fact}

\subsection{\label{subsec:Preliminaries-Nullstellensatz}The Nullstellensatz
proof system}

The Nullstellensatz proof system is a method for certifying that a
set of polynomials does not have a common root. Formally, let $\F$
be a field, and let $P=\left\{ p_{i}:\F^{\ell}\to\F\right\} _{i\in\left[m\right]}$
be a set of polynomials. It is not hard to see that a sufficient condition
for the polynomials $p_{1},\ldots,p_{m}$ to not have a common root
is the existence of polynomials $q_{1},\ldots,q_{m}:\F^{\ell}\to\F$
such that the following equality holds syntactically:
\begin{equation}
p_{1}\cdot q_{1}+\ldots+p_{m}\cdot q_{m}=1.\label{eq:nullstellensatz}
\end{equation}
We refer to such polynomials~$q_{1},\ldots,q_{m}$ as a \emph{Nullstellensatz}
\emph{refutation of~$P$. }The \emph{degree} \emph{of} \emph{the
refutation }is the maximal degree of the polynomial $p_{i}\cdot q_{i}$
over all $i\in\left[m\right]$. The \emph{Nullstellensatz degree}
\emph{of (refuting)} $P$ is the minimum degree of any Nullstellensatz
refutation of~$P$ (assuming one exists).

The Nullstellensatz proof system can be used to certify that a CNF
formula is unsatisfiable. Let $\phi$ be a CNF formula over variables
$x_{1},\ldots,x_{\ell}$. Given a clause~$C$ of $\phi$, we define
the \emph{polynomial encoding} of $C$ as the polynomial that is obtained
by multiplying $1-x_{i}$ for every positive literal $x_{i}$ that
appears in~$C$, and multiplying by $x_{i}$ for every negative literal
$\neg x_{i}$ that appears in~$C$. Let $P_{\phi}$ denote the set
of polynomials that consists of the polynomial encodings of all the
clauses of~$\phi$, and of the polynomials $x_{1}^{2}-x_{1},\ldots,x_{\ell}^{2}-x_{\ell}$.
Clearly, $\phi$~is unsatisfiable if and only if the set $P_{\phi}$
does not have a common root. Moreover, a slight extension of Hilbert's
Nullstellensatz shows that the set $P_{\phi}$ does not have a common
root if and only if $P_{\phi}$~has a Nullstellensatz refutation.
This leads to the following natural definition of the Nullstellensatz
degree of a CNF contradiction.
\begin{definition}
Let $\phi$ be a CNF contradiction, and let $\F$ be a field. The
\emph{Nullstellensatz degree} \emph{of~$\phi$ over} $\F$, denoted
$\NSF(\phi)$, is the Nullstellensatz degree of the set $P_{\phi}$
(where the polynomials in~$P_{\phi}$ are viewed as polynomials over
the field~$\F$).
\end{definition}

\subsection{\label{subsec:Preliminaries-Lifting-theorems}Lifting theorems}

Lifting theorems relate the complexity of a search problem~$S$ in
a weak model to the complexity of the composed search problem $S\d\gd$
in a strong model. Formally, given a search problem~$S\subseteq\B^{\ell}\times\cO$
and a ``gadget'' function $\gd:\D\times\D\to\B$, the \emph{lifted
search problem }$S\d\gd\subseteq\D^{\ell}\times\D^{\ell}\times\cO$
is the communication problem defined by
\[
S\d\gd\left((x_{1},\ldots,x_{\ell}),(y_{1},\ldots,y_{\ell})\right)\eqdef S\left(\gd(x_{1},y_{1}),\ldots,\gd(x_{\ell},y_{\ell})\right).
\]
Lifting theorems lower bound the complexity of $S\d\gd$ in terms
of the complexity of~$S$. The first theorems of this kind were proven
by Raz and McKenzie \cite{RM99}, Shi and Zhou \cite{SZ09}, and Sherstov
\cite{S11}. The recent years have seen a flurry of results on lifting
theorems and their applications (see, e.g., \cite{GP18,GLMWZ16,GPW15,GPW17,RNV16,RPRC16,CKLM19,PR17,WYY17,HHL18,PR18,CFKMP19}).
In this work, we use a theorem of \cite{CFKMP19} for lifting query
complexity (discussed in \ref{subsec:Lifting-Query} below), and a
theorem of \cite{RMNPRV20} for lifting Nullstellensatz degree (discussed
in \ref{subsec:Lifting-Nullstellensatz}).

\subsubsection{\label{subsec:Lifting-Query}Lifting from query complexity}

It is not hard to see that for every search problem~$S$ it holds
that $\C(S\d\gd)\le\Q(S)\cdot\C(\gd)$. This upper bound is obtained
by the protocol that simulates an optimal decision tree for~$S$
on the string $\gd(x_{1},y_{1}),\ldots,\gd(x_{\ell},y_{\ell})$, and
answers the queries of the tree by invoking an optimal protocol for~$\gd$.
The first lifting theorem, due to Raz and McKenzie~\cite{RM99},
established that if the gadget $\gd$ is the index function over sufficiently
large inputs, then this upper bound is essentially tight, that is,
\[
\C(S\d\gd)=\Omega\left(\Q(S)\cdot\C(\gd)\right).
\]
In other words, the theorem ``lifts'' the query complexity of~$S$
to a lower bound on the communication complexity of~$S\d\gd$. This
theorem was recently generalized to other choices of the gadget~$\gd$
by \cite{CKLM19,WYY17,CFKMP19}. In this paper, we use the latter
work of Chattopadhyay et al. \cite{CFKMP19}, which proved a lifting
theorem for every gadget $\gd$ that has a sufficiently low discrepancy.
Below, we define discrepancy, and state the relevant theorem of \cite{CFKMP19}.
\begin{definition}
Let $\D$ be a finite set, let $\gd:\D\times\D\to\B$ be a function,
and let $\boldsymbol{u},\boldsymbol{v}$ be independent random variables
that are uniformly distributed over~$\D$. Given a combinatorial
rectangle $R\subseteq\D\times\D$, the \emph{discrepancy of~$\gd$
with respect to~$R$}, denoted $\disc(\gd,R)$, is defined as follows:
\[
\disc(\gd,R)\eqdef\left|\Pr\left[\gd(\boldsymbol{u},\boldsymbol{v})=0\text{ and }(\boldsymbol{u},\boldsymbol{v})\in R\right]-\Pr\left[\gd(\boldsymbol{u},\boldsymbol{v})=1\text{ and }(\boldsymbol{u},\boldsymbol{v})\in R\right]\right|.
\]
The \emph{discrepancy of~$\gd$}, denoted $\disc(\gd)$, is defined
as the maximum of $\disc(\gd,R)$ over all combinatorial rectangles~$R\subseteq\D\times\D$.
\end{definition}

\begin{theorem}[\cite{CFKMP19}]
\label{lifting-query-thm}For every $\eta>0$ there exists $c\in\N$
for which the following holds: Let $S$ be a search problem that takes
inputs from~$\B^{\ell}$, and let $\gd:\B^{t}\times\B^{t}\to\B$
be an arbitrary function such that $\disc(\gd)\le2^{-\eta\cdot t}$
and $t\ge c\cdot\log\ell$. Then 
\[
\C(S\d\gd)=\Omega\left(\Q(S)\cdot t\right).
\]
\end{theorem}

\subsubsection{\label{subsec:Lifting-Nullstellensatz}Lifting from Nullstellensatz
degree}

Let $\phi$ be a $q$-CNF contradiction, i.e., $\phi$ is an unsatisfiable
Boolean formula in CNF in which every clause contains at most $q$~literals.
The \emph{search problem~$S_{\phi}$ that corresponds to $\phi$}
is the following problem: given an assignment for $\phi$, find a
clause that is violated by the assignment. A series of works \cite{RPRC16,PR17,PR18}
show that for appropriate gadgets~$\gd$, the communication complexity
of $S_{\phi}\d\gd$ can be lower bounded in terms of the Nullstellensatz
degree of~$\phi$. In fact, they actually prove lower bounds on the
Razborov rank measure of $S_{\phi}\d\gd$, which is a stronger result.

In a recent joint work with Marc Vinyals~\cite{RMNPRV20}, we generalized
the latter theorems to work for every gadget~$\gd$ that has a large
rank when viewed as a matrix. Formally, we have the following result.
\begin{theorem}[\cite{RMNPRV20}]
\label{lifting-ns-thm}Let $\phi$ be a $q$-CNF contradiction over
$\ell$~variables, and $S_{\phi}$ be its corresponding search problem.
If $\F$ is a field and $\gd:\D\times\D\to\B$ is a gadget such that
$\rkf(\gd)\ge4$, then
\[
\log\rzf(S_{\phi}\d\gd)\ge\NSF(\phi)\cdot\log\left(\frac{\NSF(\phi)\cdot\rkf(\gd)}{e\cdot\ell}\right)-\frac{6\cdot\ell\cdot\log e}{\rkf(\gd)}-\log q.
\]
\end{theorem}

In particular, when $\gd$ is the equality function with input length
$t\ge2\log\ell$, we obtain the following result.
\begin{corollary}
\label{lifting-ns-corollary}Let $\phi$ be a CNF contradiction over
$\ell$~variables, and $S_{\phi}$ be its corresponding search problem.
If $\F$ is a field and $\eq:\B^{t}\times\B^{t}\to\B$ is the equality
function such that $t\ge2\log\ell$, then
\[
\log\rzf(S_{\phi}\d\eq)=\Omega\left(\NSF(\phi)\cdot t\right).
\]
\end{corollary}

\subsection{Min-entropy}

Given a random variable $\v$ that takes values from a finite set
$\cV$, the \emph{min-entropy} of~$\v$, denoted $\Hm(\v)$, is the
largest number $k\in\R$ such that $\Pr\left[\v=v\right]\le2^{-k}$
holds for every $v\in\cV$. In other words,
\[
\Hm(\v)\eqdef\min_{v\in\cV}\left\{ \log\frac{1}{\Pr\left[\v=v\right]}\right\} .
\]
Min-entropy has the following easy-to-prove properties.
\begin{fact}
\label{min-entropy-upper-bound}$\Hm(\v)\le\log\left|\cV\right|$.
\end{fact}

\begin{fact}
\label{min-entropy-conditioning}Let $\cE\subseteq\cV$ be an event.
Then, $\Hm(\v\mid\cE)\ge\Hm(\v)-\log\frac{1}{\Pr\left[\cE\right]}$.
\end{fact}

\begin{fact}
\label{min-entrpoy-projecting}Let $\v_{1},\v_{2}$ be random variables
taking values from finite sets~$\cV_{1},\cV_{2}$ respectively. Then,
$\Hm(\v_{1})\ge H_{\infty}(\v_{1},\v_{2})-\log\left|\cV_{2}\right|$.
\end{fact}

\subsection{Prefix-free codes}

A set of strings $C\subseteq\B^{*}$ is called a \emph{prefix-free
code} if no string in~$C$ is a prefix of another string in~$C$.
Given a string $w\in\B^{*}$, we denote its length by~$\left|w\right|$.
We use the following corollary of Kraft's inequality. A simple proof
of this fact can be found in \cite[Fact 2.8]{CFKMP19}.
\begin{fact}[Corollary of Kraft's inequality]
\label{simplified-kraft}Let $C\subseteq\B^{*}$ be a finite prefix-free
code, and let $\w$ be a random string taking values from~$C$. Then,
there exists a string $w\in C$ such that $\Pr\left[\w=w\right]\ge\frac{1}{2^{\left|w\right|}}$.
\end{fact}

\subsection{\label{subsec:average-degree}Degrees of sets of strings}

We use a framework of \cite{EIRS01} for measuring the uncertainty
of coordinates of strings. As a motivation, consider a set $\cW\subseteq\Lambda^{N}$
and an unknown string~$w\in\cW$. We would like to measure how much
uncertainty we have about~$w$. Perhaps the simplest way to measure
it is the following notion of \emph{density}.
\begin{definition}
The \emph{density} of a set of strings $\cW\subseteq\D^{N}$ is
\[
\dns(\cW)\eqdef\frac{\left|\cW\right|}{\left|\D^{N}\right|}.
\]
\end{definition}

We would also like to measure the uncertainty we have about certain
coordinates of~$w$, \emph{conditioned on the other coordinates}.
The framework of \cite{EIRS01} measures this uncertainty using the
following notion of \emph{degree}.
\begin{definition}
Let $\cW\subseteq\D^{N}$, and let $I\subseteq\left[N\right]$ be
a set of coordinates. The \emph{degree of a string $w'\in\D^{\left[N\right]-I}$
in $\cW$}, denoted $\deg(w',\cW)$, is the number of extensions of
$w'$ to strings in~$\cW$. The \emph{average degree of $I$ in $\cW$},
denoted $\AD_{I}(\cW)$, is the average degree over all strings $w'\in\cW_{\left[N\right]-I}$.
If $I=\left\{ i\right\} $ is a singleton, we denote the average degree
of $I$ by $\AD_{i}(\cW)$.
\end{definition}

Intuitively, the degree of $w'$ measures how much uncertainty we
have about $w_{I}$ if we know that $w_{\left[n\right]-I}=w'$. The
average degree of~$I$ in $\cW$ is a way to capture how much uncertainty
we have about $w_{I}$ conditioned on the other coordinates. It will
be more convenient to work with the \emph{relative} average degree,
i.e., the ratio between the average degree and the largest possible
degree, defined as follows.
\begin{definition}
Let $\cW$ and $I$ be as before. The \emph{relative average degree
of~$I$ in~$\cW$} is
\[
\rAD_{I}(\cW)\eqdef\frac{\AD_{I}(\cW)}{\left|\D\right|^{\left|I\right|}}.
\]
\end{definition}

One useful property of average degree is that it behaves nicely when
additional information is revealed about~$\cW$:
\begin{fact}[\cite{EIRS01}]
\label{conditioning-average-degree}Let $\cW'\subseteq\cW\subseteq\D^{N}$
be sets of strings and let $I\subseteq\left[N\right]$. Then, $\rAD_{I}(\cW')\ge\frac{\left|\cW'\right|}{\left|\cW\right|}\cdot\rAD_{I}(\cW)$.
\end{fact}

Another useful property is that, when we remove a set of coordinates
$I\subseteq\left[N\right]$ with a small average degree, the density
of~$\cW$ \emph{increases}. Intuitively, this means that when we
drop coordinates about which a lot is known, the relative uncertainty
increases.
\begin{fact}[\cite{RM99}]
\label{removing-coordinates-average-degree}Let $\cW\subseteq\D^{N}$
and let $I\subseteq\left[N\right]$. Then
\[
\dns(\cW|_{\left[N\right]-I})=\frac{1}{\rAD_{I}(\cW)}\cdot\dns(\cW).
\]
\end{fact}

Average degree also satisfies the following useful ``chain rule''.
\begin{fact}[Implicit in \cite{EIRS01}]
\label{chain-rule-average-degree}Let $\cW\subseteq\D^{N}$, and
let $I,J\subseteq\left[N\right]$ be disjoint sets of coordinates.
Then
\[
\rAD_{I\cup J}(\cW)=\rAD_{I}(\cW)\cdot\rAD_{J}(\cW_{\left[N\right]-I}).
\]
\end{fact}

Finally, average degree is a lower bound on another measure of uncertainty,
namely, min-entropy:
\begin{fact}[\cite{KM18}, following \cite{EIRS01}]
\label{degree-to-min-entropy}Let $\cW\subseteq\D^{N}$, and let
$\w$ be a random variable that is uniformly distributed over~$\cW$.
Then, for every $I\subseteq\left[N\right]$ it holds that 
\[
\Hm(\w_{I})\ge\log\AD_{I}(\cW)=\left|I\right|\cdot\log\left|\D\right|-\log\frac{1}{\rAD_{I}(\cW)}.
\]
\end{fact}

\subsection{Kronecker product}

In what follows, we define the Kronecker product and state some of
its useful properties. We note that all matrices here are over an
arbitrary, but fixed, field $\F$.
\begin{definition}
Let $A$ and~$B$ be $m\times n$ and $m'\times n'$ matrices respectively.
The \emph{Kronecker product} of $A$ and~$B$, denoted $A\otimes B$,
is an $(m\cdot m')\times(n\cdot n')$ matrix whose rows and columns
are indexed by pairs in $\left[m\right]\times\left[m'\right]$ and
$\left[n\right]\times\left[n'\right]$ respectively, such that for
every $i\in\left[m\right]$, $i'\in\left[m'\right]$, $j\in\left[n\right]$,
and $j'\in\left[n'\right]$ it holds that
\[
\left(A\otimes B\right)_{(i,i'),(j,j')}=A_{i,j}\cdot B_{i',j'}.
\]
\end{definition}

We use the following easy-to-prove facts about the Kronecker product.
\begin{fact}
\label{Kronecker-mixed-product-property}For every four matrices $A,B,C,D$
it holds that
\[
(A\otimes B)\cdot(C\otimes D)=(A\cdot C)\otimes(B\cdot D).
\]
\end{fact}

\begin{fact}
\label{Kronecker-distributivity}For every three matrices $A,B,C$
it holds that $A\otimes(B+C)=A\otimes B+A\otimes C$.
\end{fact}

\begin{fact}
\label{Kronecker-rank}For every two matrices $A,B$ it holds that
$\rkf(A\otimes B)=\rkf(A)\cdot\rkf(B)$.
\end{fact}

\begin{fact}
\label{Kronecker-block-matrix}Let $A$ and $B$ be block matrices
that can be written as
\[
A=\left(\begin{array}{ccc}
K_{1,1} & \ldots & K_{1,q}\\
\vdots & \ddots & \vdots\\
K_{p,1} & \ldots & K_{p,q}
\end{array}\right),B=\left(\begin{array}{ccc}
L_{1,1} & \ldots & L_{1,q'}\\
\vdots & \ddots & \vdots\\
L_{p',1} & \ldots & L_{p',q'}
\end{array}\right),
\]
where $K_{i,j},L_{i',j'}$ denote the blocks. Then, the matrix $A\otimes B$
is a block matrix that can be written as
\[
A\otimes B=\left(\begin{array}{ccccc}
K_{1,1}\otimes L_{1,1} &  & \ldots &  & K_{1,q}\otimes L_{1,q'}\\
 & \ddots\\
\vdots &  & K_{i,j}\otimes L_{i',j'} &  & \vdots\\
 &  &  & \ddots\\
K_{p,1}\otimes L_{p',1} &  & \ldots &  & K_{p,q}\otimes L_{p',q'}
\end{array}\right).
\]
 
\end{fact}

\section{\label{sec:Monotone-composition}The monotone composition theorem}

In this section we prove our monotone composition theorem (\ref{monotone-composition-informal}),
which can be stated formally as follows:
\begin{theorem}
\label{monotone-composition-theorem}For every $\eta>0$ there exists
$c\in\N$ such that the following holds: Let $f:\B^{m}\to\B$ and
$g:\B^{n}\to\B$ be non-constant monotone functions. Suppose that
there exists a search problem $S\subseteq\B^{\ell}\times\cO$, and
a function $\gd:\B^{t}\times\B^{t}\to\B$ of input length $t\ge c\cdot\log(m\cdot\ell)$
and discrepancy at most $2^{-\eta\cdot t}$, such that the lifted
search problem $S\d\gd$ reduces to $\mKW_{g}$. Then, 
\[
\log\L(\mKW_{f}\d\mKW_{g})\ge\log\L(\mKW_{f})+\Omega(\Q(S)\cdot t).
\]
\end{theorem}

Let $\eta$, $f$, $g$, $S$, and $\gd$ be as in the theorem. We
will choose the parameter~$c$ at the end of the proof. For convenience,
we let $\sgd=S\d\gd$, and let $\D\eqdef\B^{t}$, so the domain of
$\gd$ is $\D\times\D$ and the domain of $\sgd$ is $\D^{\ell}\times\D^{\ell}$.

Recall the communication problem $\mKW_{f}\d\mKW_{g}$: Alice and
Bob get as inputs $m\times n$~binary matrices $X$ and~$Y$ respectively.
Let $a,b\in\B^{m}$ denote the column vectors that are obtained by
applying~$g$ to each row of~$X$ and~$Y$ respectively. Then,
$f(a)=1$ and $f(b)=0$, and the players are required to find an entry
$(i,j)$ such that $X_{i,j}>Y_{i,j}$. The rest of this section is
organized as follows.
\begin{itemize}
\item We start by proving that, without loss of generality, it can be assumed
that the players always output an entry $(i,j)$ such that $a_{i}>b_{i}$.
This is done in \ref{subsec:KRW-reduction}.
\item Then, in \ref{subsec:monotone-simpler-problem}, we show that it suffices
to prove a lower bound on a simpler communication problem, denoted
$\fsgd$.
\item We prove the lower bound on $\fsgd$ using a structure theorem, which
intuitively says that the obvious protocol for $\fsgd$ is the only
efficient protocol for $\fsgd$. In \ref{subsec:structure-thm}, we
state this structure theorem, prove it based on two lemmas, and use
it to derive the lower bound on $\fsgd$.
\item Finally, we prove the latter two lemmas in \Cref{subsec:hard-collection-lemma,subsec:second-stage-lemma}
respectively.
\end{itemize}

\subsection{\label{subsec:monotone-reductions}Reductions}

\subsubsection{\label{subsec:KRW-reduction}The observation of \cite{KRW95}}

We define the following variant of $\mKW_{f}\d\mKW_{g}$, denoted
$\mKW_{f}\dm\mKW_{g}$: The players get the same inputs as before,
but now they are required to find an entry $(i,j)$ that satisfies
both $a_{i}>b_{i}$ and $X_{i,j}>Y_{i,j}$ (rather than just $X_{i,j}>Y_{i,j}$).
Karchmer et al. \cite{KRW95} implicitly observed that $\mKW_{f}\dm\mKW_{g}$
reduces to $\mKW_{f}\d\mKW_{g}$. This means that in order to prove
\ref{monotone-composition-theorem}, it suffices to prove a lower
bound on $\mKW_{f}\dm\mKW_{g}$. We now make this observation explicit.
\begin{theorem}
The problem $\mKW_{f}\dm\mKW_{g}$ reduces to $\mKW_{f}\d\mKW_{g}$.
\end{theorem}

\begin{myproof}
We describe functions $R_{A},R_{B},\rout$ as in the definition of
a reduction (\ref{reduction}). Given a matrix $X\in\B^{m\times n}$
that is an input for Alice in $\mKW_{f}\dm\mKW_{g}$, the function~$R_{A}$
constructs an input $X'\in\B^{m\times n}$ for Alice in $\mKW_{f}\d\mKW_{g}$
as follows: For every row index~$i\in\left[m\right]$, if the $i$-th
row $X_{i}$ satisfies $g(X_{i})=1$, then we leave it intact ---
i.e., we set $X_{i}'=X_{i}$; otherwise, we set $X_{i}'$ to be the
all-zeros string. Similarly, the function $R_{B}$ takes an input
matrix~$Y\in\B^{m\times n}$ and constructs a new matrix~$Y'$ by
setting $Y_{i}'=Y_{i}$ if $g(Y_{i})=0$, and setting $Y_{i}'$ to
be the all-ones string otherwise. Finally, the function $\rout$ is
the identity function: it leaves the solution $(i,j)$ for $\mKW_{f}\d\mKW_{g}$
intact.

To prove that the reduction works, we show that if $(i,j)$ is a solution
for $\mKW_{f}\d\mKW_{g}$ on $(X',Y')$, then it is also a solution
for $\mKW_{f}\dm\mKW_{g}$ on $(X,Y)$. Let $(i,j)$ be a solution
for $\mKW_{f}\d\mKW_{g}$ on $(X',Y')$. This means that $X_{i,j}'>Y_{i,j}'$.
In particular, $X_{i}'$ is not the all-zeros string, and $Y_{i}'$
is not the all-ones string. By the definition of $R_{A},R_{B}$, it
follows that $X_{i}'=X_{i}$ and $Y_{i}'=Y_{i}$, and also that $g(X_{i})=1$
and $g(Y_{i})=0$. Therefore, $(i,j)$ is an entry that satisfies
both $a_{i}>b_{i}$ and $X_{i,j}>Y_{i,j}$. Hence, $(i,j)$ is a solution
for $\mKW_{f}\dm\mKW_{g}$ on $(X,Y)$, as required.
\end{myproof}
\begin{remark}
As discussed in the introduction, this reduction is a key technique
that works in the monotone setting but not in the non-monotone and
the semi-monotone settings. It is perhaps the main reason why it is
easier to prove composition theorems in the monotone setting.
\end{remark}

\subsubsection{\label{subsec:monotone-simpler-problem}The problem $\protect\fsgd$}

In this section, we define a new communication problem $\fsgd$ and
show that it reduces to $\mKW_{f}\dm\mKW_{g}$. Informally, the problem
$\fsgd$ is defined similarly to $\mKW_{f}\dm\mKW_{g}$, except that
the players need to solve $\sgd$ on the~$i$-th row rather than
$\mKW_{g}$. The reason that this problem is useful is that it is
more convenient to prove a lower bound on $\fsgd$ rather than directly
on~$\mKW_{f}\dm\mKW_{g}$, since $\sgd$ is a lifted search problem
and thus has a structure that we can use. For the following definition,
recall that the domain of $\sgd$ is $\D^{\ell}$, and its range is
$\cO$.
\begin{definition}
The communication problem $\fsgd$ is defined as follows: Alice gets
a matrix $X\in\D^{m\times\ell}$ and a column vector $a\in f^{-1}(1)$,
Bob gets a matrix $Y\in\D^{m\times\ell}$ and a column vector $b\in f^{-1}(0)$,
and their goal is to find a pair $(i,o)\in\left[m\right]\times\cO$
such that $a_{i}>b_{i}$ and $o\in\sgd(X_{i},Y_{i})$ (i.e., $o$
is a solution for $\sgd$ on the $i$-th rows of $X$ and~$Y$).
\end{definition}

\begin{proposition}
$\fsgd$ reduces to $\mKW_{f}\dm\mKW_{g}$.
\end{proposition}

\begin{myproof}
By assumption, $\sgd$ reduces to $\mKW_{g}$. Let $R_{A}:\D^{\ell}\to g^{-1}(1)$,
$R_{B}:\D^{\ell}\to g^{-1}(0)$, and $\rout:\left[n\right]\to\cO$
be the functions that witness the reduction. We construct a reduction
from $\fsgd$ to $\mKW_{f}\dm\mKW_{g}$ by describing appropriate
functions $R_{A}'$, $R_{B}'$, and $\rout'$.

Given an input $X\in\D^{m\times\ell}$ and $a\in f^{-1}(1)$ for Alice
in $\fsgd$, the function $R_{A}'$ constructs an input $X'\in\B^{m\times n}$
for Alice in $\mKW_{f}\dm\mKW_{g}$ as follows: for every $i\in\left[m\right]$,
we set $X_{i}'$ to $R_{A}(X_{i})$ if $a_{i}=1$ and to the all-zeros
string otherwise. The function $R_{B}'$ is defined similarly on an
input $Y\in\D^{m\times\ell}$ and $b\in f^{-1}(0)$, by setting $Y_{i}'$
to be $R_{B}(Y_{i})$ if $b_{i}=0$ and to the all-ones string otherwise.
Observe that if we apply $g$ to the rows of $X'$ and $Y'$ we get
the column vector~$a$ and~$b$ respectively. Finally, the function
$\rout'$ takes a solution $(i,j)$ for $\mKW_{f}\dm\mKW_{g}$ and
translates it to an output $(i,o)$ for $\fsgd$ by keeping $i$ intact
and setting $o=\rout(j)$.

To prove that the reduction works, we show that if $(i,j)$ is a solution
for $\mKW_{f}\dm\mKW_{g}$ on $(X',Y')$, then $(i,o)$ is also a
solution for $\fsgd$ on $\left((X,a),(Y,b)\right)$. Let $(i,j)$
be a solution for $\mKW_{f}\dm\mKW_{g}$ on $(X',Y')$. This implies
that $j$ is a solution for $\mKW_{g}$ on $(X_{i}',Y_{i}')$, and
that $a_{i}>b_{i}$. Since $a_{i}>b_{i}$, it holds that $a_{i}=1$
and $b_{i}=0$, and hence, $X_{i}'=R_{A}(X_{i})$ and $Y_{i}'=R_{B}(Y_{i})$.
It follows that $j$ is a solution for $\mKW_{g}$ on $\left(R_{A}(X_{i}),R_{B}(Y_{i})\right)$,
and therefore $o=\rout(j)$ is a solution for $\sgd$ on $(X_{i},Y_{i})$
by the definition of reduction. Thus, $(i,o)$ is a solution for $\fsgd$,
as required.
\end{myproof}

\subsection{\label{subsec:structure-thm}The structure theorem}

We turn to proving the desired lower bound on $\fsgd$. Let $q\eqdef\Q(S)$
and $\D\eqdef\B^{t}$. We prove that
\begin{equation}
\log\L(\fsgd)\ge\log\L(\mKW_{f})+\Omega(q\cdot t).\label{eq:lower-bound-on-fsgd}
\end{equation}
Observe that there is an obvious protocol for solving $\fsgd$: The
players first solve $\mKW_{f}$ on the column vectors $a,b$, thus
obtaining a coordinate $i\in\left[m\right]$ such that $a_{i}>b_{i}$.
Then, they solve $\sgd$ on $X_{i},Y_{i}$ and obtain a solution~$o$
for $\sgd$. Finally, they output the pair~$(i,o)$. The communication
complexity of this protocol is $\C(\mKW_{f})+\C(\sgd)$, and the logarithm
of its size is
\begin{align*}
\log\L(\mKW_{f})+\log\L(\sgd) & \le\log\L(\mKW_{f})+\C(\sgd)\\
 & \le\log\L(\mKW_{f})+q\cdot t.
\end{align*}
Thus, our goal is to prove that the obvious protocol is optimal in
terms of size, up to the constant factor of the $q\cdot t$ term.

We prove this bound by showing that every efficient protocol must
behave like the obvious protocol, in the sense that it must solve
$\mKW_{f}$ on $a,b$ before it starts solving $\sgd$ on the rows
$X_{i},Y_{i}$. A bit more formally, our result says that for every
protocol~$\Pi$ for $\fsgd$ the following holds: at any given point
during the execution of~$\Pi$ in which the players have not solved
$\mKW_{f}$ yet, the protocol must transmit at least another $\Omega\left(q\cdot t\right)$
bits in order to solve $\fsgd$. We refer to this result as \emph{the
structure theorem}. We state it formally below in \ref{subsec:structure-thm-statement},
and show how to use it to prove \ref{eq:lower-bound-on-fsgd} in \ref{subsec:lower-bound-on-fsgd}.
Then, we prove it based on two lemmas in \ref{subsec:proof-of-structure-theorem}.

\subsubsection{\label{subsec:structure-thm-statement}Statement of the structure
theorem}

In order to formalize the structure theorem, we need to define what
we mean when we say ``the players have not solved $\mKW_{f}$ yet''
at a given point in time. To this end, we show that the protocol~$\Pi$
contains, in a sense, a protocol for $\mKW_{f}$. Specifically, for
a fixed matrix $W\in\D^{m\times\ell}$, we define the following protocol
$\Pi_{W}$ for $\mKW_{f}$: On inputs $a,b$ for $\mKW_{f}$, the
protocol $\Pi_{W}$ invokes the protocol $\Pi$ on inputs $(W,a)$
and $(W,b)$, thus obtaining a pair $(i,o)$ such that $a_{i}>b_{i}$
and $o$ is a solution for $\sgd$ on $(W_{i},W_{i})$. Then, the
protocol $\Pi_{W}$ outputs $i$ as its solution for $\mKW_{f}$.
It is not hard to see that $\Pi_{W}$ is indeed a protocol for $\mKW_{f}$.

Now, let $\pi$ be a partial transcript that was obtained by invoking~$\Pi$
on inputs $(W,a)$ and $(W,b)$, and observe that $\pi$ can also
be viewed as a partial transcript of $\Pi_{W}$ for every $W\in\D^{m\times\ell}$.
Informally, we say that the protocol $\Pi$ has not yet solved $\mKW_{f}$
at the transcript $\pi$ if, for an average matrix~$W\in\D^{m\times\ell}$,
the protocol $\Pi_{W}$ has not solved $\mKW_{f}$ yet at $\pi$.
For short, we say that such a transcript is \emph{alive.}

We proceed to formalize this intuition. Let $\pi$ be a partial transcript
of the protocol, and let $W\in\D^{m\times\ell}$ be a matrix. We denote
by $\cX_{\pi}\times\cY_{\pi}$ the rectangle of inputs that is associated
with~$\pi$, and define
\begin{align*}
\cA_{\pi,W} & =\left\{ a\in f^{-1}(1):(W,a)\in\cX_{\pi}\right\} \\
\cB_{\pi,W} & =\left\{ b\in f^{-1}(0):(W,b)\in\cY_{\pi}\right\} .
\end{align*}
In other words, $\cA_{\pi,W}\times\cB_{\pi,W}$ is the rectangle of
inputs that is associated with $\pi$ \emph{when viewed as a transcript
of~$\Pi_{W}$}. We measure how close $\Pi_{W}$ is to solving $\mKW_{f}$
using the complexity measure 
\[
\mL(\cA_{\pi,W}\times\cB_{\pi,W})=\L(\mKW_{\cA_{\pi,W}\times\cB_{\pi,W}}).
\]
We then determine how close $\Pi$ is to solving $\mKW_{f}$ by averaging
this measure over all matrices~$W$. Formally,
\begin{definition}
\label{live-transcript}Fix a protocol~$\Pi$ for $\fsgd$. For a
transcript~$\pi$ of $\Pi$, we define
\[
\gamma(\pi)\eqdef\frac{1}{\left|\D^{m\times\ell}\right|}\cdot\sum_{W\in\D^{m\times\ell}}\mL(\cA_{\pi,W}\times\cB_{\pi,W}).
\]
We say that $\pi$ is \emph{alive }if $\gamma(\pi)\ge4m^{2}.$
\end{definition}

We are finally ready to state the structure theorem. Informally, it
says that if the protocol~$\Pi$ is currently at a live transcript,
then it must transmit at least another $\Omega\left(q\cdot t\right)$
bits in order to solve $\fsgd$. Formally, we have the following result.
\begin{theorem}[Structure theorem for $\fsgd$]
\label{structure-thm}Fix a protocol~$\Pi$ for $\fsgd$. For every
live transcript~$\pi_{1}$ of $\Pi$, there exists a suffix $\pi_{2}$
of length at least $\Omega(q\cdot t)$ such that the concatenation
$\pi_{1}\circ\pi_{2}$ is a transcript of $\Pi$.
\end{theorem}

\begin{remark}
It may seem odd that in the definition of the protocol $\Pi_{W}$
above, we give the matrix~$W$ to both players as an input, since
there is no particular reason to give the players an identical matrix.
Indeed, this requirement is made solely for convenience: We could
have worked with two matrices --- a matrix~$X$ for Alice and a
matrix~$Y$ for Bob --- but that would have been more cumbersome.
The same goes for the definition of the measure~$\gamma$: we could
have averaged over all pairs of matrices $X,Y\in\D^{m\times\ell}$
and considered the rectangle~$\cA_{\pi,X}\times\cB_{\pi,Y}$, but
using a single matrix~$W$ simplifies the presentation.
\end{remark}

\subsubsection{\label{subsec:lower-bound-on-fsgd}The lower bound on $\protect\fsgd$}

We now prove the lower bound on $\fsgd$ using the structure theorem.
Fix a protocol $\Pi$ that solves $\fsgd$.

\paragraph*{Communication complexity lower bound.}

As a warm-up, we start by proving a lower bound on the communication
complexity of $\Pi$, namely,
\begin{equation}
\C(\Pi)\ge\log\L(\mKW_{f})+\Omega(q\cdot t).\label{eq:lower-bound-on-fsgd-cc}
\end{equation}
To this end, we use the following lemma, which establishes the existence
of a relatively long live transcript.
\begin{lemma}
\label{first-stage-cc}$\Pi$ has either a live transcript of length
$\left\lfloor \log\L(\mKW_{f})-2\log m-2\right\rfloor $, or a live
transcript that is a leaf.
\end{lemma}

\begin{myproof}
The idea of the proof is the following: At the beginning of the protocol,
the complexity of solving $\mKW_{f}$ is $\log\L(\mKW_{f})$. After
the protocol transmits $\log\L(\mKW_{f})-2\log m-2$ bits, we expect
the complexity to go down to $2\log m+2$. This means that we expect
the measure~$\gamma$ to become $2^{2\log m+2}=4m^{2}$, which implies
that the corresponding transcript is alive.

This intuition is formalized using the fact that the measure $\gamma(\pi)$
of \ref{live-transcript} is subadditive on the protocol tree of~$\Pi$.
To see why, note that each of the individual terms $\mL(\cA_{\pi,W}\times\cB_{\pi,W})$
is subadditive (see \ref{subsec:Preliminaries-Karchmer-Wigderson}),
and therefore their sum is also subadditive. Next, let $M$ be the
set of vertices of $\Pi$ that are
\begin{itemize}
\item either of depth exactly $\left\lfloor \log\L(\mKW_{f})-2\log m-2\right\rfloor $;
\item or a leaf of depth at most $\left\lfloor \log\L(\mKW_{f})-2\log m-2\right\rfloor $.
\end{itemize}
It is not hard to see that $M$ is a separating set of~$\Pi$ (as
per \ref{separating-set}), and that
\[
\left|M\right|\le2^{\left\lfloor \log\L(\mKW_{f})-2\log m-2\right\rfloor }\le\L(\mKW_{f})/2^{2\log m+2}.
\]
Observe that $\gamma$ assigns to the root of $\Pi$ the value~$\L(\mKW_{f})$.
By \ref{subadditive-measure-averaging-on-cutset}, there exists a
vertex $\pi_{1}\in M$ such that
\[
\gamma(\pi_{1})\ge\frac{\L(\mKW_{f})}{\left|M\right|}\ge\frac{\L(\mKW_{f})}{\L(\mKW_{f})/2^{2\log m+2}}\ge4m^{2}.
\]
This means that $\pi_{1}$ is a live transcript of~$\Pi$, as required.
\end{myproof}
By combining \ref{first-stage-cc} with the structure theorem, we
immediately obtain the desired lower bound on the communication complexity
of $\Pi$. Indeed, \ref{first-stage-cc} says that $\Pi$ has a live
transcript~$\pi_{1}$ that is either of length~$\left\lfloor \log\L(\mKW_{f})-2\log m-2\right\rfloor $
or a leaf. The structure theorem says that there is a suffix~$\pi_{2}$
of length at least $\Omega(q\cdot t)$ such that the concatenation
$\pi_{1}\circ\pi_{2}$ is a transcript of $\Pi$. This implies in
particular that $\pi_{1}$ is not a leaf (or otherwise $\pi_{1}\circ\pi_{2}$
would not be a legal transcript of~$\Pi$), and hence $\pi_{1}$
is a partial transcript of length exactly $\left\lfloor \log\L(\mKW_{f})-2\log m-2\right\rfloor $.
It follows that $\pi_{1}\circ\pi_{2}$ is a full transcript of~$\Pi$
of length at least
\[
\left\lfloor \log\L(\mKW_{f})-2\log m-2\right\rfloor +\Omega(q\cdot t)\ge\log\L(\mKW_{f})+\Omega(q\cdot t),
\]
where the inequality uses the fact that $t\gg\log m$. Hence, the
communication complexity of~$\Pi$ is at least $\log\L(\mKW_{f})+\Omega(q\cdot t)$
as required.

\paragraph*{Protocol size lower bound.}

While the above argument proves a lower bound on $\C(\fsgd)$, our
actual goal is to obtain a lower bound on the \emph{protocol size}
of $\fsgd$, which is a stronger statement. That is, we would like
to prove that 
\[
\log\L(\Pi)\ge\log\L(\mKW_{f})+\Omega(q\cdot t).
\]
We stress that we cannot derive this lower bound from \ref{eq:lower-bound-on-fsgd-cc}
directly using protocol balancing (\ref{protocol-balancing}), since
that would lose a constant factor in the term $\log\L(\mKW_{f})$
and we cannot afford that loss. Nevertheless, we can afford to apply
protocol balancing to the structure theorem, since we can afford to
lose a constant factor in the $\Omega(q\cdot t)$ term. This leads
to the following corollary, which will be used to prove the lower
bound on~$\L(\Pi)$.
\begin{corollary}
\label{structure-thm-for-size}For every live transcript~$\pi_{1}$
of $\Pi$, there exist at least $2^{\Omega(q\cdot t)}$ suffixes $\pi_{2}$
such that the concatenation $\pi_{1}\circ\pi_{2}$ is a full transcript
of $\Pi$.
\end{corollary}

\begin{myproof}
Let $\pi_{1}$ be a live transcript of~$\Pi$, and let $\Pi_{2}$
be the sub-tree of~$\Pi$ that is rooted in $\pi_{1}$. We prove
that $\L(\Pi_{2})\ge2^{\Omega(q\cdot t)}$, and this implies the desired
claim. By \ref{protocol-balancing}, there exists a protocol $\Pi_{2}'$
that is equivalent to~$\Pi_{2}$ and has communication complexity
at most $4\log\L(\Pi_{2})$. Let $\Pi'$ be the protocol obtained
from $\Pi$ by replacing $\Pi_{2}$ with $\Pi_{2}'$. 

Now, $\Pi'$ is a protocol that solves $\fsgd$, and $\pi_{1}$ is
a live transcript of~$\Pi'$, so by \ref{structure-thm} there exists
a suffix $\pi_{2}$ of length at least $\Omega(q\cdot t)$ such that
the concatenation $\pi_{1}\circ\pi_{2}$ is a transcript of $\Pi'$.
This means that $\pi_{2}$ is a transcript of $\Pi_{2}'$ that has
length at least $\Omega(q\cdot t)$, and therefore $\C(\Pi_{2}')\ge\Omega(q\cdot t)$.
It follows that
\begin{align*}
4\log\L(\Pi_{2}) & \ge\C(\Pi_{2}')\ge\Omega(q\cdot t)\\
\log\L(\Pi_{2}) & \ge\Omega(q\cdot t),
\end{align*}
as required.
\end{myproof}
We now prove the lower bound on~$\L(\Pi)$. Ideally, we would have
liked to prove that if $\Pi$ did not have many leaves, then there
would have to be at least one live transcript~$\pi_{1}$ that does
not have many leaves in its rooted sub-tree. Since the existence of
such~$\pi_{1}$ contradicts \ref{structure-thm-for-size}, this would
prove that $\Pi$ must have many leaves.

The latter ``ideal claim'' about $\Pi$ is not true in general.
However, \cite{KM18} observed that $\Pi$ can be transformed into
an equivalent protocol~$\Pi'$ that does satisfy that claim, and
is not much larger than~$\Pi$. We can therefore use the foregoing
argument to show that $\Pi'$ has many leaves, and then argue that
since $\Pi'$ is not much larger than~$\Pi$, the protocol~$\Pi$
must have many leaves as well. The transformation of~$\Pi$ is done
by the following lemma of \cite{KM18}.
\begin{lemma}[\cite{KM18}, following \cite{T14}]
\label{protocol-decomposition}Let $\Pi$ be a protocol, and let
$s\in\N$ be a parameter such that $s\le\L(\Pi)$. Then there exists
an equivalent protocol $\Pi'$ that satisfies the following: the protocol
tree $\Pi'$ has a separating set $\pi_{1},\dots,\pi_{k}$ where $k\leq\frac{36\cdot\L(\Pi)}{s}$,
such that for every $i\in\left[k\right]$, the subtree rooted at $\pi_{i}$
has at most $s$ leaves.
\end{lemma}

By \ref{structure-thm-for-size}, there exists some $L=2^{\Omega(q\cdot t)}$
such that every live transcript~$\pi_{1}$ has at least $L$~suffixes.
We prove that
\begin{equation}
\log\L(\Pi)\ge\log\L(\mKW_{f})+\log L-2\log m-9,\label{eq:protocol-size-lb-fsgd}
\end{equation}
and this would imply that $\log\L(\fsgd)\ge\log\L(\mKW_{f})+\Omega(q\cdot t)$,
as required. Suppose for the sake of contradiction that \ref{eq:protocol-size-lb-fsgd}
does not hold, that is, we assume that
\[
\L(\Pi)<\frac{\L(\mKW_{f})\cdot L}{512\cdot m^{2}}.
\]
Let $\Pi'$ be the protocol that is obtained by applying \ref{protocol-decomposition}
to~$\Pi$ with $s=L/2$. Then, the protocol tree $\Pi'$ has a separating
set $\pi_{1},\dots,\pi_{k}$ such that
\[
k\le\frac{36\cdot\L(\Pi)}{L/2}<\frac{\L(\mKW_{f})}{4\cdot m^{2}},
\]
and such that for every $i\in\left[k\right]$, the subtree rooted
at $\pi_{i}$ has at most $L/2$ leaves. Now, recall that the measure
$\gamma(\pi)$ is subadditive on the protocol tree of~$\Pi'$. Moreover,
recall that $\gamma$ assigns to the root of $\Pi'$ the value~$\L(\mKW_{f})$.
Thus, by \ref{subadditive-measure-averaging-on-cutset}, there exists
a transcript $\pi_{i}$ in the separating set such that
\[
\gamma(\pi_{i})\ge\frac{\L(\mKW_{f})}{k}>\frac{\L(\mKW_{f})}{\L(\mKW_{f})/4m^{2}}=4m^{2}.
\]
This means that $\pi_{i}$ is alive, and therefore by \ref{structure-thm-for-size}
there are at least $L$~leaves in the sub-tree of $\Pi'$ that is
rooted in~$\pi_{i}$. However, this contradicts the fact that there
are at most $L/2$ such leaves, and hence \ref{eq:protocol-size-lb-fsgd}
holds.

\subsubsection{\label{subsec:proof-of-structure-theorem}Proof of structure theorem
from lemmas}

Let $\Pi$ be a protocol that solves $\fsgd$. Our goal is to prove
that if the protocol reaches a live transcript~$\pi_{1}$, then it
still has to transmit at least $\Omega(q\cdot t)$ bits in order to
solve $\fsgd$. The intuition for the proof is the following: The
goal of the players is to solve $\sgd$ on some row~$i$ where $a_{i}>b_{i}$.
By assumption, it is necessary to transmit $\Omega(q\cdot t)$ bits
in order to solve~$\sgd$ \emph{from scratch}. However, it could
be the case that the transcript~$\pi_{1}$ contains information that
helps in solving~$\sgd$ on some rows, which means that the players
may need to transmit less than $\Omega(q\cdot t)$~bits in order
to solve~$\sgd$ on those rows. The crucial point is that since at
$\pi_{1}$ the players have not yet solved $\KW_{f}$ on $a,b$, they
do not know on which row of $X,Y$ they should be solving $\sgd$.
Thus, the information that the players communicated about~$X,Y$
in~$\pi_{1}$ is likely to be wasted on irrelevant rows where $a_{i}\le b_{i}$.
Hence, we might as well assume that the players have not made progress
toward solving~$\sgd$ in~$\pi_{1}$, so they still have to transmit
$\Omega(q\cdot t)$~bits in order to solve $\sgd$ on some row.

This intuition is formalized as follows. Given a live transcript~$\pi_{1}$,
we partition the rows of the matrices $X,Y$ into two types:
\begin{itemize}
\item ``Revealed rows'', about which the transcript~$\pi_{1}$ reveals
a lot of information (i.e., more than two bits of information).
\item ``Unrevealed rows'', about which the transcript~$\pi_{1}$ reveals
only a little information (i.e., at most two bits of information).
\end{itemize}
Intuitively, if the protocol chooses to solve~$\sgd$ on an \emph{unrevealed}
row, then it has to send $\Omega(q\cdot t)$~additional bits, since
it barely made any progress on this row in~$\pi_{1}$. Thus, it suffices
to show that we can prevent the protocol from solving $\sgd$ on the
revealed rows. This corresponds to our previous intuition that if
the players communicate about some rows before solving $\mKW_{f}$,
then this communication is wasted.

In order to force the protocol to solve $\sgd$ on the unrevealed
rows, we show that we can find a subset of the inputs that are consistent
with~$\pi_{1}$ and that satisfy that $a_{i}\le b_{i}$ holds for
every revealed row~$i$. This means that on those inputs, the protocol
is not allowed to output a solution to~$\sgd$ in any revealed row.
The reason that we can find such a subset of inputs is that we assumed
that at $\pi_{1}$ the players have not solved $\mKW_{f}$ yet, and
hence at this point they do not know any row~$i$ for which $a_{i}>b_{i}$.
Therefore, when the protocol is invoked on this subset of inputs,
it must solve $\sgd$ on an unrevealed row, and therefore must transmit
about $\Omega(q\cdot t)$ additional bits, as required. The following
definition captures the subset of inputs that we would like to construct.
\begin{definition}
\label{hard-collection}A \emph{collection} consists of a set of matrices~$\cW\subseteq\D^{m\times\ell}$,
and of column vectors $a^{W}\in f^{-1}(1)$ and $b^{W}\in f^{-1}(0)$
for each matrix $W\in\cW$. We say that a transcript~$\pi_{1}$ of
$\Pi$ with a corresponding rectangle $\cX_{\pi_{1}}\times\cY_{\pi_{1}}$
\emph{supports} \emph{the collection} if for every matrix $W\in\cW$,
it holds that $(W,a^{W})\in\cX_{\pi_{1}}$ and $(W,b^{W})\in\cY_{\pi_{1}}$.
We say that the collection is \emph{hard} if there exists a set $R\subseteq\left[m\right]$
of ``revealed rows'' that satisfies the following:
\begin{itemize}
\item For every set $I\subseteq\left[m\right]-R$:
\[
\rAD_{I\times\left[\ell\right]}\left(\cW|_{(\left[m\right]-R)\times\left[\ell\right]}\right)\ge\frac{1}{4^{\left|I\right|}}
\]
(i.e., at most $2\left|I\right|$ bits of information were revealed
on every set~$I$ of unrevealed rows).
\item For every $W,W'\in\cW$, it holds that $a^{W}|_{R}\le b^{W'}|_{R}$.
\end{itemize}
\end{definition}

\smallskip{}

\noindent We now state two lemmas: the first lemma says that we can
always find a hard collection of inputs, and the second lemma says
that the complexity of solving $\fsgd$ on such a collection is $\Omega(q\cdot t)$.
Together, those two lemmas imply the structure theorem, and they are
proved in Sections~\oldref{subsec:hard-collection-lemma} and~\oldref{subsec:second-stage-lemma}
respectively.
\begin{lemma}
\label{hard-collection-lemma}Every live transcript of~$\Pi$ supports
a hard collection.
\end{lemma}

\begin{lemma}
\label{second-stage-lemma}If a transcript $\pi_{1}$ supports a hard
collection, then there exists a suffix~$\pi_{2}$ of length at least
$\Omega(q\cdot t)$ such that $\pi_{1}\circ\pi_{2}$ is a transcript
of~$\Pi$.
\end{lemma}

\noindent The structure theorem follows immediately by combining the
two lemmas.

\subsection{\label{subsec:hard-collection-lemma}Proof of \ref{hard-collection-lemma}}

Fix a protocol $\Pi$ that solves $\fsgd$, and let $\pi_{1}$ be
a live transcript of $\Pi$. Our goal is to construct a hard collection
that is supported by $\pi_{1}$. To this end, we identify a set of
matrices~$\cW$, a set of revealed rows~$R$, and column vectors
$a^{W}\in\cA_{\pi_{1},W}$ and $b^{W}\in\cB_{\pi_{1},W}$. We then
show that $a^{W}|_{R}\le b^{W'}|_{R}$ holds for every~$W,W'\in\cW$.
Our proof is a straightforward adaptation of an argument of \cite{KM18}
to the monotone setting.

Our assumption that $\pi_{1}$ is alive means that $\mL(\cA_{\pi_{1},W}\times\cB_{\pi_{1},W})$
is sufficiently large for the average matrix~$W$. In order to carry
out our argument, we need to start from a stronger assumption, namely,
that there is a significant number of matrices~$W$ for which $\mL(\cA_{\pi_{1},W}\times\cB_{\pi_{1},W})$
is sufficiently large. This can be proved by a standard averaging
argument. Formally, in \ref{subsec:hard-collection-lemma-averaging}
below we prove the following result.
\begin{proposition}
\label{hard-collection-lemma-averaging}There exists a number $p\in\N$
and a set of matrices $\cW_{0}\subseteq\D^{m\times\ell}$ such that
$\dns(\cW_{0})\ge2^{-p}$, and such that for every $W\in\cW_{0}$:
\begin{equation}
\log\mL(\cA_{\pi_{1},W}\times\cB_{\pi_{1},W})>p+\log m.\label{eq:hard-collection-lemma-overview-complexity-assumption}
\end{equation}
\end{proposition}

Recall that the transcript~$\pi_{1}$ is obtained by invoking the
protocol~$\Pi$ on inputs of the form $(W,a)$ and~$(W,b)$. Intuitively,
\ref{hard-collection-lemma-averaging} means that when we restrict
ourselves to~$\cW_{0}$, the transcript~$\pi_{1}$ reveals at most
$p$ bits of information about the matrix~$W$, and still it has
to transmit more than $p+\log m$ bits to solve $\mKW_{f}$ on $(a,b)$. 

\paragraph*{Warm-up.}

Before we explain the construction of the hard collection, we first
present a simplified version of the argument. Let $R\subseteq\left[m\right]$
denote the set of rows of~$W$ on which $\pi_{1}$ reveals more than
two bits of information. Since $\pi_{1}$ reveals at most $p$~bits
of information about the whole matrix~$W$, it follows that $\left|R\right|\le p/2$.

We would now like to choose column vectors $a^{W}\in\cA_{\pi_{1},W}$
and $b^{W}\in\cB_{\pi_{1},W}$, such that for every two matrices $W,W'$
in the collection we have that $a^{W}|_{R}\le b^{W'}|_{R}$. We start
by choosing, for every $W\in\cW_{0}$, a pair of column vectors $a^{W},b^{W}$
that satisfy $a^{W}|_{R}\le b^{W}|_{R}$ only for~$W$. To see why
this is possible, let $W\in\cW_{0}$, and suppose that such column
vectors $a^{W},b^{W}$ did not exist for~$W$. We claim that in this
case, it is possible to solve $\mKW_{f}$ on the rectangle $\cA_{\pi_{1},W}\times\cB_{\pi_{1},W}$
by communicating at most 
\begin{equation}
\left|R\right|+\log m<p+\log m\label{eq:hard-collection-lemma-overview-protocol-complexity}
\end{equation}
bits, contradicting \ref{eq:hard-collection-lemma-overview-complexity-assumption}.
This is done as follows: By our assumption, for every $a\in\cA_{\pi_{1},W}$
and $b\in\cB_{\pi_{1},W}$, it holds that $a_{i}>b_{i}$ for some~$i\in R$.
Alice will send $a_{R}$ to Bob, and Bob will reply with the corresponding
coordinate $i\in R$, thus solving $\mKW_{f}$ using at most $\left|R\right|+\log m$
bits.

Hence, we can choose for every matrix~$W\in\cW_{0}$ a pair of column
vectors $a^{W},b^{W}$ such that $a^{W}|_{R}\le b^{W}|_{R}$. It remains
to enforce the condition $a^{W}|_{R}\le b^{W'}|_{R}$ for every \emph{two}
\emph{matrices}~$W,W'$. To this end, let us denote by $\alpha_{R}$
the most popular value of $a^{W}|_{R}$ over all matrices~$W\in\cW_{0}$.
We take our hard collection~$\cW$ to be the subset of matrices $W\in\cW_{0}$
for which $a^{W}|_{R}=\alpha_{R}$, and discard all the other matrices.
It now holds for every $W,W'\in\cW$ that
\[
a^{W}|_{R}=\alpha_{R}=a_{R}^{W'}\le b_{R}^{W'},
\]
as required.

It might seem as if the collection $\cW$ satisfies our requirements.
Indeed, we have a set of revealed rows~$R$, and $a^{W}|_{R}\le b^{W'}|_{R}$
holds for every \emph{$W,W'\in\cW$. }However, the above reasoning
suffers from the following issue: When we moved from $\cW_{0}$ to~$\cW$,
we revealed additional bits of information about the matrices~$W$.
This newly leaked information may create new revealed rows that do
not belong to~$R$, thus violating the definition of a hard collection.

\paragraph*{The actual proof.}

We resolve the latter issue by repeating the foregoing argument iteratively:
We start by setting $\cW=\cW_{0}$ and~$R=\emptyset$. Then, in each
iteration, we identify a set~$I$ of revealed rows, add it to~$R$,
and move to a subset of~$\cW$ in which all the column vectors~$a^{W}$
have the same value $\alpha_{I}$. The process ends when there are
no more revealed rows. In \ref{subsec:hard-collection-lemma-the-iterative-procedure}
below, we show that this process yields the following.
\begin{proposition}
\label{hard-collection-lemma-the-iterative-procedure}There exists
a set of matrices~$\cW\subseteq\cW_{0}$, a set of revealed rows~$R\subseteq\left[m\right]$,
and for each matrix~$W$, a set $\cA^{W}\subseteq\cA_{\pi_{1},W}$
of candidates for~$a^{W}$ such that the following properties are
satisfied:
\begin{enumerate}
\item \label{enu:hard-collection-lemma-degree-bound}For every $I\subseteq\left[m\right]-R$:
\[
\rAD_{I\times\left[\ell\right]}(\cW|_{(\left[m\right]-R)\times\left[\ell\right]})\ge\frac{1}{4^{\left|I\right|}}.
\]
\item \label{enu:hard-collection-lemma-agreement-on-R}All the candidate
vectors in~$\cA^{W}$ for all the matrices~$W\in\cW$ agree on the
coordinates in~$R$.
\item \label{enu:hard-collection-lemma-complexity-bound}For every $W\in\cW$,
it holds that $\mL(\cA^{W}\times\cB_{\pi_{1},W})>m$.
\end{enumerate}
\end{proposition}

Let $\cW$, $R$, and $\cA^{W}$ be the sets obtained from \ref{hard-collection-lemma-the-iterative-procedure}.
We choose $\cW$ to be the set of matrices in our hard collection.
At this point, we know that the set $\cW$ satisfies the first condition
in the definition of a hard collection due to Property~\ref{enu:hard-collection-lemma-degree-bound}
above. We now explain how to choose the column vectors~$a^{W}\in\cA_{\pi_{1},W}$
and~$b^{W}\in\cB_{\pi_{1},W}$ to satisfy $a^{W}|_{R}\le b^{W'}|_{R}$
for every \emph{$W,W'\in\cW$}, and this will complete the proof of
\ref{hard-collection-lemma}.

For every matrix~$W\in\cW$, we choose $a^{W}$ arbitrarily from~$\cA^{W}$.
By Property~\ref{enu:hard-collection-lemma-agreement-on-R}, all
the column vectors~$a^{W}$ of all the matrices~$W$ agree on the
coordinates in~$R$; let us denote this agreed value by~$\alpha_{R}$.
Finally, we choose the column vectors~$b^{W}$ using the following
result.
\begin{claim}
\label{hard-collection-lemma-choosing-b}For every matrix~$W\in\cW$,
there exists a column vector $b^{W}\in\cB_{\pi_{1},W}$ such that
$b^{W}|_{R}\ge\alpha_{R}$.
\end{claim}

\begin{myproof}
Let $W\in\cW$. Suppose for the sake of contradiction that there exists
no column vector $b^{W}\in\cB_{\pi_{1},W}$ such that $b^{W}|_{R}\ge\alpha_{R}$.
We show that in this case there exists a protocol that solves $\mKW_{f}$
on $\cA^{W}\times\cB_{\pi_{1},W}$ using $\log m$ bits, which contradicts
the fact that $\log\mL(\cA^{W}\times\cB_{\pi_{1},W})>\log m$ by Property
\ref{enu:hard-collection-lemma-complexity-bound}.

We use the following protocol: Alice gets a column vector~$a\in\cA^{W}$,
and Bob gets a column vector $b\in\cB_{\pi_{1},W}$. Note that $a_{R}=\alpha_{R}$
by the definition of~$\alpha_{R}$. Moreover, by our assumption,
it does not hold that $b_{R}\ge\alpha_{R}$, and therefore there exists
some coordinate~$i\in R$ such that $(\alpha_{R})_{i}>b_{i}$. We
know that $a_{i}=(\alpha_{R})_{i}$, so $a_{i}>b_{i}$, and therefore~$i$
is a solution for $\mKW_{f}$ on $\cA^{W}\times\cB_{\pi_{1},W}$.
Furthermore, Bob knows $b$, and also knows~$\alpha_{R}$ (since
it does not depend on Alice's input), and therefore he can deduce
$i$. Hence, Bob can send $i$ to~Alice, thus solving the problem.
It is easy to see that this protocol sends at most $\log m$ bits,
so we reached the desired contradiction.
\end{myproof}
We conclude by showing that the column vectors~$a^{W},b^{W}$ that
we chose satisfy that $a^{W}|_{R}\le b^{W'}|_{R}$ for every $W,W'\in\cW$.
Let $W,W'\in\cW$. Then, by \ref{hard-collection-lemma-choosing-b},
\[
a^{W}|_{R}=\alpha_{R}\le b^{W'}|_{R},
\]
as required.

\subsubsection{\label{subsec:hard-collection-lemma-averaging}The initial set~$\protect\cW_{0}$}

We now prove \ref{hard-collection-lemma-averaging}, which constructs
the initial set~$\cW_{0}$ for our argument.
\begin{restated}{\ref{hard-collection-lemma-averaging} (restated)}
There exists a number $p\in\N$ and a set of matrices $\cW_{0}\subseteq\D^{m\times\ell}$
such that $\dns(\cW_{0})\ge2^{-p}$, and such that for every $W\in\cW_{0}$:
\[
\log\mL(\cA_{\pi_{1},W}\times\cB_{\pi_{1},W})>p+\log m.
\]
\end{restated}

\begin{myproof}
By assumption, the transcript $\pi_{1}$ is alive, and therefore
\[
\gamma(\pi_{1})=\frac{1}{\left|\D^{m\times\ell}\right|}\cdot\sum_{W\in\D^{m\times\ell}}\mL(\cA_{\pi_{1},W}\times\cB_{\pi_{1},W})\ge4\cdot m^{2}.
\]
In other words, 
\[
\sum_{W\in\D^{m\times\ell}}\mL(\cA_{\pi_{1},W}\times\cB_{\pi_{1},W})\ge4\cdot m^{2}\cdot\left|\D^{m\times\ell}\right|.
\]
We partition the matrices~$W$ into $m-\log m$ buckets as follows:
the first bucket~$\cV_{1}$ consists of all matrices~$W$ for which
\[
\mL(\cA_{\pi_{1},W}\times\cB_{\pi_{1},W})\le2m,
\]
and for every $k>1$, the $k$-th bucket~$\cV_{k}$ consists of all
matrices~$W$ for which
\[
2^{k-1}\cdot m<\mL(\cA_{\pi_{1},W}\times\cB_{\pi_{1},W})\le2^{k}\cdot m.
\]
For every $k\in\left[m-\log m\right]$, we define the \emph{weight
of a bucket~$\cV_{k}$} to be the sum
\[
\sum_{W\in\cV_{k}}\mL(\cA_{\pi_{1},W}\times\cB_{\pi_{1},W}).
\]
Our assumption that $\pi_{1}$ is alive says that the total weight
of all the buckets together is at least $4\cdot m^{2}\cdot\left|\D^{m\times\ell}\right|$.
Moreover, it is easy to see that the weight of $\cV_{1}$ is at most
$2\cdot m\cdot\left|\D^{m\times\ell}\right|$. Hence, the total weight
of all buckets except the first bucket is at least
\[
4\cdot m^{2}\cdot\left|\D^{m\times\ell}\right|-2\cdot m\cdot\left|\D^{m\times\ell}\right|\ge2\cdot m^{2}\cdot\left|\D^{m\times\ell}\right|.
\]
By an averaging argument, there exists $k\in\left[m-\log m\right]-\left\{ 1\right\} $
such that the weight of~$\cV_{k}$ is at least
\[
\frac{2\cdot m^{2}\cdot\left|\D^{m\times\ell}\right|}{m-\log m-1}\ge2\cdot m\cdot\left|\D^{m\times\ell}\right|.
\]
We choose $\cW_{0}\eqdef\cV_{k}$ and $p\eqdef k-1$. By definition,
for every $W\in\cW_{0}$ we have
\[
\mL(\cA_{\pi_{1},W}\times\cB_{\pi_{1},W})>2^{k-1}\cdot m=2^{p}\cdot m
\]
and hence
\[
\log\mL(\cA_{\pi_{1},W}\times\cB_{\pi_{1},W})>p+\log m.
\]
It remains to lower bound the size of~$\cW_{0}$. To this end, recall
that the weight of $\cW_{0}$ is at least $2\cdot m\cdot\left|\D^{m\times\ell}\right|$.
On the other hand, for every $W\in\cW_{0}$:
\[
\mL(\cA_{\pi_{1},W}\times\cB_{\pi_{1},W})\le2^{k}\cdot m=2^{p+1}\cdot m.
\]
Hence, the number of elements in~$\cW_{0}$ must be at least
\[
\frac{2\cdot m\cdot\left|\D^{m\times\ell}\right|}{2^{p+1}\cdot m}=2^{-p}\cdot\left|\D^{m\times\ell}\right|,
\]
as required.
\end{myproof}

\subsubsection{\label{subsec:hard-collection-lemma-the-iterative-procedure}The
iterative procedure}

We conclude the proof of the lemma by proving \ref{hard-collection-lemma-the-iterative-procedure},
restated next.
\begin{restated}{\ref{hard-collection-lemma-the-iterative-procedure} (restated)}
There exists a set of matrices~$\cW\subseteq\cW_{0}$, a set of
revealed rows~$R\subseteq\left[m\right]$, and for each matrix~$W$,
a set $\cA^{W}\subseteq f^{-1}(1)$ of candidates for~$a^{W}$ such
that properties are satisfied:
\begin{enumerate}
\item For every $I\subseteq\left[m\right]-R$:
\[
\rAD_{I\times\left[\ell\right]}(\cW|_{(\left[m\right]-R)\times\left[\ell\right]})\ge\frac{1}{4^{\left|I\right|}}.
\]
\item All the candidate vectors in~$\cA^{W}$ for all the matrices~$W\in\cW$
agree on the coordinates in~$R$.
\item For every $W\in\cW$, it holds that $\mL(\cA^{W}\times\cB_{\pi_{1},W})>m$.
\end{enumerate}
\end{restated}

In order to streamline the presentation, we denote the set of unrevealed
rows by $U\eqdef\left[m\right]-R$. For convenience, throughout the
procedure we will maintain the property that every submatrix~$W'\in\cW|_{U\times\left[\ell\right]}$
has a unique extension to a matrix~$W\in\cW$. Intuitively, this
property is convenient since only the value of the unrevealed rows
of a matrix matters. We refer to this invariant as the \emph{unique
extension property}.

Let $\cW_{0}$ be the set of matrices obtained from~\ref{hard-collection-lemma-averaging}.
The procedure starts by setting $\cW=\cW_{0}$, $R=\emptyset$, and
$\cA^{W}=\cA_{\pi_{1},W}$ for every $W\in\cW$. Now, as long as there
exists a non-empty set $I\subseteq U$ such that 
\[
\rAD_{I\times\left[\ell\right]}(\cW|_{U\times\left[\ell\right]})<\frac{1}{4^{\left|I\right|}},
\]
we perform the following steps:
\begin{enumerate}
\item \label{enu:hard-collection-lemma-adding-I}We add $I$ to~$R$ (and
remove $I$ from~$U$).
\item \label{enu:hard-collection-lemma-unique-extension}We restore the
unique extension invariant by choosing for every submatrix $W'\in\cW|_{U\times\left[\ell\right]}$
a single extension $W\in\cW$, and removing all the other extensions
of~$W'$ from~$\cW$.
\item \label{enu:hard-collection-lemma-fixing-a_I-individual-matrix}For
every $W\in\cW$, we make sure that all column vectors in~$\cA^{W}$
agree on the coordinates in~$I$ as follows:
\begin{enumerate}
\item For each $W\in\cW$, we partition~$\cA^{W}$ into buckets $\left\{ \cA^{W,v}\right\} _{v\in\B^{I}}$,
such that the bucket~$\cA^{W,v}$ contains the column vectors~$a\in\cA^{W}$
that satisfy $a_{I}=v$. 
\item Let $v_{W}$ be the value that maximizes $\mL(\cA^{W,v}\times\cB_{\pi_{1},W})$. 
\item We replace $\cA^{W}$ with the bucket~$\cA^{W,v_{W}}$ .
\end{enumerate}
\item \label{enu:hard-collection-lemma-fixing-a_I-all-matrices}Finally,
we make sure that all column vectors \emph{of all matrices} agree
on the coordinates in~$I$ as follows:
\begin{enumerate}
\item Let $\alpha_{I}$ be the most popular value among all the $v_{W}$'s.
\item We replace $\cW$ with the subset of matrices $W$ for which $v_{W}=\alpha_{I}.$
\end{enumerate}
\end{enumerate}
By definition, when the procedure ends, Property \ref{enu:hard-collection-lemma-degree-bound}
of \ref{hard-collection-lemma-the-iterative-procedure} is satisfied.
Moreover, it is easy to see that Property~\ref{enu:hard-collection-lemma-agreement-on-R}
is satisfied.

It remains to show that Property~\ref{enu:hard-collection-lemma-complexity-bound}
is satisfied. To this end, recall that when the procedure starts,
every $W\in\cW$ satisfies $\mL(\cA_{\pi_{1},W}\times\cB_{\pi_{1},W})>2^{p}\cdot m$
by the definition of~$\cW_{0}$. Next, observe that in every iteration,
Step~\ref{enu:hard-collection-lemma-fixing-a_I-individual-matrix}
decreases $\mL(\cA^{W}\times\cB_{\pi_{1},\cW})$ by a factor of at
most $2^{\left|I\right|}$ by the subadditivity of $\mL(\cA^{W}\times\cB_{\pi_{1},\cW})$.
All the other steps of the procedure do not affect $\mL(\cA^{W}\times\cB_{\pi_{1},\cW})$
at all. Hence, by the time the procedure halts, the value $\mL(\cA^{W}\times\cB_{\pi_{1},\cW})$
has decreased by a factor of at most $2^{\left|R\right|}$, so $\mL(\cA_{\pi_{1},W}\times\cB_{\pi_{1},W})>2^{p-\left|R\right|}\cdot m$.
Thus, to prove that $\mL(\cA^{W}\times\cB_{\pi_{1},W})>m$, it suffices
to show that $\left|R\right|\le p$, which we establish next.
\begin{claim}
\label{hard-collection-lemma-bound-on-revealed-rows}When the procedure
halts, $\left|R\right|\le p$.
\end{claim}

\begin{myproof}
We upper bound the size of~$R$ using a potential argument. Intuitively,
the potential function is the amount of information the players know
about the rows in~$U$. At the beginning of the process, $U=\left[m\right]$,
and the players know $p$ bits of information about all the rows together.
For every revealed row~$i$ that is added to~$R$, the potential
is decreased by at least two, since the two bits that the players
knew about the row~$i$ are discarded. Then, when the value $a_{i}$
is fixed to a constant~$\alpha_{i}$, it reveals at most one bit
of information, thus increasing the potential by at most one. All
in all, each revealed row that is added to~$R$ decreases the potential
function by at least one. Since the potential starts from~$p$ and
is always non-negative, it follows that the number of revealed rows
will never surpass~$p$, which is what we wanted to prove.

Formally, our potential function is the density of $\cW|_{U\times\left[\ell\right]}$.
Recall that at the beginning of this procedure, this density is at
least $2^{-p}$ by the definition of~$\cW_{0}$. We prove that in
every iteration, the density of~$\cW|_{U\times\left[\ell\right]}$
increases by a factor of at least~$2^{\left|I\right|}$, where $I$
is the set of rows that is added to~$R$ at the iteration. Note that
this implies the claim, since the density of a set can never exceed~$1$,
and $R$ consists of the union of all the sets~$I$.

Fix a single iteration. By assumption, at the beginning of the iteration
we have
\[
\rAD_{I\times\left[\ell\right]}(\cW|_{U\times\left[\ell\right]})<\frac{1}{4^{\left|I\right|}}.
\]
In Step~\ref{enu:hard-collection-lemma-adding-I}, the procedure
removes~$I$ from~$U$. To see how this step affects the density
of $\cW|_{U\times\left[\ell\right]}$, observe that \ref{removing-coordinates-average-degree}
implies that

\[
\dns\left(\cW|_{(U-I)\times\left[\ell\right]}\right)\ge\frac{1}{\rAD_{I\times\left[\ell\right]}(\cW|_{U\times\left[\ell\right]})}\cdot\dns(\cW|_{U\times\left[\ell\right]})>4^{\left|I\right|}\cdot\dns(\cW|_{U\times\left[\ell\right]}).
\]
Thus, Step~\ref{enu:hard-collection-lemma-adding-I} increases the
density at least by a factor of~$4^{\left|I\right|}$. Steps~\ref{enu:hard-collection-lemma-unique-extension}
and~\ref{enu:hard-collection-lemma-fixing-a_I-individual-matrix}
do not affect the density of~$\cW|_{U\times\left[\ell\right]}$ at
all. Finally, it is not hard to see that Step~\ref{enu:hard-collection-lemma-fixing-a_I-all-matrices}
decreases the size of~$\cW|_{U\times\left[\ell\right]}$ by a factor
of at most~$2^{\left|I\right|}$. All in all, at the end of the iteration,
the density of $\cW|_{U\times\left[\ell\right]}$ is increased by
at least a factor of $2^{\left|I\right|}$, as required.
\end{myproof}
This concludes the proof of \ref{hard-collection-lemma}.

\subsection{\label{subsec:second-stage-lemma}Proof of \ref{second-stage-lemma}}

In this section, we prove \ref{second-stage-lemma}. Let $\pi_{1}$
be a transcript that supports a hard collection~$\cW$, and let $\cX_{\pi_{1}}\times\cY_{\pi_{1}}$
be its associated rectangle. Our goal is to prove that the communication
complexity of solving $\fsgd$ on the inputs in $\cX_{\pi_{1}}\times\cY_{\pi_{1}}$
is at least $\Omega(q\cdot t)$. We use the following proof strategy:
We observe that solving $\fsgd$ on $\cX_{\pi_{1}}\times\cY_{\pi_{1}}$
amounts to solving sub-problem~$H$ of some lifted problem $S'\d\gd$.
Then, we apply to~$H$ our generalized lifting theorem, which deals
with sub-problems of lifted search problems, thus obtaining a lower
bound on $\fsgd$. More details follow.

Let $R$ be the set of revealed rows of the hard collection~$\cW$,
and let $U\eqdef\left[m\right]-R$ denote the set of unrevealed rows.
Let $\cW'$ denote the projection of the matrices in $\cW$ to the
rows in~$U$. The communication problem~$H$ is defined as follows:
Alice gets a matrix $X'\in\cW'$, Bob gets a matrix $Y'\in\cW'$,
and their goal is to output $(i,o)\in U\times\cO$ such that $o\in\sgd(X_{i}',Y_{i}')$.
We have the following observation.
\begin{proposition}
$H$ reduces to solving $\fsgd$ on the inputs in $\cX_{\pi_{1}}\times\cY_{\pi_{1}}$.
\end{proposition}

\begin{myproof}
We define the functions $R_{A},R_{B},\rout$ of the reduction. Given
an input $X'\in\cW'$ of Alice in~$H$, the function $R_{A}$ translates
it to an input $(X,a^{X})$ of Alice in $\fsgd$, where $X\in\cW$
is an arbitrary fixed extension of $X'$ to a matrix in~$\cW$. We
define $R_{B}(Y')\eqdef(Y,b^{Y})$ similarly. Finally, we set $\rout$
to be the identity function.

Observe that the outputs $(X,a^{X})$ and $(Y,b^{Y})$ of this reduction
are indeed inputs in $\cX_{\pi_{1}}\times\cY_{\pi_{1}}$, since $\pi_{1}$
supports the collection~$\cW$. It remains to show that if $(i,o)$
is a solution for $\fsgd$ on inputs $(X,a^{X})$ and $(Y,b^{Y})$,
then it is a solution for $H$ on $(X',Y')$. First, recall that the
assumption that $(i,o)$ is a solution for $\fsgd$ implies that $a_{i}^{X}>b_{i}^{Y}$
and that $o\in\sgd(X_{i},Y_{i})$. In particular, it must hold that
$i\in U$, since by assumption $a_{i}^{X}\le b_{i}^{Y}$ for every
$i\in R$. Therefore, $(i,o)$ is a solution for~$H$ on $(X',Y')$,
as required. 
\end{myproof}
It remains to prove a lower bound of $\Omega(q\cdot t)$ on~$\C(H)$.
To this end, we show that $H$ is (a sub-problem of) a lifted search
problem~$S'\d\gd$. Consider the following search problem~$S'$:
given a matrix~$Z\in\B^{U\times\left[\ell\right]}$, we would like
to find a pair $(i,o)$ such that $o$ is a solution for~$S$ on~$Z_{i}$
(i.e., $o\in S(Z_{i})$). Now, consider the corresponding lifted search
problem $\sgd'\eqdef S'\d\gd$, and observe that it can be described
as follows: Alice gets a matrix $X'\in\D^{U\times\left[\ell\right]}$,
Bob gets a matrix $Y'\in\D^{U\times\left[\ell\right]}$, and their
goal is to find a pair $(i,o)\in U\times\cO$ such that $o\in\sgd(X_{i},Y_{i})$.
Hence, the problem $H$ is simply the restriction of the lifted search
problem $S'\d\gd$ to input matrices that come from the set~$\cW'$. 

It is not hard to see that the query complexity of the problem~$S'$
is at least~$q\eqdef\Q(S)$: indeed, if we had a decision tree~$T$
that solves $S'$ using less than~$q$ queries, we could have used~$T$
to solve~$S$ with less than~$q$ queries by invoking~$T$ on matrices
whose rows are all equal. The lifting theorem of \cite{CFKMP19} (\ref{lifting-query-thm})
implies that $\C(S'\d\gd)\ge\Omega(q\cdot t)$. In order to prove
a similar lower bound for~$H$, we use our generalized lifting theorem,
to be proved in \ref{sec:generalized-lifting-thm}. This generalization
applies to lifted search problems when restricted to sets of inputs
that have sufficiently large average degree.
\begin{restated}{\ref{generalized-lifting-thm}}
For every $\eta>0$ and $d\in\N$ there exist $c\in\N$ and $\kappa>0$
such that the following holds: Let $S$ be a search problem that takes
inputs from~$\B^{\ell}$, and let $\gd:\B^{t}\times\B^{t}\to\B$
be an arbitrary function such that $\disc(\gd)\le2^{-\eta\cdot t}$
and such that $t\ge c\cdot\log\ell$. Let $\cX,\cY\subseteq\left(\B^{t}\right)^{\ell}$
be such that for every $I\subseteq\left[\ell\right]$  both $\rAD_{I}(\cX)$
and $\rAD_{I}(\cY)$ are at least $1/(d\cdot\ell^{d})^{\left|I\right|}$.
Then the communication complexity of solving $S\d\gd$ on inputs from
$\cX\times\cY$ is at least $\kappa\cdot\Q(S)\cdot t$.
\end{restated}

We apply \ref{generalized-lifting-thm} to $H$ by viewing the input
matrices of the players as strings in~$\D^{\left|U\right|\cdot\ell}$.
To this end, we need to lower bound the average degree of every set
of entries $K\subseteq U\times\left[\ell\right]$ in~$\cW'$. 
\begin{claim}
\label{hard-collection-average-degree-of-coordinates}For every set
of entries~$K\subseteq U\times\left[\ell\right]$, it holds that
$\rAD_{K}(\cW')\ge\frac{1}{4^{\left|K\right|}}$.
\end{claim}

Before proving the claim, we show it implies the lower bound on~$H$.
We apply \ref{generalized-lifting-thm} with $S=S'$, $\cX=\cY=\cW'$,
$\eta=\eta$, and $d=4$. We choose the constant $c$ to be the corresponding
constant that is obtained from the application of \ref{generalized-lifting-thm}.
\ref{hard-collection-average-degree-of-coordinates} shows that the
average degrees of~$\cX$ and~$\cY$ are sufficienltly large to
apply the theorem. It now follows that $\C(H)\ge\kappa\cdot q\cdot t$
for some constant~$\kappa>0$, which completes the proof of \ref{second-stage-lemma}.
\begin{myproof}[Proof of \ref{hard-collection-average-degree-of-coordinates}. ]
Intuitively, we need to prove that for every set $K\subseteq U\times\left[\ell\right]$
of \emph{entries}, the players know at most $2\left|K\right|$ bits
of information. By the assumption that $\cW$ is a hard collection,
we know that on any set~$I\subseteq U$ of \emph{rows}, the players
know at most $2\left|I\right|$ bits of information. Since every set
of entries~$K$ is contained in at most $\left|K\right|$~rows,
the claim follows. We now formalize this intuition.

Let $K\subseteq U\times\left[\ell\right]$ be a set of entries, and
let $I\subseteq U$ be the set of rows that contain entries from~$K$.
By the assumption that $\cW$ is a hard collection, 
\[
\rAD_{I\times\left[\ell\right]}(\cW')\ge\frac{1}{4^{\left|I\right|}}.
\]
By the ``chain rule'' for average degree (\ref{chain-rule-average-degree})
it holds that
\[
\rAD_{I\times\left[\ell\right]}(\cW')=\rAD_{K}(\cW')\cdot\rAD_{I\times\left[\ell\right]-K}(\cW'_{U\times\left[\ell\right]-K}),
\]
and since relative average degree is always at most~$1$ it follows
that
\[
\rAD_{K}(\cW')\ge\rAD_{I\times\left[\ell\right]}(\cW')\ge\frac{1}{4^{\left|I\right|}}\ge\frac{1}{4^{\left|K\right|}},
\]
as required.
\end{myproof}

\section{\label{sec:Semi-monotone-composition}The semi-monotone composition
theorem}

In this section we prove our semi-monotone composition theorem (\ref{semi-monotone-composition-informal}),
which can be stated formally as follows:
\begin{theorem}[semi-monotone composition theorem]
\label{semi-monotone-composition-thm}Let $m\in\N$ and let $g:\B^{n}\to\B$
be a non-constant monotone function, and let $\eq$~be the equality
function on strings of length~$t$. Suppose that there exists a CNF
contradiction $\phi$ over $\ell$ variables, such that the lifted
search problem $S_{\phi}\d\eq$ reduces to $\mKW_{g}$ via an injective
reduction and such that $t\ge2\log\ell$ . Then, 
\begin{equation}
\log\L(U_{m}\d\mKW_{g})\ge m+\Omega(\NS(\phi)\cdot t).\label{eq:semi-monotone-composition-lb}
\end{equation}
\end{theorem}

The rest of this section is organized as follows. We start by setting
up some notation. Then, we define a sub-problem of $U_{m}\d\mKW_{g}$,
denoted $\uxy$. Finally, we prove the desired lower bound on $\uxy$
using three propositions, which are proved in turn in Sections \oldref{subsec:rank-of-M},
\oldref{subsec:rank-of-monochromatic-rectangles}, and \oldref{subsec:Razborov-of-mKWg}.

Let $m,g,\eq,\phi,S_{\phi}$ be as in the theorem. For simplicity
of notation, let $\D\eqdef\B^{t}$, so that the domain of the lifted
search problem $\seq$ is $\D^{\ell}\times\D^{\ell}$. Let $R_{A}:\D^{\ell}\to g^{-1}(1)$,
$R_{B}:\D^{\ell}\to g^{-1}(0)$, and $\rout:\left[n\right]\to\cO$
be the functions that witness the reduction from $\sgd\d\eq$ to $\mKW_{g}$,
and recall that the functions $R_{A}$ and $R_{B}$ are injective.
Let $\cX\eqdef R_{A}(\D^{\ell})$ and $\cY\eqdef R_{B}(\D^{\ell})$
denote the images of $R_{A}$ and~$R_{B}$ respectively, and observe
that $\left|\cX\right|=\left|\cY\right|=\left|\D^{\ell}\right|$.
For conciseness, let $K\eqdef\left|\D^{\ell}\right|$. For every $p\in\N$,
we denote by $I_{p}$ the identity matrix of order~$p$, and we denote
by $I\eqdef I_{K}$ the identity matrix of order~$K$. Finally, let
$\cW\subseteq\B^{m\times n}$ be the set of $m\times n$ matrices~$W$
such that all the rows of~$W$ belong to~$\cX\cup\cY$. 

We turn to define the sub-problem $\uxy$. Recall that in the introduction
the communication problem~$U_{m}\d\mKW_{g}$ was defined as follows:
Alice and Bob get matrices $X,Y\in\B^{m\times n}$, and denote by
$a$ and~\textbf{$b$} the column vectors that are obtained by applying
$g$ to the rows of $X$ and~$Y$ respectively. The players are promised
that $a\ne b$, and they should either solve $\mKW_{g}$ on a row
where $a_{i}\ne b_{i}$ or find $(i,j)$ such that $a_{i}=b_{i}$
and $X_{i,j}\ne Y_{i,j}$. 

In the sub-problem $\uxy$, we restrict the input matrices of the
players to come from the set~$\cW$. We also change the problem a
bit as follows: we do not promise the players that $a\ne b$, but
rather, if the players find that $a=b$ they are allowed to declare
failure. It is not hard to see that this modification changes the
complexity of the problem by at most two bits (see \cite{HW93}),
and it makes the problem easier to analyze since it ensures that the
domain of the problem is a combinatorial rectangle. Let us make this
definition formal.
\begin{definition}
The communication problem $\uxy$ is defined as follows: The inputs
of Alice and Bob are matrices $X,Y\in\cW$ respectively. Let $a$
and $b$ denote the column vectors that are obtained by applying~$g$
to the rows of~$X$ and~$Y$ respectively. The goal of the players
is to find an entry $(i,j)$ that satisfies one of the following three
options:
\begin{itemize}
\item $a_{i}>b_{i}$ and $X_{i,j}>Y_{i,j}$.
\item $a_{i}<b_{i}$ and $X_{i,j}<Y_{i,j}$.
\item $a_{i}=b_{i}$ and~$X_{i,j}\ne Y_{i,j}$.
\end{itemize}
In addition, if $a=b$ then players are allowed to output the failure
symbol~$\bot$ instead of an entry $(i,j)$.
\end{definition}

\begin{myproof}[Proof of \ref{semi-monotone-composition-thm}.]
We prove the theorem by establishing a lower bound on the Razborov
rank measure of $\uxy$ (see \ref{subsec:Preliminaries-Razborov}
for the definition). To this end, we construct a matrix~$M\in\F_{2}^{\cW\times\cW}$,
and show that 
\[
\log\rz(\uxy,M)\ge m+\Omega(\NS(\phi)\cdot t).
\]
As a building block for~$M$, we use the matrix~$A\in\F_{2}^{\cX\times\cY}$
that is given by the following proposition, which is proved in \ref{subsec:Razborov-of-mKWg}.
\begin{proposition}
\label{Razborov-of-mKWg}There exists a symmetric matrix $A\in\F_{2}^{\cX\times\cY}$
such that 
\[
\log\rz(\xy,A)\ge\Omega(\NS(\phi)\cdot t),
\]
and such that $A^{2}=I$.
\end{proposition}

We now describe how the matrix~$M$ is constructed. Recall that the
rows and columns of~$M$ are indexed by matrices $X,Y\in\cW$. We
order the indices $X,Y\in\cW$ according to the vectors $a,b\in\B^{m}$
obtained when applying $g$ to the rows of $X,Y$. In this way, we
view $M$ as a block matrix consisting of $2^{m}\cdot2^{m}$ blocks,
each labeled by a pair $(a,b)$. The blocks that correspond to pairs
where $a=b$ are all-zeros. For every other block, we take the Kronecker
product of $m$~matrices, where the $i$-th matrix is $A$ (if $a_{i}\ne b_{i}$)
or $I$ (if $a_{i}=b_{i}$). More formally, for any two bits $\gamma,\delta\in\B$,
let
\[
A^{\gamma,\delta}\eqdef\begin{cases}
A & \text{if }\gamma\ne\delta\\
I & \text{otherwise.}
\end{cases}
\]
Then, for every $a,b\in\B^{m}$, the block of~$M$ that corresponds
to the pair $(a,b)$ is
\[
\begin{cases}
A^{a_{1},b_{1}}\otimes A^{a_{2},b_{2}}\otimes\cdots\otimes A^{a_{m},b_{m}} & a\ne b\\
\text{all zeroes} & a=b
\end{cases}.
\]
Intuitively, on rows where $a_{i}\ne b_{i}$, the players should solve
$\xy$, so we put the matrix~$A$ which is ``hard'' for $\xy$.
Similarly, on rows where $a_{i}=b_{i}$, the players should verify
the inequality of strings from $\cX\cup\cY$, so we put the matrix~$I$
which is ``hard'' for this task.

We turn to prove the lower bound on $\rz(\uxy,M)$. To this end, we
prove a lower bound on the ratio $\rk(M)/\rk(M|_{R})$ over all the
monochromatic rectangles $R$ of~$\uxy$. This is done in the following
two propositions, which bound the numerator and denominator in the
latter ratio, and are proved in Sections \oldref{subsec:rank-of-M}
and \oldref{subsec:rank-of-monochromatic-rectangles} respectively.
\begin{proposition}
\label{rank-of-M}The matrix~$M$ has full rank, i.e., $\log\rk(M)=\log\left|\cW\right|$.
\end{proposition}

\begin{proposition}
\label{rank-of-monochromatic-rectangles}For every monochromatic rectangle~$R$
of $\uxy$, 
\[
\log\rk(M|_{R})\le\log\left|\cW\right|-m-\log\rz(\xy,A).
\]
\end{proposition}

\noindent Together, the above two propositions immediately imply the
desired lower bound on $\rz(\uxy,M)$, and hence, \ref{semi-monotone-composition-thm}.
\end{myproof}
We now establish some notation that will be used in the proofs of
both \ref{rank-of-M} and \ref{rank-of-monochromatic-rectangles}.
First, we define an auxiliary matrix~$M'\in\F_{2}^{\cW}$ in a similar
way as~$M$, except that the blocks where $a=b$ are not treated
differently. In other words, $M'$ is a block matrix that, for every
$a,b\in\B^{m}$, has the block $A^{a_{1},b_{1}}\otimes A^{a_{2},b_{2}}\otimes\cdots\otimes A^{a_{m},b_{m}}$.
Observe that the blocks where $a=b$ are equal to~$I$, and that
those blocks are placed along the main diagonal of~$M'$. Thus,  $M'=M+I_{\left|\cW\right|}$.

We denote by $M_{(m-1)}$ and $M_{(m-1)}'$ the versions of~$M$
and~$M'$ that are defined for~$m-1$ rather than~$m$ --- in
other words, those are the matrices $M$ and~$M'$ that we would
define for $U_{m-1}\d\xy$.

\subsection{\label{subsec:rank-of-M}The rank of $M$}

We start by proving \ref{rank-of-M}, which says that $M$ has full
rank. We first claim that
\begin{equation}
M=\left(\begin{array}{cc}
I\otimes M_{(m-1)} & A\otimes M_{(m-1)}'\\
A\otimes M_{(m-1)}' & I\otimes M_{(m-1)}
\end{array}\right).\label{eq:recursive-M}
\end{equation}
The equality holds for the following reason: The upper and lower halves
of~$M$ correspond to the cases where $a_{1}=0$ and $a_{1}=1$ respectively,
and the left and right halves of~$M$ correspond to the cases where
$b_{1}=0$ and $b_{1}=1$. Applying \ref{Kronecker-block-matrix}
with~$I$ being the ``block matrix'' that has a single block, the
matrix $I\otimes M_{(m-1)}$ is the block matrix that is obtained
by taking the Kronecker product of $I$ with each block of $M_{(m-1)}$,
and these are exactly the blocks of~$M$ that correspond to $a_{1}=b_{1}$.
Similarly, the matrix $A\otimes M_{(m-1)}'$ is the block matrix that
is obtained by taking the Kronecker product of $A$ with each block
of $M_{(m-1)}'$, and these are exactly the blocks of~$M$ that correspond
to $a_{1}\ne b_{1}$. In the latter case, we used $M_{(m-1)}'$ rather
than $M_{(m-1)}$ since all those blocks satisfy $a\ne b$, and therefore
we do not want to zero out the blocks when $a_{-1}=b_{-1}$ (where
$a_{-1},b_{-1}$ denote the column vectors~$a,b$ without the first
coordinate).

We prove that $M$ has full rank by applying row and column operations
to \ref{eq:recursive-M}. Let $I'$ be the identity matrix of the
same order as $M_{(m-1)}$, and recall that $M_{(m-1)}'=M_{(m-1)}+I'$.
Since we are working over~$\F_{2}$, the latter equality can also
be written as~$M_{(m-1)}=M_{(m-1)}'+I'$. By substituting the latter
equality in \ref{eq:recursive-M}, we obtain the matrix
\[
\left(\begin{array}{cc}
I\otimes(M_{(m-1)}'+I') & A\otimes M_{(m-1)}'\\
A\otimes M_{(m-1)}' & I\otimes(M_{(m-1)}'+I')
\end{array}\right)=\left(\begin{array}{cc}
I\otimes M_{(m-1)}'+I\otimes I' & A\otimes M_{(m-1)}'\\
A\otimes M_{(m-1)}' & I\otimes M_{(m-1)}'+I\otimes I'
\end{array}\right).
\]
Next, we subtract from the left half the product of $A\otimes I'$
and the right half, and get
\begin{equation}
\left(\begin{array}{cc}
I\otimes M_{(m-1)}'+I\otimes I'-(A\otimes I')\cdot(A\otimes M_{(m-1)}') & A\otimes M_{(m-1)}'\\
A\otimes M_{(m-1)}'-(A\otimes I')\cdot(I\otimes M_{(m-1)}')-(A\otimes I')\cdot(I\otimes I') & I\otimes M_{(m-1)}'+I\otimes I'
\end{array}\right).\label{eq:semi-monotone-composition-rank-of-M-complicated-expression}
\end{equation}
We now use \ref{Kronecker-mixed-product-property} to determine each
of the matrix products that appear in the last expression. Recall
that $A^{2}=I$ by \ref{Razborov-of-mKWg}. Then,
\begin{align}
(A\otimes I')\cdot(A\otimes M_{(m-1)}') & =(A\cdot A)\otimes(I'\cdot M_{(m-1)}')=I\otimes M_{(m-1)}'\label{eq:semi-monotone-composition-rank-of-M-mixed-products}\\
(A\otimes I')\cdot(I\otimes M_{(m-1)}') & =(A\cdot I)\otimes(I'\cdot M_{(m-1)}')=A\otimes M_{(m-1)}'\nonumber \\
(A\otimes I')\cdot(I\otimes I') & =(A\cdot I)\otimes(I'\cdot I')=A\otimes I'.\nonumber 
\end{align}
By substituting the latter equalities in the matrix of \ref{eq:semi-monotone-composition-rank-of-M-complicated-expression},
we obtain the matrix
\begin{align*}
 & \left(\begin{array}{cc}
I\otimes M_{(m-1)}'+I\otimes I'-I\otimes M_{(m-1)}' & A\otimes M_{(m-1)}'\\
A\otimes M_{(m-1)}'-A\otimes M_{(m-1)}'-A\otimes I' & I\otimes M_{(m-1)}'+I\otimes I'
\end{array}\right)\\
= & \left(\begin{array}{cc}
I\otimes I' & A\otimes M_{(m-1)}'\\
A\otimes I' & I\otimes M_{(m-1)}'+I\otimes I'
\end{array}\right)
\end{align*}
where in the last equality we replaced $-A\otimes I'$ with $A\otimes I'$
by using the fact that we are working over~$\F_{2}$. We now subtract
the product of $A\otimes I'$ and the upper half from the lower half,
and obtain the matrix
\[
\left(\begin{array}{cc}
I\otimes I' & A\otimes M_{(m-1)}'\\
A\otimes I'-(A\otimes I')\cdot(I\otimes I') & I\otimes M_{(m-1)}'+I\otimes I'-(A\otimes I')\cdot(A\otimes M_{(m-1)}')
\end{array}\right).
\]
By substituting the equalities of \ref{eq:semi-monotone-composition-rank-of-M-mixed-products}
in the latter expression, we obtain the matrix
\begin{align*}
 & \left(\begin{array}{cc}
I\otimes I' & A\otimes M_{(m-1)}'\\
A\otimes I'-A\otimes I' & I\otimes M_{(m-1)}'+I\otimes I'-I\otimes M_{(m-1)}'
\end{array}\right)\\
= & \left(\begin{array}{cc}
I\otimes I' & A\otimes M_{(m-1)}'\\
0 & I\otimes I'
\end{array}\right).
\end{align*}
The latter matrix is an upper triangular matrix that has ones on its
main diagonal, and therefore has full rank, as required.

\subsection{\label{subsec:rank-of-monochromatic-rectangles}The rank of monochromatic
rectangles}

We turn to prove \ref{rank-of-monochromatic-rectangles}, which upper
bounds the rank of monochromatic rectangles. Let $R\subseteq\cW\times\cW$
be a monochromatic rectangle of~$\uxy$. We prove that
\[
\rk(M|_{R})\le\frac{\left|\cW\right|}{2^{m}\cdot\rz(\xy,A)}.
\]
Recall that $R$ can be one of four types:
\begin{enumerate}
\item \label{enu:semi-monotone-composition-monotone-solution}It could correspond
to a solution $(i,j)$ where $a_{i}>b_{i}$ and $X_{i}>Y_{i}$.
\item \label{enu:semi-monotone-composition-monotone-solution-2}It could
correspond to a solution $(i,j)$ where $a_{i}<b_{i}$ and $X_{i}<Y_{i}$.
\item \label{enu:semi-monotone-composition-nonmonotone-solution}It could
correspond to a solution $(i,j)$ where $a_{i}=b_{i}$ and $X_{i,j}\ne Y_{i,j}$.
\item \label{enu:enu:semi-monotone-composition-failure-solution}It could
correspond to the failure symbol~$\bot$, which means that $a=b$.
\end{enumerate}
We consider each of the types separately, starting with the simpler
Types~\ref{enu:semi-monotone-composition-nonmonotone-solution} and~\ref{enu:enu:semi-monotone-composition-failure-solution}.
If $R$ is of Type~\ref{enu:enu:semi-monotone-composition-failure-solution},
every entry $(X,Y)\in R$ satisfies $a=b$, and by the definition
of~$M$, this implies that $M_{X,Y}=0$. Hence, $M|_{R}$ is the
all-zeros matrix and therefore $\rk(M|_{R})=0$.

If $R$ is of Type~\ref{enu:semi-monotone-composition-nonmonotone-solution},
there exist some $i\in\left[m\right]$ and $j\in\left[n\right]$ such
that every entry $(X,Y)\in R$ satisfies $a_{i}=b_{i}$ and $X_{i,j}\ne Y_{i,j}$.
We show that in this case, $M|_{R}$ is again the all-zeros matrix.
Without loss of generality, assume that $i=1$. If $a=b$, then again
$M_{X,Y}=0$. Otherwise, by the definition of~$M$, the block that
corresponds to $(a,b)$ is equal to
\[
I\otimes A^{a_{2},b_{2}}\otimes\cdots\otimes A^{a_{m},b_{m}},
\]
and thus the entry that corresponds to $(X,Y)$ is equal to
\[
M_{X,Y}=I_{X_{1},Y_{1}}\cdot\prod_{i=2}^{m}(A^{a_{i},b_{i}})_{X_{i},Y_{i}}.
\]
Since $X_{1}\ne Y_{1}$, we have $I_{X_{1},Y_{1}}=0$ and thus $M_{X,Y}=0$
as well. Hence, $M|_{R}$ is the all-zeros matrix and therefore $\rk(M|_{R})=0$.

The bulk of the proof is devoted to the case where $R$ is of Type~\ref{enu:semi-monotone-composition-monotone-solution}
(the case where $R$ is of Type~\ref{enu:semi-monotone-composition-monotone-solution-2}
can be dealt with similarly since $A$ is symmetric). Assume that
$R$ corresponds to a solution $(i,j)$ where $a_{i}>b_{i}$ and $X_{i,j}>Y_{i,j}$.
Without loss of generality, assume that $i=1$. Moreover, without
loss of generality, we may assume that $R$~is maximal, since extending~$R$
can only increase the rank of~$M|_{R}$. This implies that $R$ can
be assumed to contain\emph{ all }inputs that satisfy $a_{i}>b_{i}$
and $X_{i,j}>Y_{i,j}$. In other words, $R$ can be written as $R=\cU\times\cV$
where: 
\begin{align*}
\cU & \eqdef\left\{ X\in\cW:a_{1}=1,X_{1,j}=1\right\} =\left\{ X\in\cW:X_{1}\in\cX,X_{1,j}=1\right\} \\
\cV & \eqdef\left\{ Y\in\cW:b_{1}=0,Y_{1,j}=0\right\} =\left\{ Y\in\cW:Y_{1}\in\cY,Y_{1,j}=0\right\} ,
\end{align*}
where the second equality in each line holds since $\cX\subseteq g^{-1}(1)$
and $\cY\subseteq g^{-1}(0)$. Now, define a rectangle $R^{*}\subseteq\cX\times\cY$
by
\begin{align*}
R^{*} & \eqdef\left\{ x\in\cX:x_{j}=1\right\} \times\left\{ y\in\cY:y_{j}=0\right\} .
\end{align*}
Then, we can write
\[
R=\left\{ (X,Y)\in\cW\times\cW:(X_{1},Y_{1})\in R^{*}\right\} .
\]
Recall that we denote by $M_{(m-1)}$ and $M_{(m-1)}'$ the versions
of~$M$ and~$M'$ for $U_{m-1}\d\xy$. It follows that
\[
M|_{R}=A|_{R^{*}}\otimes M_{(m-1)}',
\]
where we use $M_{(m-1)}'$ rather than $M_{(m-1)}$ since $a\ne b$
for all the entries in~$R$. In order to bound the rank of this matrix,
we use the following proposition, whose proof is deferred to the end
of this section.
\begin{proposition}
\label{rank-of-M'}It holds that $\rk(M')=K^{m}$.
\end{proposition}

Observe that $\left|\cW\right|=(2K)^{m}$: to see why, recall that
$\cW$ consists of all $m\times n$~matrices whose rows come from~$\cX\cup\cY$.
The sets $\cX,\cY$ are disjoint and satisfy $\left|\cX\right|=\left|\cY\right|=K$,
and hence $\left|\cW\right|=(\left|\cX\cup\cY\right|)^{m}=(2K)^{m}$.
Moreover, observe that $\rk(A)=K$, since $A^{2}=I$ and so $A$ has
full rank. It follows that
\begin{align*}
\rk(M|_{R}) & =\rk(A|_{R^{*}})\cdot\rk(M_{(m-1)}') & \text{(by \ref{Kronecker-rank})}\\
 & =\rk(A|_{R^{*}})\cdot K^{m-1} & \text{(by \ref{rank-of-M'})}\\
 & \le\frac{\rk(A)}{\rz(\xy,A)}\cdot K^{m-1} & \text{(by definition of \ensuremath{\rz})}\\
 & =\frac{K}{\rz(\xy,A)}\cdot K^{m-1} & (\ensuremath{\rk(A)=K)}\\
 & =\frac{K^{m}}{\rz(\xy,A)}\\
 & =\frac{\left|\cW\right|}{2^{m}\cdot\rz(\xy,A)}. & (\left|\cW\right|=(2K)^{m})
\end{align*}
This concludes the proof.
\begin{myproof}[Proof of \ref{rank-of-M'}.]
Let $B$ denote the block matrix
\[
B\eqdef\left(\begin{array}{cc}
I & A\\
A & I
\end{array}\right).
\]
We claim that $M'=\underbrace{B\otimes\cdots\otimes B}_{m\text{ times}}$.
To see why, note that the upper and lower halves of $B$ correspond
to the cases where $a_{i}=0$ and $a_{i}=1$ respectively, and the
left and right halves correspond to the cases where $b_{i}=0$ and
$b_{i}=1$. Hence, by \ref{Kronecker-block-matrix}, when we take
the Kronecker product of $m$~copies of~$B$ we get all the possible
blocks of the form~$A^{a_{1},b_{1}}\otimes A^{a_{2},b_{2}}\otimes\cdots\otimes A^{a_{m},b_{m}}.$

It therefore suffices to prove that $\rk(B)=K$, since that will imply
that $\rk(M')=K^{m}$ by \ref{Kronecker-rank}. To this end, we subtract
the product of $A$ with the upper half of~$B$ from the lower half
of~$B$, and obtain the matrix
\[
\left(\begin{array}{cc}
I & A\\
A-A\cdot I & I-A^{2}
\end{array}\right)=\left(\begin{array}{cc}
I & A\\
A-A & I-I
\end{array}\right)=\left(\begin{array}{cc}
I & A\\
0 & 0
\end{array}\right),
\]
where the first equality holds since $A^{2}=I$ by \ref{Razborov-of-mKWg}.
The matrix on the right-hand size clearly has rank~$K$ (since $I\eqdef I_{K}$
is the identity matrix of order~$K$). This implies $\rk(B)=K$,
as required.
\end{myproof}

\subsection{\label{subsec:Razborov-of-mKWg}The existence of the matrix $A$}

Finally, we prove \ref{Razborov-of-mKWg}, restated next.
\begin{restated}{\ref{Razborov-of-mKWg} (restated)}
There exists a symmetric matrix $A\in\F_{2}^{\cX\times\cY}$ such
that 
\begin{equation}
\log\rz(\xy,A)\ge\Omega(\NS(\phi)\cdot t),\label{eq:Razborov-of-mKWg-lb}
\end{equation}
and such that $A^{2}=I$.
\end{restated}

To this end, we use the lifting theorem of \cite{RMNPRV20} (\ref{lifting-ns-thm}).
Recall that $\phi$~is a CNF contradiction over $\ell$~variables
and that $\eq$~is the equality gadget over $t\ge2\log\ell$ bits.
By applying that theorem to the lifted search problem $\seq$, we
obtain a matrix $A\in\F_{2}^{\D^{\ell}\times\D^{\ell}}$ that satisfies
the lower bound of \ref{eq:Razborov-of-mKWg-lb} for $\seq$. Our
goal is to prove that $A$ satisfies this lower bound for $\xy$,
and to prove that $A$ is symmetric and satisfies $A^{2}=I$.

We start by tackling the following minor technical issue: By its definition,
the rows and columns of~$A$ are indexed by $\D^{\ell}$, whereas
in order to lower bound $\rz(\xy)$, we need a matrix whose rows and
columns are indexed by $\cX$ and~$\cY$ respectively. To this end,
recall that $\cX\eqdef R_{A}(\D^{\ell})$ and $\cY\eqdef R_{B}(\D^{\ell})$,
where $R_{A}$ and~$R_{B}$ are the injective functions of the reduction
from $\seq$ to~$\mKW_{g}$. Thus, $R_{A}$ and $R_{B}$ are bijections
from $\D^{\ell}$ to $\cX$ and~$\cY$ respectively. It follows that
we can view the rows and columns of~$A$ as being indexed by~$\cX$
and~$\cY$ respectively by using $R_{A}$ and $R_{B}$ to translate
the indices.

Now, in order to prove that $A$ gives the desired lower bound on
$\rz(\xy)$, we show that every monochromatic rectangle $T\subseteq\cX\times\cY$
of $\xy$ is also a monochromatic rectangle of $\seq$ (when interpreted
as a rectangle in $\D^{\ell}\times\D^{\ell}$ via $R_{A}^{-1},R_{B}^{-1}$).
Let $T\subseteq\cX\times\cY$ be a monochromatic rectangle of~$\xy$,
and suppose that it is labeled with a solution $j\in\left[n\right]$.
Let $o\eqdef\rout(j)$, where $\rout$ is the function of the reduction
from $\seq$ to~$\mKW_{g}$. Then, by the definition of $\rout$,
for every $(x,y)\in T$ it holds that $o$ is a solution for $\seq$
on $\left(R_{A}^{-1}(x),R_{B}^{-1}(y)\right)$. Thus, $T$ can be
viewed as an $o$-monochromatic rectangle of $\seq$. It follows that
\begin{align*}
\log\rz(\xy,A) & \eqdef\log\rk(A)-\max_{\text{\ensuremath{\begin{array}{c}
\text{monochromatic rectangle}\\
\text{\ensuremath{T} of \ensuremath{\xy}}
\end{array}}}}\log\rk(A|_{T})\\
 & \ge\log\rk(A)-\max_{\text{\ensuremath{\begin{array}{c}
\text{monochromatic rectangle}\\
\text{\ensuremath{T} of \ensuremath{\seq}}
\end{array}}}}\log\rk(A|_{T})\\
 & \eqdef\log\rz(\seq,A)\\
 & \ge\Omega(\NS(\phi)\cdot t), & \text{(\ref{lifting-ns-corollary})}
\end{align*}
as required.

It remains to prove that $A$ is symmetric and satisfies $A^{2}=I$.
To this end, we take a closer look at how the matrix~$A$ is constructed.
The proof of \cite{RMNPRV20} (following \cite{S11,RPRC16,PR17})
chooses the matrix~$A$ to be a \emph{pattern matrix}, that is, for
every two inputs $x,y\in\D^{\ell}$ it holds that 
\begin{equation}
A_{x,y}\eqdef p(\eq(x_{1},y_{1}),\ldots,\eq(x_{\ell},y_{\ell})),\label{eq:definition-of-A}
\end{equation}
where $p:\F_{2}^{\ell}\to\F_{2}$ is a multi-linear polynomial of
degree~$\ell$. This immediately implies that $A$ is symmetric,
since it is easy to see that the right-hand side of \ref{eq:definition-of-A}
remains the same if we swap $x$ and~$y$. In order to show that
$A^{2}=I$, we write~$A$ as a sum of Kronecker products: For every
set $T\subseteq\left[\ell\right]$, we denote by $\hat{p}(T)$ the
coefficient of $p$ at the monomial $\prod_{i\in T}x_{i}$. Let $\one_{\left|\D\right|}$
denote the all-ones matrix of order $\left|\D\right|\times\left|\D\right|$,
and for every $T\subseteq\left[\ell\right]$ and $i\in\left[\ell\right]$,
let
\[
Q_{i,T}=\begin{cases}
I_{\left|\D\right|} & \text{if }i\in T,\\
\one_{\left|\D\right|} & \text{if }i\notin T.
\end{cases}
\]
Robere~\cite[Sec. 5.1]{R18} showed that $A$ can be written as follows:
\[
A=\sum_{T\subseteq\left[\ell\right]}\hat{p}(T)\cdot Q_{1,T}\otimes\cdots\otimes Q_{\ell,T}.
\]
Essentially, the latter identity holds since for every $i\in T$,
the value of $I_{\left|\D\right|}$ at the entry $x_{i},y_{i}$ is
$\eq(x_{i},y_{i})$, whereas for every $i\notin T$, multiplying by
$\one_{\left|\D\right|}$ does not change the value of the product.
It follows that
\begin{align*}
A^{2} & =\left(\sum_{T\subseteq\left[\ell\right]}\hat{p}(T)\cdot Q_{1,T}\otimes\cdots\otimes Q_{\ell,T}\right)^{2}\\
 & =\left(\sum_{T\subseteq\left[\ell\right]:\hat{p}(T)=1}Q_{1,T}\otimes\cdots\otimes Q_{\ell,T}\right)^{2} & \text{ (we are working over \ensuremath{\F_{2}})}\\
 & =\sum_{T,T'\subseteq\left[\ell\right]:\hat{p}(T)=\hat{p}(T')=1}\left(Q_{1,T}\otimes\cdots\otimes Q_{\ell,T}\right)\cdot\left(Q_{1,T'}\otimes\cdots\otimes Q_{\ell,T'}\right)\\
 & =\sum_{T,T'\subseteq\left[\ell\right]:\hat{p}(T)=\hat{p}(T')=1}\left(Q_{1,T}\cdot Q_{1,T'}\right)\otimes\cdots\otimes\left(Q_{\ell,T}\cdot Q_{\ell,T'}\right) & \text{ (\ref{Kronecker-mixed-product-property}).}
\end{align*}
Next, observe that for every two distinct sets $T,T'\subseteq\left[\ell\right]$,
the last sum contains two terms:
\[
(Q_{1,T}\cdot Q_{1,T'})\otimes\cdots\otimes(Q_{\ell,T}\cdot Q_{\ell,T'})\quad\text{and}\quad(Q_{1,T'}\cdot Q_{1,T})\otimes\cdots\otimes(Q_{\ell,T'}\cdot Q_{\ell,T})
\]
We now claim that those two terms are equal and therefore cancel each
other. To this end, we claim that for every $i\in\left[\ell\right]$
the matrices $Q_{i,T}$ and $Q_{i,T'}$ commute: the reason is that
either both matrices are equal to $\one_{\left|\D\right|}$ (and then
they clearly commute) or one of those matrices is $I_{\left|\D\right|}$
(and then again they clearly commute). It follows that for every two
distinct sets $T,T'\subseteq\left[\ell\right]$, the above terms are
equal and thus cancel each other. Hence, we remain only with the terms
that correspond to $T=T'$, so
\[
A^{2}=\sum_{T\subseteq\left[\ell\right]:\hat{p}(T)=1}Q_{1,T}^{2}\otimes\cdots\otimes Q_{\ell,T}^{2}.
\]
Finally, observe that $\left|\D\right|=2^{t}$ is even, and thus $\left(\one_{\left|\D\right|}\right)^{2}$
is the all-zeros matrix. Hence, every term in the above sum in which
one of the matrices $Q_{i,T}$ is equal to $\one_{\left|\D\right|}$
zeros out. The only term that remains is therefore the term that corresponds
to~$T=\left[\ell\right]$. Furthermore, the degree of~$p$ is~$\ell$,
and therefore $\hat{p}(\left[\ell\right])=1$. It follows that
\begin{align*}
A^{2} & =Q_{1,\left[\ell\right]}^{2}\otimes\cdots\otimes Q_{\ell,\left[\ell\right]}^{2}\\
 & =\underbrace{I_{\left|\D\right|}^{2}\otimes\cdots\otimes I_{\left|\D\right|}^{2}}_{\text{\ensuremath{\ell} times}}\\
 & =\underbrace{I_{\left|\D\right|}\otimes\cdots\otimes I_{\left|\D\right|}}_{\text{\ensuremath{\ell} times}}\\
 & =I_{\left|\D^{\ell}\right|}\eqdef I.
\end{align*}
Hence, we have shown that $A$ is symmetric and that $A^{2}=I$, as
required.

\section{\label{sec:generalized-lifting-thm}A generalized lifting theorem}

In this section, we prove our generalization of the lifting theorem
of~\cite{CFKMP19} (\ref{lifting-query-thm}). The latter theorem
says that if a search problem~$S\subseteq\B^{\ell}\times\cO$ is
lifted with an appropriate gadget~$\gd:\B^{t}\times\B^{t}\to\B$,
then $\C(S\d\gd)=\Omega(\Q(S)\cdot t)$. Essentially, our theorem
says that this lower bound remains intact even if we restrict the
inputs of $S\d\gd$ to a rectangle~$\cX\times\cY$, as long as the
relative average degree of any coordinate in~$\cX$ and~$\cY$ is
at least $\frac{1}{\poly(\ell)}$. Formally, we have the following
result.
\begin{theorem}
\label{generalized-lifting-thm}For every $\eta>0$ and $d\in\N$
there exist $c\in\N$ and $\kappa>0$ such that the following holds:
Let $S$ be a search problem that takes inputs from~$\B^{\ell}$,
and let $\gd:\B^{t}\times\B^{t}\to\B$ be an arbitrary function such
that $\disc(\gd)\le2^{-\eta\cdot t}$ and such that $t\ge c\cdot\log\ell$.
Let $\cX,\cY\subseteq\left(\B^{t}\right)^{\ell}$ be such that for
every $I\subseteq\left[\ell\right]$  both $\rAD_{I}(\cX)$ and $\rAD_{I}(\cY)$
are at least $1/(d\cdot\ell^{d})^{\left|I\right|}$. Then the communication
complexity of solving $S\d\gd$ on inputs from $\cX\times\cY$ is
at least $\kappa\cdot\Q(S)\cdot t$.
\end{theorem}

We believe that it is possible to prove similar generalizations of
the lifting theorems of~\cite{RM99,GPW15,CKLM19,WYY17}, which in
turn would extend our monotone composition theorem to work with those
theorems.

Let $\eta,d,S,\gd$ be as in the theorem. We will choose the constants~$c$
and~$\kappa$ at the end of the proof to be sufficiently large and
sufficiently small respectively so that the various inequalities hold.
For convenience, for every set of coordinates $I\subseteq\left[\ell\right]$
we denote by $\gd^{I}$ the function that takes $\left|I\right|$~independent
inputs to~$\gd$ and computes~$\gd$ on all of them. In particular,
let $G\eqdef\gd^{\left[\ell\right]}$, so we can write $S\d\gd=S\circ G$.

Let $\Pi$ be a protocol that solves $S\d\gd$ using $C$ bits of
communication. We construct a decision tree~$T$ that solves~$S$
using $O(\frac{C}{t})$ queries, which implies the desired result.
The rest of this section is organized as follows: In \ref{subsec:generalized-lifting-overview},
we provide an overview of the proof. In \ref{subsec:generalized-lifting-machinery},
we state the background that we need from the lifting literature.
Then, in \ref{subsec:construction}, we describe the decision tree~$T$
and prove its correctness. Finally, in \ref{subsec:generalized-lifting-query-complexity},
we upper bound the query complexity of~$T$.

\subsection{\label{subsec:generalized-lifting-overview}Proof overview}

We start with an overview of the proof of \cite{CFKMP19}. Their proof
works by a simulation argument: Given an input~$z\in\B^{\ell}$,
the tree~$T$ constructs a full transcript~$\pi$ of~$\Pi$, such
that the rectangle $\cX_{\pi}\times\cY_{\pi}$ contains an input $(x,y)\in G^{-1}(z)$,
and returns the output of~$\pi$. Clearly, the transcript~$\pi$
must output the correct solution for~$z$, since $S\circ G(x,y)=S(z)$.

The tree~$T$ constructs the transcript~$\pi$ by simulating~$\Pi$
message-by-message. Throughout the simulation, the tree~$T$ maintains
random variables~$\x,\y$ that are distributed over~$\cX_{\pi}\times\cY_{\pi}$.
Let $\z\eqdef G(\x,\y)$. The goal of the tree~$T$ is to make sure
that when the simulation of~$\Pi$ halts, the input~$z$ is in the
support of~$\z$. 

When the simulation starts, we set $\x,\y$ to be uniformly distributed
over all inputs, and therefore $\z$ is uniformly distributed over
$\B^{\ell}$. As the simulation progresses, the transcript~$\pi$
reveals more and more information about $\x,\y$, until at some point
there are coordinates~$I\subseteq\left[\ell\right]$ about which
a lot of information has been revealed. At this point, there is a
danger that the value of~$\z_{I}$ might get fixed to a value different
than~$z_{I}$. Before this happens, the tree~$T$ queries $z_{I}$,
and conditions the random variables~$\x,\y$ on the event~$\z_{I}=z_{I}$.
This conditioning is repeated whenever a significant amount of information
is revealed about some coordinates, where ``a significant amount''
is $\alpha\cdot t$ bits of information per coordinate in~$I$ for
some constant~$\alpha>0$.

Eventually, the simulation halts. At this point, we know that $\z$
is consistent with~$z$ in all its fixed coordinates. Moreover, we
can show that since only a little information has been revealed about
all the other coordinates, the value of~$\z$ in the rest of the
coordinates is uniformly distributed. Hence, $z$ must be in the support
of~$\z$, as required.

The final step is to upper bound the query complexity of~$T$. On
the one hand, the tree~$T$ queries~$z$ once for each coordinate
on which the transcript revealed $\alpha\cdot t$~bits of information.
On the other hand, we know that the transcript~$\pi$ reveals at
most $C$~bits of information about~$\x,\y$, since this is the
communication complexity of~$\Pi$. Thus, there are at most $\frac{C}{\alpha\cdot t}$
coordinates about which $\pi$~reveals $\alpha\cdot t$~bits of
information, so the query complexity of~$T$ is $O(\frac{C}{t})$,
as required.

We now give some more details on how the query complexity is bounded,
since we will need those details shortly. We bound the query complexity
of~$T$ using a potential argument. Let $U$ be the set of unfixed
coordinates. Our potential function is the sum $\Hm(\x_{U})+\Hm(\y_{U}).$
At the beginning of the simulation, $\x,\y$ are uniformly distributed
over all inputs and $U=\left[\ell\right]$, so the potential is $2\cdot t\cdot\ell$.
After $C$~bits were transmitted and $q$~queries have been made,
it is possible to show that the potential is decreased by at most
$C+(2-\alpha)\cdot t\cdot q$. On the other hand, the potential is
always upper bounded by~$2\cdot t\cdot\left|U\right|$, and since
$\left|U\right|=\ell-q$ it follows that
\begin{align}
2\cdot t\cdot\ell-C-(2-\alpha)\cdot t\cdot q & \le2\cdot t\cdot\left|U\right|=2\cdot t\cdot(\ell-q).\label{eq:generalized-lifting-overview-potential-argument}
\end{align}
from which we obtain the bound $q=O(C/t)$ after rearranging.

\paragraph*{Our contribution.}

Our proof follows a similar outline, but at the beginning of the simulation,
we set $\x,\y$ to be uniformly distributed over $\cX$ and $\cY$
respectively. This difference results in two issues. The first issue
is that if some coordinate~$i$ of~$\x,\y$ starts with relatively
low min-entropy, then there is a danger that $\z_{i}$ will be fixed
too early. Fortunately, such a situation can never happen since we
assumed that $\cX,\cY$ have high average degrees, which lower bounds
the min-entropy (by \ref{degree-to-min-entropy}).

The second issue is that the foregoing potential argument becomes
slightly more complicated. Specifically, the initial potential is
now $\log\left|\cX\right|+\log\left|\cY\right|$ rather than $2\cdot t\cdot\ell$,
and the upper bound on the potential is now $\log\left|\cX_{U}\right|+\log\left|\cY_{U}\right|$
rather than $2\cdot t\cdot\left|U\right|$. Thus, \ref{eq:generalized-lifting-overview-potential-argument}
is replaced with the equation
\[
\log\left|\cX\right|+\log\left|\cY\right|-C-(2-\alpha)\cdot t\cdot q\le\log\left|\cX_{U}\right|+\log\left|\cY_{U}\right|.
\]
In order to derive a bound on~$q$ from the latter equation, we need
to lower bound the difference
\[
\left(\log\left|\cX\right|+\log\left|\cY\right|\right)-\left(\log\left|\cX_{U}\right|+\log\left|\cY_{U}\right|\right).
\]
To this end, we observe that 
\[
\log\left(\left|\cX\right|\right)-\log\left(\left|\cX_{U}\right|\right)=\log\left(\frac{\left|\cX\right|}{\left|\cX_{U}\right|}\right)=\log\left(\AD_{\left[\ell\right]-U}\left(\cX\right)\right),
\]
and a similar equality holds for~$\cY$. We now get the desired lower
bound by using our assumed bound on the average degrees of $\cX$
and~$\cY$.

\subsection{\label{subsec:generalized-lifting-machinery}Lifting machinery}

As explained above, a key part of the simulation is keeping track
of the coordinates on which the protocol did \emph{not} transmit a
lot of information. We model a string about which not much information
has been revealed using the following notion of a \emph{dense} random
variable (not to be confused with the notion of \emph{density} from
\ref{subsec:average-degree}).
\begin{definition}[\cite{GLMWZ16}]
Let $n\in\N$ and $\delta>0$, and let $\x$ be a random variable
taking values in~$\D^{n}$. We say that $\x$ is $\delta$-dense
if for every set of coordinates $I\subseteq\left[n\right]$ it holds
that $\Hm(\x_{I})\ge\delta\cdot t\cdot\left|I\right|$.
\end{definition}

We will keep track of which coordinates of~$\z$ have been fixed
and which are still free using the standard notion of restriction.
\begin{definition}
A \emph{restriction~$\rho$} is a string in $\left\{ 0,1,*\right\} ^{\ell}$.
We say that a coordinate~$i\in\left[\ell\right]$ is \emph{free}
in~$\rho$ if $\rho_{i}=*$, and otherwise we say that $i$~is \emph{fixed}.
Given a restriction $\rho\in\left\{ 0,1,*\right\} ^{\ell}$, we denote
by $\free(\rho)$ and $\fix(\rho)$ the sets of free and fixed coordinates
of~$\rho$ respectively. We say that a string $z\in\B^{\ell}$ is
\emph{consistent} with $\rho$ if $z_{\fix(\rho)}=\rho_{\fix(\rho)}$.
\end{definition}

Our decision tree will maintain the following invariant, which captures
the idea that $\z=G(\x,\y)$ is fixed in some coordinates, and not
too much information has been revealed on the other coordinates.
\begin{definition}[\cite{GLMWZ16,GPW17}]
\label{structured-variables}Let $\rho\in\left\{ 0,1,*\right\} ^{\ell}$
be a restriction, let $\tau>0$, and let $\x,\y$ be independent random
variables taking values in~$\D^{\ell}$. We say that $\x$~and~$\y$
are $(\rho,\tau)$\emph{-structured} if there exist $\delta_{x},\delta_{y}>0$
such that $\x_{\free(\rho)}$ and $\y_{\free(\rho)}$ are $\delta_{x}$-dense
and $\delta_{y}$-dense respectively, $\delta_{x}+\delta_{y}\ge\tau$,
and
\[
\gd^{\fix(\rho)}\left(\x_{\fix(\rho)},\y_{\fix(\rho)}\right)=\rho_{\fix(\rho)}.
\]
\end{definition}

The following results use the assumption that $\gd$ has input length
$t\ge c\cdot\log\ell$ and discrepancy at least~$2^{-\eta\cdot t}$.
A key property of structured variables $\x,\y$ is that in all the
free coordinates, the random variable $\z_{\free(\rho)}=G(\x,\y)$
has full support. This property is formalized by the following result.
\begin{proposition}[{special case of \cite[Prop 3.10]{CFKMP19}}]
\label{generalized-lifting-full-support}There exists a universal
constant~$h$ such that the following holds: Let $\x,\y$ be random
variables that are $(\rho,\tau)$-structured for $\tau>2+\frac{h}{c}-\eta$.
Then, the support of the random variable $\gd^{\free(\rho)}(\x_{\free(\rho)},\y_{\free(\rho)})$
is $\B^{\free(\rho)}$.
\end{proposition}

Whenever the protocol transmits so much information that $\x$ or
$\y$ cease to be dense, we wish to fix some coordinates in order
to restore their density. This is done by the following folklore fact.
\begin{proposition}[see, e.g., \cite{GPW17}]
\label{restoring-density}Let $n\in\N$, let $\delta>0$, and let
$\x$ be a random variable taking values in~$\D^{n}$. Let $I\subseteq\left[n\right]$
be a maximal subset of coordinates such that $H_{\infty}(\x_{I})<\delta\cdot t\cdot|I|$,
and let $x_{I}\in\D^{I}$ be a value such that $\Pr\left[\x_{I}=x_{I}\right]>2^{-\delta\cdot t\cdot|I|}$.
Then, the random variable $\x_{\left[n\right]-I}\mid\x_{I}=x_{I}$
is $\delta$-dense.
\end{proposition}

\ref{restoring-density} allows us to restore the density of~$\x$
by fixing~$\x$ on some set of coordinates~$I$. In order to maintain
the invariant that $\x$ and~$\y$ are structured, we also need to
ensure that $\gd^{I}\left(x_{I},\y_{I}\right)=\rho_{I}$. To this
end, we condition~$\y$ on the latter event. However, this conditioning
reveals information about~$\y$, which may have two harmful effects:
\begin{itemize}
\item \textbf{Leaking:} As discussed in \ref{subsec:generalized-lifting-overview},
our analysis of the query complexity assumes that the transcript~$\pi$
reveals at most $O(C)$ bits of information. It is important not to
reveal more information than that, or otherwise our query complexity
may increase arbitrarily. On average, we expect that conditioning
on the event $\gd^{I}\left(x_{I},\y_{I}\right)=\rho_{I}$ would reveal
only $\left|I\right|$~bits of information, which is sufficiently
small for our purposes. However, there could be values of $x_{I}$
and $\rho_{I}$ for which much more information is leaked. In this
case, we say that the conditioning is \emph{leaking}.
\item \textbf{Sparsifying:} Even if the conditioning reveals only $\left|I\right|$
bits of information about~$\y$, this could still ruin the density
of~$\y$ if the set~$I$ is large. In this case, we say that the
conditioning is \emph{sparsifying}.
\end{itemize}
We refer to values of~$\x$ that may lead to those effects as \emph{dangerous},
and define them as follows.
\begin{definition}[\cite{CFKMP19}]
\label{dangerous-value}Let $n\in\N$ and let $\y$ be a random variable
taking values from~$\D^{n}$. We say that a value~$x\in\Lambda^{n}$
is \emph{leaking} if there exists a set $I\subseteq\left[\ell\right]$
and an assignment $z_{I}\in\B^{I}$ such that 
\[
\Pr\left[\gd^{I}(x_{I},\y_{I})=z_{I}\right]<2^{-\left|I\right|-1}.
\]
Let $\delta,\varepsilon>0$, and suppose that $\y$ is $\delta$-dense.
We say that a value $x\in\D^{n}$ is \emph{$\varepsilon$-sparsifying}
if there exists a set $I\subseteq\left[n\right]$ and an assignment
$z_{I}\in\B^{I}$ such that the random variable
\[
\y_{\left[n\right]-I}\mid\gd^{I}(x_{I},\y_{I})=z_{I}
\]
is not $(\delta-\varepsilon)$-dense. We say that a value $x\in\D^{n}$
is $\varepsilon$\emph{-dangerous} if it is either leaking or $\varepsilon$-sparsifying.
\end{definition}

Chattopadhyay et al. \cite{CFKMP19} deal with this issue by upper
bounding the probability of dangerous values:
\begin{lemma}[{special case of \cite[Lemma 3.9]{CFKMP19}}]
\label{dangerous-values-lemma}There exists a universal constant~$h$
such that the following holds: Let $0<\gamma,\varepsilon,\tau\le1$
be such that $\tau\ge2+\frac{h}{c\cdot\varepsilon}-\eta$ and $\varepsilon\ge\frac{4}{t}$,
and let $\x,\y$ be $(\rho,\tau)$-structured random variables. Then,
the probability that $\x_{\free(\rho)}$ takes a value that is $\varepsilon$-dangerous
for~$\y_{\free(\rho)}$ is at most~$\frac{1}{2}$.
\end{lemma}

\subsection{\label{subsec:construction}The construction of the decision tree~$T$}

Let $h$ be the maximum among the universal constants of \ref{generalized-lifting-full-support}
and \ref{dangerous-values-lemma}, and let $\varepsilon\eqdef\frac{2h}{c\cdot\eta}$,
$\delta\eqdef1-\frac{\eta}{4}+\frac{\varepsilon}{2}$, and $\tau\eqdef2\cdot\delta-\varepsilon$.
The tree~$T$ constructs a transcript~$\pi$ by simulating the protocol~$\Pi$
round-by-round, each time adding a single message to~$\pi$. Throughout
the simulation, the tree maintains two independent random variables
$\x$ and~$\y$ that are distributed over $\cX_{\pi}$ and~$\cY_{\pi}$
respectively. The tree will maintain the invariant that $\x$ and~$\y$
are $(\rho,\tau)$-structured, where $\rho$ is a restriction that
keeps track of the queries the tree has made to~$z$ so far. In fact,
the tree will maintain a more specific invariant: whenever it is Alice's
turn to speak, $\x_{\free(\rho)}$ is $(\delta-\varepsilon)$-dense
and $\y_{\free(\rho)}$ is $\delta$-dense, and whenever it is Bob's
turn to speak, the roles of $\x$ and~$\y$ are reversed.

When the tree~$T$ starts the simulation, it sets the transcript
$\pi$ to be the empty string, the restriction $\rho$ to~$\left\{ *\right\} ^{\ell}$,
and the variables $\x$ and $\y$ to be uniformly distributed over
$\cX$ and~$\cY$ respectively. We first note that at this point,
$\x$ and~$\y$ are both $\delta$-dense, and thus satisfy the invariant.
Indeed, let $I\subseteq\left[\ell\right]$ be any set of coordinates.
We show that $\Hm(\x_{I})\ge\delta\cdot t\cdot\left|I\right|$, and
the proof for~$\y$ is analogous. Recall that by \ref{degree-to-min-entropy},
the logarithm of average degree is a lower bound on min-entropy. Thus,
the assumed lower bound on the relative average degrees of~$\cX$
implies that
\begin{align}
\Hm(\x_{I}) & \ge t\cdot\left|I\right|-\log\frac{1}{\rAD_{I}(\cX)}\label{eq:generalized-lifting-simulation-initial-density}\\
 & \ge\left(t-d\log\ell-\log d\right)\cdot\left|I\right|\nonumber \\
 & =\left(1-\frac{d\log\ell}{t}-\frac{\log d}{t}\right)\cdot t\cdot\left|I\right|\nonumber \\
 & \ge\left(1-\frac{d+\log d}{c}\right)\cdot t\cdot\left|I\right| & \text{(\ensuremath{t\ge c\cdot\log\ell}).}\nonumber 
\end{align}
Since $c$~can be chosen to be arbitrary large, and may depend on~$d$
and~$\eta$, we can ensure that the last expression is at least $\delta\cdot t\cdot\left|I\right|$,
as required. We now explain how $T$ simulates a single round of the
protocol while maintaining the invariant. Suppose that the invariant
holds at the beginning of the current round, and assume without loss
of generality that it is Alice's turn to speak. The tree~$T$ performs
the following steps:
\begin{enumerate}
\item \label{enu:simulation-dangerous-values}The tree conditions~$\x_{\free(\rho)}$
on not taking a value that is $\varepsilon$-dangerous for $\y_{\free(\rho)}$.
\item \label{enu:simulation-choosing-message}The tree~$T$ chooses an
arbitrary message~$M$ of Alice with the following property: the
probability of Alice sending $M$ on input~$\x$ is at least $2^{-\left|M\right|}$
(the existence of~$M$ will be justified soon). The tree adds~$M$
to the transcript~$\pi$, and conditions $\x$ on the event of sending~$M$.
\item \label{enu:simulation-taking-heavy-value}Let $I\subseteq\free(\rho)$
be a maximal set that violates the $\delta$-density of~$\x_{\free(\rho)}$
(i.e., $H_{\infty}(\x_{I})<\delta\cdot t\cdot\left|I\right|$), and
let $x_{I}\in\D^{I}$ be a value that satisfies $\Pr\left[\x_{I}=x_{I}\right]>2^{-\delta\cdot t\cdot\left|I\right|}$.
The tree conditions $\x$~on $\x_{I}=x_{I}$. By \ref{restoring-density},
the variable~$\x_{\free(\rho)-I}$ is now $\delta$-dense.
\item \label{enu:simulation-query}The tree queries~$z_{I}$, and sets
$\rho_{I}=z_{I}$.
\item \label{enu:simulation-consistent-with-queries}The tree conditions
$\y$~on $\gd^{I}(x_{I},\y_{I})=\rho_{I}$. Due to Step~\ref{enu:simulation-dangerous-values},
the variable~$\x_{\free(\rho)}$ must take a value that is not $\varepsilon$-dangerous,
and therefore $\y_{\free(\rho)}$ is necessarily $(\delta-\varepsilon)$-dense.
\end{enumerate}
After those steps take place, it is Bob's turn to speak, and indeed,
$\x_{\free(\rho)}$ and $\y_{\free(\rho)}$ are $\delta$-dense and
$(\delta-\varepsilon)$-dense respectively. Thus, the invariant is
maintained. In order for the foregoing steps to be well-defined, it
remains to explain three points:
\begin{itemize}
\item First, we should explain why Step~\ref{enu:simulation-dangerous-values}
conditions $\x$ on an event with a non-zero probability. To this
end, we note that $\tau$ is larger than $2+\frac{h}{c\cdot\varepsilon}-\eta$
(see \ref{eq:generalized-lifting-query-complexity-bound-on-tau} below
for a detailed calculation). Hence, by \ref{dangerous-values-lemma},
the variable $\x_{\free(\rho)}$ has a non-zero probability of taking
a value that is not $\varepsilon$-dangerous for~$\y_{\free(\rho)}$.
\item Second, we should explain why the message~$M$ in Step~\ref{enu:simulation-choosing-message}
exists. To see why, observe that the set of Alice's possible messages
forms a prefix-free code --- otherwise, Bob would not be able to
tell when Alice finished speaking and his turn starts. Hence, by \ref{simplified-kraft},
it follows that there exists a message~$M$ with probability at least~$2^{-\left|M\right|}$.
\item Third, we should explain why Step~\ref{enu:simulation-consistent-with-queries}
conditions $\y$ on an event with a non-zero probability. To this
end, recall that $\x$ must take a value that is not $\varepsilon$-dangerous
for~$\y$, and in particular, the value of~$\x$ is necessarily
not leaking. This means that the string $\gd^{I}(x_{I},\y_{I})$ has
a non-zero probability of being equal to $\rho_{I}$.
\end{itemize}
Finally, when the protocol halts, the tree~$T$ outputs the solution
of the transcript~$\pi$. We claim that this solution is a correct
solution for~$z$. Indeed, recall that since $\x$ and~$\y$ are
consistent with~$\pi$, the transcript~$\pi$ outputs a solution
for $S\d\gd$ that is correct for every pair $(x,y)$ in the support
of~$(\x,\y)$. Thus, it suffices to show that there exists some pair
$(x,y)$ in the support of~$(\x,\y)$ such that $G(x,y)=z$. In other
words, it suffices to show that $\Pr\left[G(\x,\y)=z\right]>0$.

Since $\x$ and~$\y$ are $(\rho,\tau)$-structured and $\rho$ is
consistent with~$z$, it holds that $\gd^{\fix(\rho)}(\x_{\fix(\rho)},\y_{\fix(\rho)})=z_{\fix(\rho)}$
with probability~$1$. It remains to deal with the free coordinates
of~$\rho$. To this end, we note that $\tau$ is larger than $2+\frac{h}{c\cdot\varepsilon}-\eta$
(see \ref{eq:generalized-lifting-query-complexity-bound-on-tau} below
for a detailed calculation). Hence, \ref{generalized-lifting-full-support}
implies that $z_{\free(\rho)}$ is in the support of~$\gd^{\free(\rho)}(\x_{\free(\rho)},\y_{\free(\rho)})$.
It follows that $\Pr\left[G(\x,\y)=z\right]>0$, as required.

\subsection{\label{subsec:generalized-lifting-query-complexity}The query complexity
of~$T$}

Let $z$~be an arbitrary input for~$T$, and let $q$ be the number
of queries that $T$~makes on input~$z$. We show that for some
constant~$\kappa$ that depends only on $\eta$ and~$d$, the number
of bits~$C$ that are transmitted by the protocol~$\Pi$ is at least~$\kappa\cdot q\cdot t$,
and this will conclude the proof of the lifting theorem. To this end,
we will prove that when the tree~$T$ halts, 
\begin{equation}
\Hm(\x_{\free(\rho)})+\Hm(\y_{\free(\rho)})\ge\log\left|\cX\right|+\log\left|\cY\right|-3\cdot C-\left(1+\delta+\frac{1}{c}\right)\cdot t\cdot q.\label{eq:generalized-lifting-query-complexity-potential-argument}
\end{equation}
We first show that \ref{eq:generalized-lifting-query-complexity-potential-argument}
implies the desired bound on~$C$. To see why, observe that by \ref{min-entropy-upper-bound}
it holds that 
\[
\Hm(\x_{\free(\rho)})+\Hm(\y_{\free(\rho)})\le\log\left|\cX_{\free(\rho)}\right|+\log\left|\cY_{\free(\rho)}\right|.
\]
By combining the two bounds, it follows that
\begin{align}
3\cdot C & \ge\log\left|\cX\right|+\log\left|\cY\right|-\log\left|\cX_{\free(\rho)}\right|-\log\left|\cY_{\free(\rho)}\right|-\left(1+\delta+\frac{1}{c}\right)\cdot t\cdot q\nonumber \\
 & =\log\frac{\left|\cX\right|}{\left|\cX_{\free(\rho)}\right|}+\log\frac{\left|\cY\right|}{\left|\cY_{\free(\rho)}\right|}-(1+\delta+\frac{1}{c})\cdot t\cdot q\nonumber \\
 & =\log\AD_{\fix(\rho)}(\cX)+\log\AD_{\fix(\rho)}(\cY)-\left(1+\delta+\frac{1}{c}\right)\cdot t\cdot q\label{eq:generalized-lifting-query-complexity-cc-bound}
\end{align}
Next, using our assumed lower bound on the relative average degrees
and noting that $q=\left|\fix(\rho)\right|$, we obtain that
\begin{equation}
\AD_{\fix(\rho)}(\cX)=2^{t\cdot q}\cdot\rAD_{\fix(\rho)}(\cX)\ge\left(\frac{2^{t}}{d\cdot\ell^{d}}\right)^{q}\label{eq:generalized-lifting-query-complexity-avg-deg-bound}
\end{equation}
and the same lower bound holds for $\AD_{\fix(\rho)}(\cY)$. By combining
\ref{eq:generalized-lifting-query-complexity-cc-bound,eq:generalized-lifting-query-complexity-avg-deg-bound},
it follows that
\begin{align*}
3\cdot C & \ge2\cdot\left(t-d\cdot\log\ell-\log d\right)\cdot q-\left(1+\delta+\frac{1}{c}\right)\cdot t\cdot q\\
 & \ge2\cdot\left(1-\frac{d+\log d}{c}\right)\cdot t\cdot q-\left(1+\delta+\frac{1}{c}\right)\cdot t\cdot q & (t\ge c\cdot\log\ell)\\
 & =\left(1-\delta-\frac{2d+2\log d+1}{c}\right)\cdot t\cdot q\\
 & =\left(\frac{\eta}{4}-\frac{h}{2\cdot c\cdot\eta}-\frac{2d+2\log d+1}{c}\right)\cdot t\cdot q & \text{\ensuremath{\left(\ensuremath{\delta\eqdef1-\frac{\eta}{4}+\frac{h}{2\cdot c\cdot\eta}}\right)}.}
\end{align*}
We now choose $\kappa\eqdef\frac{\eta}{4}-\frac{h}{2\cdot c\cdot\eta}-\frac{2d+2\log d+1}{c}$
and observe that we can make sure that $\kappa>0$ by choosing $c$
to be sufficiently large. 

It remains to prove \ref{eq:generalized-lifting-query-complexity-potential-argument}.
Observe that when the tree starts the simulation,  $\free(\rho)=\left[\ell\right]$
and $\x,\y$ are uniformly distributed over $\cX,\cY$ respectively,
and hence
\[
\Hm(\x_{\free(\rho)})+\Hm(\y_{\free(\rho)})=\log\left|\cX\right|+\log\left|\cY\right|.
\]
We will show that in every round of the simulation, the sum $\Hm(\x_{\free(\rho)})+\Hm(\y_{\free(\rho)})$
decreases by at most $3\cdot\left|M\right|+(1+\delta+\frac{1}{c})\cdot t\cdot\left|I\right|$,
where $M$ is the message sent and $I$ is the set of queries made
at that round. Since the sum of the lengths of all the messages~$M$
is at most~$C$, and the sum of the sizes of all sets~$I$ is~$q$,
this will imply \ref{eq:generalized-lifting-query-complexity-potential-argument}.

Fix a round of the simulation, let $M$ and~$I$ as above, and assume
without loss of generality that the message is sent by Alice. We analyze
the effect on $\Hm(\x_{\free(\rho)})+\Hm(\y_{\free(\rho)})$ of each
of the steps individually:
\begin{itemize}
\item In Step~\ref{enu:simulation-dangerous-values}, the tree conditions~$\x_{\free(\rho)}$
on taking values that are not $\varepsilon$-dangerous for~$\y_{\free(\rho)}$.
We show that this step decreases $\Hm(\x_{\free(\rho)})$ by at most
one bit. Recall that at this point $\x$ and~$\y$ are $(\rho,\tau)$-structured,
where
\begin{align}
\tau & \eqdef2\cdot\delta-\varepsilon\label{eq:generalized-lifting-query-complexity-bound-on-tau}\\
 & =2\cdot\left(1-\frac{\eta}{4}+\frac{\varepsilon}{2}\right)-\varepsilon & \text{(by definition of\,\ensuremath{\delta})}\nonumber \\
 & =2-\frac{\eta}{2}\nonumber \\
 & =2+\frac{\eta}{2}-\eta\nonumber \\
 & =2+\frac{h}{c\cdot\varepsilon}-\eta & \text{(since \ensuremath{\varepsilon\eqdef\frac{2h}{c\cdot\eta}})}.\nonumber 
\end{align}
Therefore, by applying \ref{dangerous-values-lemma}, it follows that
the probability that $\x_{\free(\rho)}$ is $\varepsilon$-dangerous
is at most~$\frac{1}{2}$. By \ref{min-entropy-conditioning}, conditioning
on that event decreases $\Hm(\x_{\free(\rho)})$ by at most one bit.
\item In Step~\ref{enu:simulation-choosing-message}, the tree conditions~$\x$
on the event of sending the message~$M$, which has probability at
least~$2^{-\left|M\right|}$. By \ref{min-entropy-conditioning},
this decreases $\Hm(\x_{\free(\rho)})$ by at most $\left|M\right|$
bits.
\item In Step \ref{enu:simulation-taking-heavy-value}, the tree conditions
on $\x$ on the event $\x_{I}=x_{I}$, which has probability greater
than~$2^{-\delta\cdot t\cdot\left|I\right|}$. By \ref{min-entropy-conditioning},
this decreases $\Hm(\x_{\free(\rho)})$ by at most $\delta\cdot t\cdot\left|I\right|$
bits.
\item In Step \ref{enu:simulation-query}, the tree removes $I$ from $\free(\rho)$.
By \ref{min-entrpoy-projecting}, this removal decreases $\Hm(\y_{\free(\rho)})$
by at most $t\cdot\left|I\right|$~bits. Moreover, this removal does
not affect $\Hm(\x_{\free(\rho)})$, since at this point $\x_{I}$
is fixed. 
\item Finally, in Step~\ref{enu:simulation-consistent-with-queries}, the
tree conditions~$\y$ on the event $\gd^{I}(x_{I},\y_{I})=\rho_{I}$.
Due to Step~\ref{enu:simulation-dangerous-values}, the value $x_{I}$
is not dangerous and hence not leaking, so the latter event has probability
at least $2^{-\left|I\right|-1}$. It follows that this conditioning
decreasing $\Hm(\y_{\free(\rho)})$ by at most $\left|I\right|+1$
bits.
\end{itemize}
Summing up, in this round the sum $\Hm(\x_{\free(\rho)})+\Hm(\y_{\free(\rho)})$
decreases by at most
\begin{align*}
 & 1+\left|M\right|+\delta\cdot t\cdot\left|I\right|+t\cdot\left|I\right|+\left|I\right|+1\\
= & \left|M\right|+\left(1+\delta+\frac{1}{t}\right)\cdot t\cdot\left|I\right|+2\\
\le & \,3\cdot\left|M\right|+\left(1+\delta+\frac{1}{t}\right)\cdot t\cdot\left|I\right|\\
\le & \,3\cdot\left|M\right|+\left(1+\delta+\frac{1}{c}\right)\cdot t\cdot\left|I\right|,
\end{align*}
as required.

\section{\label{sec:classic-functions}Composition theorems for classical
functions}

In this section, we show that our composition theorems can be applied
to three classical functions, namely: $s\text{-}t$-connectivity~\cite{KW90},
clique~\cite{GH92,RW92}, and generation~\cite{RM99}. Recall that
if $\phi$ is a CNF contradiction, we denote by $S_{\phi}$ its corresponding
search problem. We prove our results by showing that for each of the
above functions, there is an injective reduction from the lifted search
problem $S_{\phi}\d\gd$ to $\mKW_{g}$ for some appropriate formula~$\phi$
and gadget~$\gd$. Specifically, for our monotone composition theorem
we choose the gadget~$\gd$ to be the inner product mod~$2$ function~$\ip$.
For our semi-monotone composition theorem we choose the gadget to
be the equality function~$\eq$. In both cases, we denote the input
length of the gadget by~$t$.

\subsection{Preliminaries}

Following \cite{GP18,O15,R18}, we construct our reductions from $S_{\phi}\d\gd$
to $\mKW_{g}$ in two steps: first, we reduce $S_{\phi}\d\gd$ to
the monotone KW relation $\mKW_{\csat}$ for a certain constraint
satisfaction problem $\csat$, and then we reduce the latter relation
to $\mKW_{g}$. We now define the constraint satisfaction problem
and the related notions.
\begin{definition}
Let $H=(L\cup R,E)$ be a bipartite graph, and let $\D$ be a finite
alphabet. For every vertex $r\in R$, we denote by $N(r)\subseteq L$
the set of neighbors of~$r$. The constraint satisfaction problem
$\csat_{H,\D}$ is the following decision problem: The input consists
of a set of predicates $P_{r}:\D^{N(r)}\to\B$ for every $r\in R$.
The answer on an input is ``yes'' if and only if there exists an
assignment $\alpha:L\to\D$ that satisfies all the predicates.
\end{definition}

\begin{definition}
Let $\phi$~be a CNF formula. The \emph{graph of}~$\phi$, denoted~$\gr(\phi)$,
is the bipartite graph whose left and right vertices are the variables
and clauses of~$\phi$ respectively, and whose edges connect each
clause with its variables.
\end{definition}

We reduce $S_{\phi}\d\gd$ to $\csat$ using the following generic
technique, due to \cite{RM99,GP18,O15} (see also \cite[Sec. 6.1]{R18}).
We note that the ``moreover'' part in the following theorem is implicit
in those works, and that its condition is satisfied by the gadgets
that we use.
\begin{theorem}
\label{reduction-to-csp-sat}For every CNF contradiction~$\phi$
and gadget function~$\gd:\cX\times\cY\to\B$, the lifted search problem
$S_{\phi}\d\gd$ reduces to the monotone KW relation of $\csat_{\gr(\phi),\cX}$.
Moreover, the reduction is injective if for every $y\in\cY$, the
function $\gd(\cdot,y):\cX\to\B$ is non-constant and determines~$y$.
\end{theorem}

\noindent In order to reduce $\mKW_{\csat}$ to $\mKW_{g}$, we reduce
the function $\csat$ to~$g$ using the following special type of
reduction.
\begin{definition}
We say that a function $\rho:\B^{n_{1}}\to\B^{n_{2}}$ is a \emph{monotone
projection} if for every $j\in\left[n_{2}\right]$, it either holds
that the $j$-th output is a constant (i.e., always~$0$ or always~$1$),
or there exists an input coordinate $i\in\left[n_{1}\right]$ such
that for every $x\in\B^{n_{1}}$ it holds that $\rho(x)_{j}=x_{i}$.
Given two monotone functions $g_{1}:\B^{n_{1}}\to\B$ and $g_{2}:\B^{n_{2}}\to\B$,
we say that there is \emph{monotone projection} from~$g_{1}$ to
$g_{2}$ if $g_{1}=g_{2}\circ\rho$ for some monotone projection $\rho:\B^{n_{1}}\to\B^{n_{2}}$.
\end{definition}

It is not hard to see that if there is a monotone projection from
$g_{1}$ to~$g_{2}$, then there is an injective reduction from $\mKW_{g_{1}}$
to $\mKW_{g_{2}}$ (we assume here that $g_{1}$ depends on all its
input bits, which is the case for all the functions we consider).
Finally, we will use the following fact to lower bound the query complexity
of search problems.
\begin{fact}[{see, e.g., \cite[Appx. C]{RMNPRV20}}]
\label{ns-vs-query-complexity}Let $\phi$ be a CNF contradiction.
Then $\Q(S_{\phi})\ge\NS(\phi)$.
\end{fact}

\subsection{The $s\text{-}t$-connectivity function}

The $s\text{-}t$-connectivity function $\stc$ takes as input the
adjacency matrix of a directed graph over $n$~vertices with two
distinguished vertices~$s,t$, and outputs whether $s$ and~$t$
are connected in the graph. Karchmer and Wigderson~\cite{KW90} proved
that $\C(\mKW_{\stc})=\Theta(\log^{2}n)$ for the case of undirected
graphs, and alternative proofs were given by \cite{GS91,P17,R18}
for the case of directed graphs.

Below, we apply our main results to derive composition theorems with
the inner function being $\stc$. Following \cite{R18}, we do this
using the induction principle of~\cite{BP98}, which is the CNF contradiction
defined as follows:
\[
\ind(z_{1},\ldots,z_{\ell})\eqdef z_{1}\wedge(\neg z_{1}\vee z_{2})\wedge(\neg z_{2}\vee z_{3})\wedge\ldots\wedge(\neg z_{\ell-1}\vee z_{\ell})\wedge\neg z_{\ell}.
\]
Buss and Pitassi~\cite{BP98} showed that $\NS(\ind)=\Theta(\log\ell)$.
We now reduce $S_{\ind}\d\gd$ to $\mKW_{\stc}$ by constructing a
monotone projection from $\csat_{\gr(\ind),\D}$ to $\stc$.
\begin{proposition}
\label{projection-stconn}For every $\ell\in\N$ and every finite
set~$\D$, there is a monotone projection from $\csat_{\gr(\ind),\D}$
to $\stc$ for $n=\ell\cdot\left|\D\right|+2$.
\end{proposition}

\begin{myproof}
We construct a projection that maps an input of $\csat_{\gr(\ind),\D}$
to an input of $\stc$. The input of $\stc$ is a layered graph~$G$
that has $\ell+2$ layers. The first layer contains only the distinguished
vertex~$s$, and the last layer contains only the distinguished vertex~$t$.
Each of the $\ell$~middle layers consists of $\left|\D\right|$~vertices,
which we label with the elements of~$\D$.

The edges of~$G$ are determined by the input of $\csat_{\gr(\ind),\D}$
as follows. Recall that the input to $\csat_{\gr(\ind),\D}$ consists
of the following predicates: a predicate $P_{z_{1}}:\D\to\B$, predicates
of the form $P_{\neg z_{i}\vee z_{i+1}}:\D^{2}\to\B$ for every $i\in\left[\ell-1\right]$,
and a predicate $P_{\neg z_{\ell}}:\D\to\B$. Now,
\begin{itemize}
\item For every vertex $v\in\D$ of the second layer, we include the edge
$(s,v)$ in~$G$ if and only if $P_{z_{1}}(v)=1$.
\item For every vertex $v\in\D$ of the second-to-last layer, we include
the edge $(v,t)$ in~$G$ if and only if $P_{\neg z_{\ell}}(v)=1$.
\item For every two middle layers~$i$ and~$i+1$, we include the edge
between a vertex $u\in\D$ of the layer~$i$ and a vertex~$v\in\D$
of the layer $i+1$ if and only if $P_{\neg z_{i}\vee z_{i+1}}(u,v)=1$.
\end{itemize}
It can be verified that this mapping from inputs of $\csat_{\gr(\ind),\D}$
to inputs of $\stc$ is a monotone projection. To see that it maps
``yes'' inputs of $\csat_{\gr(\ind),\D}$ to ``yes'' instances
of $\stc$ and vice versa, observe that every satisfying assignment
for the input of~$\csat_{\gr(\ind),\D}$ specifies a path from~$s$
to~$t$ in~$G$ and vice versa.
\end{myproof}
We now apply our main results to obtain monotone and semi-monotone
composition theorems with the inner function being $g=\stc$.
\begin{theorem}
For every non-constant monotone function~$f:\B^{m}\to\B$ and every
sufficiently large~$n\in\N$ it holds that 
\begin{align*}
\log\L(\mKW_{f}\d\mKW_{\stc}) & =\log\L(\mKW_{f})+\Omega(\log\L(\mKW_{\stc}))\\
\C(U_{m}\d\mKW_{\stc}) & \ge m+\Omega\left(\C(\mKW_{\stc})\right).
\end{align*}
\end{theorem}

\begin{myproof}
Let $f:\B^{m}\to\B$ be a non-constant monotone function, and let
$c$ be the maximum between $2$ and the constant obtained from our
monotone composition theorem for $\eta=\frac{1}{2}$ (since the discrepancy
of $\ip$ is $2^{-\frac{1}{2}t}$). We first show that the theorem
holds for an input length~$n$ of the form~$n=\ell\cdot2^{t}+2$,
where $\ell$ is a natural number such that $\ell\ge m$, and $t\eqdef\left\lceil c\cdot\log(m\cdot\ell)\right\rceil $.
We will then show how to derive the theorem for every sufficiently
large~$n\in\N$ using padding.

Let $\ell\in\N$ be such that $\ell\ge m$, let $t\eqdef\left\lceil c\cdot\log(m\cdot\ell)\right\rceil $,
and let $n\eqdef\ell\cdot2^{t}+2$. By combining \ref{projection-stconn}
with \ref{reduction-to-csp-sat}, we obtain injective reductions from
$S_{\ind}\d\ip$ and $S_{\ind}\d\eq$ to $\mKW_{\stc}$. By the aforementioned
result of~\cite{BP98} it holds that $\NS(\ind)=\Theta(\log\ell)$,
and this implies that $\Q(\ind)\ge\Omega(\log\ell)$ by \ref{ns-vs-query-complexity}.
It now follows by our monotone composition theorem that
\begin{align*}
\log\L(\mKW_{f}\d\mKW_{\stc}) & \ge\log\L(\mKW_{f})+\Omega(\Q(\ind)\cdot t)\\
 & =\log\L(\mKW_{f})+\Omega\left(\log\ell\cdot\log(m\cdot\ell)\right)\\
 & =\log\L(\mKW_{f})+\Omega\left(\log^{2}n\right)\\
 & =\log\L(\mKW_{f})+\Omega\left(\log\L(\mKW_{\stc})\right).
\end{align*}
Similarly, our semi-monotone composition theorem implies that
\begin{align*}
\C(U_{m}\d\mKW_{\stc}) & \ge m+\Omega(\NS(\ind)\cdot t)\\
 & =m+\Omega\left(\C(\mKW_{\stc})\right),
\end{align*}
as required.

We turn to prove the theorem for a general value of~$n$. Let $n\in\N$
be a sufficiently large number, and let $n'\in\N$ be the largest
number in~$\left[n\right]$ of the form $n'=\ell\cdot2^{t}+2$ (where
$\ell$ and~$t$ are as above). Next, observe that there is a monotone
projection from $\textsc{stConn}_{n'}$ to $\stc$: in order to get
an instance of $\stc$ from an instance of $\textsc{stConn}_{n'}$,
just add $(n-n')$~isolated vertices to the graph. Therefore, by
the above proof, it holds that
\begin{align*}
\log\L(\mKW_{f}\d\mKW_{\stc}) & \ge\log\L(\mKW_{f}\d\mKW_{\textsc{stConn}_{n'}})\\
 & =\log\L(\mKW_{f})+\Omega\left(\log^{2}n'\right).
\end{align*}
Finally, it is not hard to show that $\frac{n'}{n}\ge\frac{1}{4}$
(for a sufficiently large~$n$), and therefore
\begin{align*}
\log\L(\mKW_{f}\d\mKW_{\stc}) & \ge\log\L(\mKW_{f})+\Omega\left(\log^{2}(\frac{n}{4})\right)\\
 & =\log\L(\mKW_{f})+\Omega\left(\log^{2}n\right)\\
 & =\log\L(\mKW_{f})+\Omega\left(\log\L(\mKW_{\stc})\right),
\end{align*}
as required. The lower bound for the semi-monotone version can be
proved similarly.
\end{myproof}

\subsection{The clique function}

We denote by $\clq$ the function that takes as an input the adjacency
matrix of an $n$-vertex graph and outputs whether it contains a $k$-clique.
Observe that for every $k,n\in\N$ it holds that $\C(\mKW_{\clq})\le O(k\log n)$,
which is witnessed by the circuit that checks all $\binom{n}{k}$~potential
cliques by brute force. Goldmann and H{\aa}stad~\cite{GH92} proved
that $\C(\mKW_{\clq})\ge\Omega(\sqrt{k})$ for every $k\le\left(n/2\right)^{2/3}$,
and Raz and Wigderson~\cite{RW92} improved this bound to $\C(\mKW_{\clq})=\Omega(k)$
that for every $k\le\frac{2}{3}n+1$. In what follows, we apply our
main results to obtain corresponding compositions theorems with the
inner function being~$g=\clq$ for $k=2^{O(\sqrt{\log n})}$.

To this end, we choose our CNF contradiction to be the bitwise pigeonhole
principle, defined as follows: For $d\in\N$, the \emph{bitwise pigeonhole
principle }$\php$ is a $2(d-1)$-CNF contradiction over $\ell\eqdef2^{d}\cdot(d-1)$
variables. The variables are partitioned into $2^{d}$ blocks of $(d-1)$~variables
each, and we view each block as encoding a number in~$\left[2^{d-1}\right]$.
The formula $\php$ contains $\binom{2^{d}}{2}$~constraints that
check that every two blocks encode different numbers. Informally,
this formula encodes the statement that $2^{d}$ pigeons can not be
mapped injectively into $2^{d-1}$~pigeonholes.

Razborov~\cite{R98_polynomial_calculus} proved that the Nullstellensatz
degree of the standard pigeonhole principle with $2^{d-1}$~holes
is at least $\Omega(2^{d})$. Using a reduction from de Rezende et
al.~\cite{RGNPRS21}, this implies that $\NS(\php)\ge\Omega(2^{d}/d)=\Omega(\ell/\log^{2}\ell)$.
We have the following monotone projection from $\csat_{\gr(\php),\D}$
to $\clq$.
\begin{proposition}
\label{projection-clique}For every $d\in\N$ and every finite set~$\D$,
there is a monotone projection from $\csat_{\gr(\php),\D}$ to $\clq$
for $n=2^{d}\cdot\left|\D\right|^{d-1}$ and $k=2^{d}$.
\end{proposition}

\begin{myproof}
We construct a monotone projection that maps an input of $\csat_{\gr(\php),\D}$
to an input of $\clq$. The input of $\clq$ is a graph~$G$ that
consists of $2^{d}$~classes of $\left|\D\right|^{d-1}$~vertices
each. Within each class, we label the vertices with strings in $\D^{d-1}$.
As defined next, all the edges of~$G$ connect different classes,
so a clique contains at most one vertex from each class.

The edges between the classes are determined by the input of $\csat_{\gr(\php),\D}$
as follows. Recall that an input of $\csat_{\gr(\php),\D}$ is a constraint
satisfaction problem over of $2^{d}\cdot(d-1)$~variables, which
are partitioned to $2^{d}$~blocks of $(d-1)$~variables each. Moreover,
the input to $\csat_{\gr(\php),\D}$ consists, for every two distinct
blocks~$i,j$, of a predicate~$P_{i,j}:\D^{d-1}\times\D^{d-1}\to\B$.
Now, for every distinct $i,j$, we include in~$G$ an edge between
a vertex $u\in\D$ of the $i$-th class and a vertex $v\in\D$ of
$j$-th class if and only if $P_{i,j}(u,v)=1$.

It can be verified that this mapping from inputs of $\csat_{\gr(\php),\D}$
to inputs of $\text{\ensuremath{\clq}}$ is a monotone projection.
To see that it maps ``yes'' inputs of $\csat_{\gr(\php),\D}$ to
``yes'' instances of $\clq$ and vice versa, observe that every
satisfying assignment of the input of~$\csat_{\gr(\php),\D}$ specifies
a clique of size $2^{d}$ in~$G$ and vice versa.
\end{myproof}
\begin{remark}
There is a minor technical subtlety that we ignored in the foregoing
proof. Recall that we defined $\php$ as a CNF formula that contains
$\binom{2^{d}}{2}$~constraints, one for each pair blocks. Hence,
in our description of $\csat_{\gr(\php),\D}$, we assumed that there
is a single predicate~$P_{i,j}$ for each constraint. However, those
constraints are not clauses: rather, each constraint can be implemented
using $2^{2(d-1)}$ clauses. Therefore, if we stick to the formal
definition of~$\csat$, an input to~$\csat_{\gr(\php),\D}$ should
contain $2^{2(d-1)}$ predicates for each constraint. However, since
all these predicates are over the same variables, they can be replaced
with a single predicate without changing the output of~$\csat$.
\end{remark}

We now apply our main results to obtain a monotone and semi-monotone
composition theorems with the inner function being $g=\clq$.
\begin{theorem}
\label{lb-clique}There exists a constant $\varepsilon>0$ such that
the following holds. For every non-constant monotone function~$f:\B^{m}\to\B$,
for every sufficiently large~$n\in\N$, and for $k\le2^{\varepsilon\cdot\sqrt{\log n}}$,
it holds that 
\begin{align*}
\log\L(\mKW_{f}\d\mKW_{\clq}) & =\log\L(\mKW_{f})+\Omega(k)\\
\C(U_{m}\d\mKW_{\clq}) & \ge m+\Omega(k).
\end{align*}
\end{theorem}

\begin{myproof}
Let $c$ be the maximum between $2$ and the constant obtained from
our monotone composition theorem for $\eta=\frac{1}{2}$ (since the
discrepancy of $\ip$ is $2^{-\frac{1}{2}t}$), and let $\varepsilon\eqdef\sqrt{\frac{1}{10\cdot c}}$.
Let $f:\B^{m}\to\B$ be a non-constant monotone function. We first
prove that the theorem holds when $n$ and~$k$ are of the forms~$n=2^{d+(d-1)\cdot t}$
and $k=2^{d}$. We will then show how to derive the theorem for every
sufficiently large~$n\in\N$ and every $k\le2^{\varepsilon\cdot\sqrt{\log n}}$
using padding.

Let $d\in\N$, let $\ell\eqdef2^{d}\cdot(d-1)$, and let $t\eqdef\left\lceil c\cdot\log(m\cdot\ell)\right\rceil $.
Let $n\eqdef2^{d+(d-1)\cdot t}$ and $k\eqdef2^{d}$, and observe
that it indeed holds that $k\le2^{\sqrt{\log n}}$ (since $t\ge d$
and hence $n\ge2^{d^{2}}$). By combining \ref{projection-clique}
with \ref{reduction-to-csp-sat}, we obtain injective reductions from
$S_{\php}\d\ip$ and $S_{\php}\d\eq$ to $\mKW_{\clq}$. By the aforementioned
result of~\cite{R98_polynomial_calculus,RGNPRS21} it holds that
$\NS(\php)\ge\Omega(\frac{2^{d}}{d})$, and this implies that $\Q(\php)=\Omega(\frac{2^{d}}{d})$
by \ref{ns-vs-query-complexity}. It now follows by our monotone composition
theorem that
\begin{align*}
\log\L(\mKW_{f}\d\mKW_{\clq}) & \ge\log\L(\mKW_{f})+\Omega(\Q(\php)\cdot t)\\
+ & =\log\L(\mKW_{f})+\Omega\left(\frac{2^{d}}{d}\cdot\log(m\cdot\ell)\right)\\
 & \ge\log\L(\mKW_{f})+\Omega\left(2^{d}\right) & \text{(since \ensuremath{\log\ell\ge d})}\\
 & =\log\L(\mKW_{f})+\Omega(k).
\end{align*}
Using a similar calculation, our semi-monotone composition theorem
implies that
\[
\C(U_{m}\d\mKW_{\clq})\ge m+\Omega(\NS(\php)\cdot t)\ge m+\Omega(k),
\]
as required.

We turn to prove the theorem for general values of~$n$ and~$k$.
Let $n\in\N$ be a sufficiently large number such that $m\le2^{\varepsilon\cdot\sqrt{\log n}}$,
and let $k\in\N$ be such that $k\le2^{\varepsilon\cdot\sqrt{\log n}}$.
Let $d\eqdef\left\lfloor \log k\right\rfloor $, and define $\ell$
and~$t$ as above. Let $n'\eqdef2^{d+(d-1)\cdot t}$ and $k'\eqdef2^{d}$.
Observe that $k/2\le k'\le k$, and that
\begin{align*}
n' & \le2^{d\cdot(1+t)}\\
 & \le2^{d\cdot\left(2+c\cdot\log(m\cdot\ell)\right)}=2^{d\cdot\left(2+c\cdot\log m+c\cdot\log\ell\right)} & \text{(since \ensuremath{t\le c\cdot\log(m\cdot\ell)+1})}\\
 & \le2^{d\cdot\left(2+c\cdot\log m+2\cdot c\cdot d\right)}\le2^{4\cdot c\cdot d^{2}+c\cdot d\cdot\log m} & \text{(since \ensuremath{\log(\ell)\le2d})}\\
 & \le2^{5\cdot c\cdot\varepsilon^{2}\cdot\log n} & \text{(since \ensuremath{d,\log m\le\varepsilon\cdot\sqrt{\log n}})}\\
 & =\sqrt{n} & \text{(since \ensuremath{\varepsilon^{2}=\frac{1}{10\cdot c}})}.
\end{align*}

Next, observe that there is a monotone projection from $\textsc{Clique}_{n',k'}$
to $\clq$. Indeed, in order to get an instance of $\clq$ from an
instance of $\textsc{Clique}_{n',k'}$, we add $n-n'$ vertices to
the graph as follows: we first add a clique of size~$k-k'$, and
connect its vertices to all the other vertices in the graph; then,
we add another $(n-n')-(k-k')$ isolated vertices. Therefore, by the
above proof, it holds that
\begin{align*}
\log\L(\mKW_{f}\d\mKW_{\clq}) & \ge\log\L(\mKW_{f}\d\mKW_{n',k'})\\
 & =\log\L(\mKW_{f})+\Omega(k),
\end{align*}
as required. The lower bound for the semi-monotone version can be
proved similarly.
\end{myproof}

\subsection{The generation function}

Let $n\in\N$. Given a set $\cT\subseteq\left[n\right]^{3}$, we say
that $\cT$ \emph{generates} a point $w\in\left[n\right]$ if $w=1$,
or if there is a triplet $(u,v,w)\in\cT$ such that $\cT$ generates
$u$ and~$v$. The generation function~$\gen$ takes as an input
a set~$\cT\subseteq\left[n\right]^{3}$ and says whether $\cT$ generates~$n$
or not. This function was introduced by Raz and McKenzie~\cite{RM99}
in order to separate the monotone $\NC$ hierarchy.

Raz and McKenzie~\cite{RM99} showed that $\C(\mKW_{\gen})=\Omega(n^{\varepsilon})$
for some constant $\varepsilon>0$ by using their lifting theorem
for query complexity. Specifically, they considered a certain $3$-CNF
contradiction~$\peb$ (namely, the pebbling contradiction of the
pyramid graph) and reduced the lifted search problem~$S_{\peb}\d\gd$
to $\mKW_{\gen}$. Robere~\cite{R18} applied their method with the
lifting theorem for Nullstellensatz degree of~\cite{RPRC16} and
obtained a bound of $\C(\mKW_{\gen})=\Omega(n^{1/6})$. The latter
bound was subsequently improved to $\C(\mKW_{\gen})=\tilde{\Omega}(n)$
by de Rezende et al.~\cite{RMNPRV20}. Below, we use our main results
to obtain corresponding composition theorems with the inner function
being $g=\gen$.

For every $h\in\N$, the formula $\peb$ has $\ell\eqdef\frac{h(h+1)}{2}$
variables. It can be shown that $\NS(\peb)=\Theta(h)$ by combining
the results of Cook~\cite{C74} and Buresh-Oppenheim et al. \cite{BCIP02}
(see \cite[Sec 6.3]{R18} for details). We use the following result
due to Robere~\cite{R18}.
\begin{proposition}[{implicit in the proof of \cite[Thm. 6.3.3]{R18}}]
\label{projection-gen}For every $h\in\N$ and every finite set~$\D$,
there is a monotone projection from $\csat_{\gr(\peb),\D}$ to $\gen$
for $n=\ell\cdot\left|\D\right|+2$.
\end{proposition}

\begin{remark}
We note that the proof of \ref{projection-gen} in \cite{R18} only
states this claim for $\D=\left[\ell^{2}\right]$, but it actually
works for every finite set~$\D$.
\end{remark}

We now apply our main results to obtain a monotone and semi-monotone
composition theorems with the inner function being $g=\gen$ that
match the lower bounds of \cite{RM99} and \cite{R18} respectively.
\begin{theorem}
\label{lb-monotone-gen}There exists $\varepsilon>0$ such that, for
every non-constant monotone function~$f:\B^{m}\to\B$ and every sufficiently
large~$n\in\N$, it holds that 
\begin{equation}
\log\L(\mKW_{f}\d\mKW_{\gen})=\log\L(\mKW_{f})+\Omega(n^{\varepsilon}).\label{eq:monotone-KRW-for-gen}
\end{equation}
\end{theorem}

\begin{myproof}
Let $f:\B^{m}\to\B$ be a non-constant monotone function, and let
$c$ be the constant obtained from our monotone composition theorem
for $\eta=\frac{1}{2}$ (since the discrepancy of $\ip$ is $2^{-\frac{1}{2}t}$).
Let $h\in\N$ and $\ell\eqdef\frac{h(h+1)}{2}$ be such that $\ell\ge m$,
and let $t\in\N$ be such that $t=\left\lceil c\cdot\log(m\cdot\ell)\right\rceil $.
We first show that \ref{eq:monotone-KRW-for-gen} holds for $n=\ell\cdot2^{t}+2=O(h^{4c+2})$,
and then we will prove it for every sufficiently large~$n\in\N$
using padding.

Let $n=\ell\cdot2^{t}+2$ where $h$, $\ell$, and~$t$ are as above.
By combining \ref{projection-gen} with \ref{reduction-to-csp-sat},
we obtain a reduction from $S_{\peb}\d\ip$ to $\mKW_{\peb}$. By
the foregoing discussion  $\NS(\peb)\ge\Omega(h)$, and this implies
that $\Q(\peb)\ge\Omega(h)$ by \ref{ns-vs-query-complexity}. It
now follows by our monotone composition theorem that
\begin{align*}
\log\L(\mKW_{f}\d\mKW_{\gen}) & \ge\log\L(\mKW_{f})+\Omega(\Q(\peb)\cdot t)\\
 & \ge\log\L(\mKW_{f})+\Omega\left(h\right)\\
 & =\log\L(\mKW_{f})+\Omega\left(n^{\frac{1}{4c+2}}\right).
\end{align*}
By choosing $\varepsilon=\frac{1}{4c+2}$, we obtain the required
result.

We turn to prove the theorem for a general value of~$n$. Let $n\in\N$
be a sufficiently large number, and let $n'\in\N$ be the largest
number in~$\left[n\right]$ of the form $n'=\ell\cdot2^{t}+2$ (where
$\ell$ and~$t$ are as above). Next, observe that there is a monotone
projection from $\textsc{Gen}_{n'}$ to $\gen$: in order to get an
instance of $\gen$ from an instance of $\textsc{Gen}_{n'}$, add
$(n-n')$~points that do not participate in any triplet, and replace
$n'$ with~$n$. Therefore, by the above proof, it holds that
\begin{align*}
\log\L(\mKW_{f}\d\mKW_{\gen}) & \ge\log\L(\mKW_{f}\d\mKW_{\textsc{Gen}_{n'}})\\
 & =\log\L(\mKW_{f})+\Omega\left((n')^{\varepsilon}\right).
\end{align*}
Finally, it is not hard to show that $\frac{n'}{n}\ge\frac{1}{4}$
(for a sufficiently large~$n$), and therefore
\begin{align*}
\log\L(\mKW_{f}\d\mKW_{\gen}) & \ge\log\L(\mKW_{f})+\Omega\left(\left(\frac{n}{4}\right)^{\varepsilon}\right)\\
 & =\log\L(\mKW_{f})+\Omega\left(n^{\varepsilon}\right),
\end{align*}
as required.
\end{myproof}
\begin{theorem}
For every $m\in\N$ and every sufficiently large~$n\in\N$ it holds
that $\C(U_{m}\d\mKW_{\gen})\ge m+\Omega\left(n^{1/6}\right)$.
\end{theorem}

\begin{myproof}
Let $m\in\N$. Let $h\in\N$, let $\ell\eqdef\frac{h(h+1)}{2}$ and
let $t=\left\lceil 2\log\ell\right\rceil $. We prove the theorem
for an input length~$n$ of the form $n=\ell\cdot2^{t}+2=\Theta(h^{6})$,
and this will imply the theorem for every sufficiently large~$n\in\N$
using padding as in the proof of \ref{lb-monotone-gen}. By combining
\ref{projection-gen} with \ref{reduction-to-csp-sat}, we obtain
an injective reduction from $S_{\peb}\d\eq$ to $\mKW_{\gen}$. Moreover,
by the foregoing discussion  $\NS(\peb)\ge\Omega(h)$. It now follows
by our semi-monotone composition theorem that
\begin{align*}
\C(U_{m}\d\mKW_{\gen}) & \ge m+\Omega(\NS(\peb)\cdot t)\\
 & \ge m+\Omega(h)\\
 & =m+\Omega(n^{1/6}),
\end{align*}
as required.
\end{myproof}

\section{\label{sec:Intro-Open-problems}Open questions}

An obvious question that arises from this work is whether we can strengthen
our semi-monotone composition theorem (\ref{semi-monotone-composition-thm})
to work for every non-constant outer function~$f$. As a starting
point, can we prove such a semi-monotone composition theorem that
holds when the inner function $g$ is the $s\text{-}t$-connectivity
function? We note that proving such a result would likely require
new ideas, since our techniques seem to be insufficient:
\begin{itemize}
\item On the one hand, we cannot prove such a result along the lines of
our monotone composition theorem, since in the semi-monotone setting
we cannot assume that the protocol outputs an entry $(i,j)$ for which
$a_{i}\ne b_{i}$ (as in the observation of \cite{KRW95} in the monotone
case).
\item On the other hand, we cannot prove such a result along the lines of
our semi-monotone composition theorem, since the Razborov rank measure
cannot prove interesting lower bounds for non-monotone KW relations
\cite{R92_complexity_measures}. In particular, we would not be able
to analyze the complexity of a non-monotone outer relation~$\KW_{f}$
using this technique.
\end{itemize}
Another interesting question is whether we can strengthen our monotone
composition theorem (\ref{monotone-composition-theorem}) even further:
Although this theorem holds for many choices of the inner functions~$g$,
there are still a few ``classical'' monotone functions that it does
not cover --- most notably the matching function~\cite{RW92}. Can
we prove a monotone composition theorem where $f$ can be any non-constant
monotone function, and $g$~is the matching function?

Finally, recall that, in the long run, our goal is to prove the KRW
conjecture for the composition~$\KW_{f}\d\MX$ (for every~$f$),
since this would imply that $\pnc$. To this end, it seems reasonable
to try to prove first the monotone and semi-monotone versions of this
conjecture. The monotone version might be within reach (see \cite{M20}
for the statement of this conjecture). Can we prove it?

\section*{Acknowledgment}

We would like to thank anonymous referees for providing numerous comments
that improved presentation of this manuscript. This work was partly
carried out while the authors were visiting the Simons Institute for
the Theory of Computing in association with the DIMACS/Simons Collaboration
on Lower Bounds in Computational Complexity, which is conducted with
support from the National Science Foundation. An extended abstract
of this paper has appeared as \cite{RMNPR20}.

\newcommand{\etalchar}[1]{$^{#1}$}

\end{document}